%% file: main_article.tex
\documentclass[preprint,3p,times]{elsarticle}

\usepackage{amssymb}
\usepackage{amsmath}

\journal{Journal of Computational Physics}

\newtheorem{theorem}{Theorem}[section]
\newtheorem{lemma}{Lemma}[section]
\newtheorem{assumption}{Assumption}[section]
\newdefinition{rmk}{Remark}
\newdefinition{cor}{Corollary}
\newproof{pf}{Proof}

\usepackage{natbib}
\usepackage{lineno}
\usepackage{algorithmic}
\usepackage{hyperref}
\usepackage{cleveref}
\usepackage{bm}
\usepackage{caption}
\usepackage{subcaption}
\usepackage{bbm}
\usepackage{mathtools}

\usepackage{adjustbox}
\usepackage{algorithm}

\usepackage[symbol]{footmisc}
\usepackage{enumitem}
\usepackage{xcolor}
\usepackage{transparent}
\usepackage{booktabs}
\usepackage{tabularx}

\crefname{definition}{Definition}{Definitions}
\crefname{theorem}{Theorem}{Theorems}
\crefname{lemma}{Lemma}{Lemmas}
\crefname{assumption}{Assumption}{Assumptions}

\crefname{equation}{Eq.}{Eqs.}
\crefname{figure}{Fig.}{Figs.}
\crefname{table}{Table}{Tables}
\crefname{algorithm}{Algorithm}{Algorithms}

\begin{document}

\begin{frontmatter} 


\title{Synchronized step multilevel Markov chain Monte Carlo}
\author[label2]{Sanjan C. Muchandimath\corref{cor1}}
\ead{sanjancm@umich.edu}
\author[label2]{Alex A. Gorodetsky}
\ead{goroda@umich.edu}
\cortext[cor1]{Corresponding author}
\affiliation[label2]{
  organization={University of Michigan},city={Ann Arbor},
  postcode={48105},
  state={MI},
  country={USA}
}

\begin{abstract}
  We propose SYNCE (synchronized step correlation enhancement), a new algorithm for coupling Markov chains within multilevel Markov chain Monte Carlo (ML-MCMC) estimators. We apply this algorithm to solve Bayesian inverse problems using multiple model fidelities. SYNCE is inspired by the concept of common random number coupling in Markov chain Monte Carlo sampling. Unlike state-of-the-art methods that rely on the overlap of level-wise posteriors, our approach enables effective coupling even when posteriors differ substantially. This overlap-independence generates significantly higher correlation between samples at different fidelity levels, improving variance reduction and computational efficiency in the ML-MCMC estimator. We prove that SYNCE admits a unique invariant probability measure and demonstrate that the coupled chains converge to this measure faster than existing overlap-dependent methods, particularly when models are dissimilar. Numerical experiments validate that SYNCE consistently outperforms current coupling strategies in terms of computational efficiency and scalability across varying model fidelities and problem dimensions.
\end{abstract}

\begin{keyword}
  Markov chain Monte Carlo \sep Bayesian inverse problems \sep multilevel Monte Carlo \sep multi-fidelity \sep coupling
  \MSC 62F15 \sep 62M05 \sep 65C05 \sep 65C40

\end{keyword}

\end{frontmatter}



\section{Introduction}\label{sec:intro}

Bayesian inference is a well established technique for estimating parameters of complex mathematical models.  In this stochastic framework, the parameters $\theta$ of a model $\mathcal{F}$ are modeled as random variables. By conditioning a prior distribution $\pi^{0}(\theta)$ on observed data $y_\mathcal{D}$, we obtain a posterior distribution $\pi(\theta \mid y_\mathcal{D})$ using Bayes' rule. Given this updated distribution over the parameters, application goals typically seek to estimate statistics of some output functional $Q$; for example, to compute the expectation $\mathbb{E}_{\theta \sim \pi(\theta \mid y_\mathcal{D})}[Q(\theta)].$ In non-linear or non-Gaussian settings, these estimates are typically computed via sampling methods such as Markov chain Monte Carlo (MCMC)~\cite{robert_monte_2004,hammersley1961monte}.

However, unlike forward uncertainty quantification (UQ) problems where input parameter distributions are often known and can be sampled directly, Bayesian inverse problems (BIP) pose a greater computational challenge. MCMC algorithms struggle with sampling from posterior distributions that are high-dimensional and exhibit complex structures such as multi-modality or strong correlations. These struggles result in high autocorrelation and reduced effective sample size. Consequently, achieving accurate estimates of quantities of interest requires a large number of posterior samples $N$ to overcome the slow $1/\sqrt{N}$ Monte Carlo convergence rate. This inefficiency is further exacerbated when the underlying simulations are computationally intensive -- each posterior sample requires a forward model evaluation for the likelihood, and every sample drawn requires evaluating a possibly expensive functional $Q$. This simultaneous demand for high sample counts and expensive model evaluations makes Bayesian inverse problems particularly challenging.

Multi-fidelity strategies alleviate these computational costs by combining outputs from a hierarchy of models, leveraging cheaper, less accurate models to achieve computational speedups while preserving accuracy~\cite{peherstorfer_survey_2018}. This paper seeks to improve these strategies to specifically target the high sampling costs inherent to BIPs. Implementing multi-fidelity methods requires two key components: (1) the availability of both low-fidelity and high-fidelity models, and (2) a model management strategy that distributes work/samples among these available models. Such hierarchies commonly arise in physical simulations -- for example, through varying mesh resolutions, simplified physics, or surrogate modeling techniques. Historically, multi-fidelity model management strategies have been realized through control variate (CV)~\cite{nelson1990control,hammersley1961monte,emsermann2002improving,geraci2015multifidelity}, approximate control variate (ACV)~\cite{gorodetsky_generalized_2020,pham2022ensemble,thomasgroupedACV2024}, multilevel Monte Carlo (MLMC)~\cite{heinrich2001multilevel,giles_multilevel_2015,cliffe_multilevel_2011,nobile2015multi} and multilevel best linear unbiased estimator (MLBLUE)~\cite{schaden2020multilevel} techniques. The efficiency of these estimators relies on the \textit{correlation} between the functional quantities of the low and high-fidelity models, with higher correlations leading to higher variance reduction. 

This correlation requirement introduces unique challenges in the context of inverse problems. Unlike forward problems where input samples are drawn from a known distribution, inverse problems require sampling from posterior distributions that depend on the model fidelity. Hence, in the inverse context, we have an additional opportunity to reduce the computational cost of generating samples by appealing to \textit{low-fidelity posteriors}. Rather than drawing samples from the high-fidelity posterior alone, we can mix and match samples from posteriors of different fidelities. However, this opportunity comes with some additional challenges. Since the posterior distributions change with model fidelity, we maintain correlation required for variance reduction by using \textit{coupled} Markov chains. Coupling in this context refers to the idea of creating a joint Markov kernel that induces a Markov chain on the product space of the individual chains (e.g., a low-fidelity and high-fidelity chain). A fundamental requirement of this coupling is that the marginals of the joint chain target the distributions of the individual chains. Beyond this single requirement, there is significant freedom in how the joint kernel is constructed. This flexibility allows us to design clever coupling strategies that can increase the correlation between samples drawn from different fidelity posteriors.

Several related methodologies have been proposed in the literature for constructing such couplings within multilevel MCMC (ML-MCMC) estimators. Dodwell et al.~\cite{dodwell2015hierarchical} integrate a delayed acceptance strategy within ML-MCMC by proposing a point from a low-fidelity (coarse) MCMC chain as a candidate for the high-fidelity (fine) MCMC chain. The foundational delayed acceptance idea was formalized by Christen and Fox~\cite{christen_markov_2005}, who demonstrate that the method is effective only when the low-fidelity model is a good approximation of the high-fidelity model. Lykkegaard et al.~\cite{lykkegaard_multilevel_2023} expand on the delayed acceptance proposal idea by creating ergodic chains using recursivity and leveraging an adaptive error model (AEM) to correct for dissimilar posteriors. Similarly, specialized proposals are explored in~\cite{DILIMultilevel2024}, which integrates delayed acceptance with dimension independent likelihood informed (DILI) MCMC~\cite{DILI2016}. Independent proposals are analyzed by Cianci et al.~\cite{madrigal-cianci_analysis_2023}, where chains are coerced to accept the same proposed point from the independent sampler. However, the authors provide restrictive conditions on the independent proposal to ensure ergodicity. To enable state dependent proposals, Cianci and Pablo~\cite{cianci_thesis} used the maximal coupling technique in the ML-MCMC framework. Recently, other interesting applications of coupling have been applied to create unbiased~\cite{jacob_unbiased_2020,heng_unbiased_2023} and hybrid~\cite{Juntao2025hybrid} estimators. 

While these methodologies differ in their specific implementations, they all share the same strategy: couple chains by proposing the same sample for both the chains. This approach is similar to the logic successfully applied in forward UQ problems~\cite{geraci2018leveraging,lam2020multifidelity}. In the forward setting, coupling methods rely on the idea that sharing samples between models (in a subspace) increases correlation between outputs, thereby reducing variance. However, this shared sample strategy has limitations in the inverse problem context. The effectiveness of these methods relies heavily on the overlap between the two posteriors targeted by the coupled chains -- when models are dissimilar, a sample from a low-fidelity posterior may fall in a region of low probability for the high-fidelity posterior. This mismatch leads to frequent rejections, decoupling the chains and reducing the correlation between samples. This limitation necessitates a more robust approach --- one that couples the evolution of the chains through synchronized steps rather than their states. In this context, we propose SYNCE (synchronized step correlation enhancement), a common random number (CRN) coupling algorithm that synchronizes the random steps of the coupled chains to maintain correlation even when the posterior distributions do not overlap. 

The key contributions of this paper are:
\begin{enumerate}
  \item We review existing methodologies for integrating MCMC with the MLMC estimator and provide insights into how these methods come under the framework of Markov chain coupling in~\Cref{sec:inverse-problem-estimators-couplings}. 
  \item  We introduce SYNCE coupling, inspired by~\cite{pinto2001improving}. We prove the coupled chains admit a unique invariant probability measure and establish geometric ergodicity with explicit convergence rates in~\Cref{thm:convergence-synce}. Our analysis in~\Cref{subsec: Theory} demonstrates that SYNCE achieves faster sampling convergence than existing methods when posteriors weakly overlap or exhibit mismatch.
  \item We propose an adaptive version of our algorithm, SYNCE-AR in~\Cref{subsec:covariance-adaptation,subsec:resynchronization} that works by adapting and resynchronizing the coupled Markov kernel based on insights from the theoretical analysis.
  \item We validate that our proposed methodology is more efficient than existing couplings in the literature in~\Cref{sec:experiments}, achieving gains in variance reduction by an order of two.
\end{enumerate}

\section{Background}\label{sec:background}
In this section, we first set up the Bayesian inverse problem and review how it is solved using the standard Metropolis-Hastings MCMC (MH-MCMC) algorithm. We then focus on multi-fidelity methods and introduce the multilevel Markov chain Monte Carlo (ML-MCMC) framework for variance reduction. 

\subsection{Bayesian inverse problems and MCMC}\label{subsec:bayesian-inverse-problems}

Let $(X, \mathcal{X})$ be a measurable space with $X \subseteq \mathbb{R}^d$, and define the forward operator for the BIP as $\mathcal{F} : \Theta \rightarrow \mathcal{Y}$. Here, $\Theta \in X$ denotes the input parameter space and $\mathcal{Y} \subseteq \mathbb{R}^m$ denotes the output space of our model. For a particular parameter vector $\theta \in \Theta$, we denote the forward model as $y = \mathcal{F}(\theta)$. Assuming additive noise, the observation model is expressed as $y = \mathcal{F}(\theta) + \epsilon$, where $\epsilon \sim \mathcal{N}(0, \Sigma)$ is Gaussian noise with covariance $\Sigma$. Given observations $y_{\mathcal{D}} \in \mathcal{Y}$, BIPs seek the posterior distribution of the parameter set $\theta$ denoted by $\pi(\theta)$. This posterior distribution is given by Bayes' rule when assuming a prior distribution $\pi^{0}(\theta)$ over the parameters:
\begin{align}\label{eq:bayes-posterior}
    \pi(\theta) = \frac{\mathcal{L}(\theta) \pi^{0}(\theta)}{Z},
\end{align}
where
\begin{align}\label[equation]{eq:likelihood} 
  \mathcal{L}(\theta) = \frac{1}{\sqrt{2 \pi}^{d}\sqrt{\Sigma}}\exp\left( -\frac{1}{2}\left(y_{\mathcal{D}} - \mathcal{F}(\theta) \right)^{T}\Sigma^{-1} \left(y_{\mathcal{D}} - \mathcal{F}(\theta) \right) \right),
\end{align}
is the likelihood under Gaussian noise and $Z$ is the normalizing constant termed the evidence. 

The posterior has no closed form solution for general cases, and sampling approaches based on MCMC are used to generate samples distributed according to $\pi$. MCMC approaches create a sequence of random variables $\left\{ \theta^i \right\}_{i \in \mathbb{Z}^+},$ where each $\theta^i$ is the state of the chain at iteration $i$ and $\mathbb{Z}^+$ denotes the set of positive integers. The chain is constructed so that its stationary distribution is the target posterior $\pi(\theta)$. Among MCMC methods, the Metropolis--Hastings algorithm~\cite{hastings_monte_1970} is the most popular and widely used. The proposal distribution $q(\cdot \mid \cdot)$ and the acceptance probability $\alpha(\cdot,\cdot)$ define the algorithm. Samples are proposed iteratively from $q$ and are accepted or rejected based on $\alpha$. The method is shown in~\Cref{alg:mhmcmc} and induces a Markov transition kernel of the form:
\begin{align}\label{eq:kernel-mhmcmc}
    p(\theta, A) = \int_A \alpha\left(\theta, \theta^*\right)q\left(\theta \mid d\theta^*\right) + \left(\int_{X} 1 - \alpha\left(\theta, \theta^*\right) q\left(\theta \mid d\theta^*\right)\right)\delta_{\theta}(A), \quad A \in \mathcal{X}, 
\end{align}
where $\delta$ is the Dirac delta function. Existence of a stationary distribution is guaranteed if the Markov kernel above satisfies detailed balance, and convergence is guaranteed if the kernel is ergodic. Further proof and reading can be found in~\cite{hastings_monte_1970,robert_monte_2004}.
\begin{algorithm}[h]
  \caption{Metropolis-Hastings MCMC~\cite{hastings_monte_1970}}
  \label{alg:mhmcmc}
    \begin{algorithmic}[1]
    \STATE{\textbf{Input}: Target density $\pi$, proposal distribution $q(\cdot \mid \cdot)$, number of samples $N$, initial guess $\theta^0$}
    \STATE{\textbf{Output}: Samples $\left\{\theta^i\right\}_{i=1}^{N}$ from $\pi$}
    \FOR{$i = 0,1,\ldots, N-1$}
      \STATE{Sample $\theta^* \sim q(\cdot \mid \theta^i)$}
      \STATE{Sample $u \sim \mathcal{U}(0,1)$}
      \STATE{Set $\theta^{i+1} = \theta^*$ if $u < \alpha$ where 
      \begin{align}~\label{eq:accept-reject}
        \alpha(\theta^i, \theta^*) = \min\left(1, \frac{\pi(\theta^*)q(\theta^i \mid \theta^*)}{\pi(\theta^i)q(\theta^* \mid \theta^i)}\right),
      \end{align}}
      \STATE{Set $\theta^{i+1} = \theta^i$ otherwise}
    \ENDFOR
    \RETURN $\left\{\theta^i\right\}_{i=1}^{N}$ 
    \end{algorithmic}
\end{algorithm}
Given the ability to draw samples from the posterior $\pi(\theta)$, we seek to model predictions by computing statistics of a quantity of interest (QoI) defined as $Q: \Theta \rightarrow \mathcal{Q}$, where $\mathcal{Q} \subseteq \mathbb{R}^k$ denotes the output space and $k \geq 1$ its dimensionality. In this work, we focus on the scalar case $k=1$ but extensions to multi-output quantities follow directly from~\cite{thomasgroupedACV2024}.

A standard Monte Carlo (MC) estimator for the posterior mean of $Q$ is given by:
\begin{align}\label{eq:mc-estimate}
  \hat{Q}  = \mathbb{E}_{\Theta}[Q(\theta)] = \int_{\theta \in \Theta} Q(\theta) \pi(\theta) d\theta \approx \frac{1}{N} \sum_{i=1}^{N} Q(\theta^{i}), \quad \theta^{i} \in \Theta, \theta^i \sim \pi,
\end{align}
where the notation $\hat{Q}$ denotes an estimator for the mean of $Q$ obtained by generating $N$ MCMC samples $\left\{\theta^i\right\}_{i=1}^N$ from the distribution $\pi$.
The first equality in \cref{eq:mc-estimate} comes from writing the estimate of the mean as an expectation, the second equality comes from writing the expectation as an integral and the approximation comes from writing the integral as a Monte Carlo estimation with $N$ samples. The estimator in \cref{eq:mc-estimate} is unbiased with respect to the model used if the MCMC samples are independent and identically distributed. However, samples generated by MCMC are inherently correlated, and the estimator is only asymptotically unbiased. While recent work has developed unbiased estimators using couplings~\cite{jacob_unbiased_2020}, in this work, we assume a sufficient burn-in period and disregard any initial bias. The error of the MC estimator is then measured by its variance, and when the variance of $Q(\theta)$ is finite, the variance of the estimator is inversely proportional to the number of samples $N$. In this paper, we do not consider the bias of the model with respect to representing some underlying truth model, and therefore our goal is to reduce the error of this procedure through variance reduction. Here, we employ multi-fidelity techniques.

\subsection{Multi-fidelity methods for variance reduction}\label{subsec:multi-fidelity-methods}
The Monte Carlo estimator in \cref{eq:mc-estimate} is computationally expensive, with improvements to accuracy requiring orders of magnitude more samples. One strategy to address this challenge is to leverage multi-fidelity techniques that combine estimates from an ensemble of models to achieve a significantly lower variance. This ensemble typically includes a high-fidelity model and several low-fidelity models. In the context of a discretized PDE parameterized by $\ell$, the low fidelity models are coarse grid approximations of the original discretization given by $\{\mathcal{F}_{\ell}\}_{\ell=0}^{L-1}$. Similarly, we obtain a hierarchy of output functionals $\{Q_{\ell}\}_{\ell=0}^{L-1}$. We assume that the evaluation of both $\mathcal{F}_{\ell}$ and $Q_{\ell}$ is a function of the discretization parameter $M_{\ell} = s^{\ell}M_0$. In the following, outputs with subscript $L$ are assumed to be the highest fidelity with respect to which we seek variance reduction.

Although a plethora of multi-fidelity techniques have been explored in the past~\cite{peherstorfer_survey_2018}, in this work, we focus on creating \textit{fusion} estimators of the type,
\begin{align}\label{eq:multi-fidelity-estimator}
    \hat{Q}^{\text{MF}}_L = f\left(\hat{Q}_{0}, \hat{Q}_{1}, \dots, \hat{Q}_{L}\right),
\end{align}
where $\hat{Q}_{\ell}$ denotes the estimator of the QoI at levels $\ell = 0,1,\dots,L$ and $f(\cdot)$ is a function that combines these estimators. The goal of the fusion estimator is to reduce the variance of the estimator $\hat{Q}^{\text{MF}}_L$ by exploiting the correlation between the estimators $\hat{Q}_{\ell}$ at different levels. Both the choice of the function $f(\cdot)$ and the correlation between the multi-fidelity estimators play a crucial role in determining the efficiency of the estimator.
\subsubsection{Multilevel Monte Carlo}\label{subsubsec:mlmc}
The multilevel Monte Carlo (MLMC) estimator~\cite{giles_multilevel_2015} is a popular multi-fidelity fusion technique that has been widely used in the context of forward uncertainty quantification. The MLMC estimator is based on the telescoping sum representation of the high-fidelity QoI: 
\begin{align}\label{eq:mlmc-estimate}
    \hat{Q}^{\text{MLMC}}_L &= \hat{Q}_{0}+ \sum_{\ell=1}^{L} \left(\hat{Q}_{\ell} - \hat{Q}_{\ell-1}\right) = \sum_{\ell=0}^{L} \hat{Y}_{\ell},
\end{align}
where the level wise estimator is defined as $\hat{Y}_{\ell} = \hat{Q}_{\ell} - \hat{Q}_{\ell-1}$ and $\hat{Y}_0 = \hat{Q}_{0}$. The MLMC estimator is unbiased and the variance of the estimator is dependent on the variance of each of the terms $\{\hat{Y}_{\ell}\}^{L}_{\ell=0}$ in \cref{eq:mlmc-estimate}. The key is then to create highly correlated estimators $\hat{Q}_{\ell}$ and $\hat{Q}_{\ell-1}$ by coupling samples that will result in $\mathbb{V}[Y_{\ell}] \rightarrow 0$ as $\ell \rightarrow \infty$. Consider the variance of the level wise outputs at a particular level $\ell$ given by:
\begin{align*}
    \mathbb{V}[Y_{\ell}] = \mathbb{V}[Q_{\ell}] + \mathbb{V}[Q_{\ell-1}] - 2\text{Cov}[Q_{\ell}, Q_{\ell-1}].
\end{align*}
Assuming $\mathbb{V}[Q_{\ell}] \approx \mathbb{V}[Q_{\ell-1}] = V$ and using the Pearson correlation coefficient, 
\begin{align}~\label{eq:pearson-correlation}
  \rho_{\ell} = \text{Cov}[Q_{\ell}, Q_{\ell-1}]/\sqrt{\mathbb{V}[Q_{\ell}]\mathbb{V}[Q_{\ell-1}]},
\end{align}
we write the variance as $\mathbb{V}[Y_{\ell}] = 2V(1-\rho_{\ell})$. The variance of the MLMC estimator can be reduced by increasing the correlation between the \textit{successive} level-dependent functional outputs $Q_{\ell}$ and $Q_{\ell-1}$. In standard MLMC applications, when exact sampling is possible for the forward problem, correlating successive level estimators is achieved by sharing samples of the highest fidelity $L$ for both the estimators at a particular level. However, for inverse problems, the distributions of each level may be different and correspond to the inverse problem at each such level.

\section{Inverse problem estimators and couplings}\label{sec:inverse-problem-estimators-couplings}
In the inverse case, where MCMC sampling is employed, achieving high correlations between the functional estimators becomes significantly complex. The dependency on the forward model introduced by MCMC methods complicates direct sharing of samples, making it {\it challenging to use the same input samples for each level, as is typical for MLMC approaches for forward problems}. These challenges necessitate studying \textit{couplings} that can efficiently establish and exploit such correlations within the MCMC framework. In this section, we will formally introduce couplings and demonstrate their critical role in variance reduction. Then, we will extend our discussion to the integration of MLMC frameworks with MCMC methods to reduce the computational cost of the estimation procedure for inverse problems.

\subsection{Couplings}\label{subsec:couplings}
Let $(X, \mathcal{X}, \mu)$ and $(Y, \mathcal{Y}, \nu)$ be two probability spaces. Coupling of two probability measures $\mu$ and $\nu$ is defined as a joint probability measure $\Gamma$ on $(\Omega, \mathcal{O}, \Gamma)$ where $\Omega = X \times Y$ and $\mathcal{O} = \mathcal{X} \otimes \mathcal{Y}$, such that the marginals of $\Gamma$ are $\mu$ and $\nu$~\cite{villani2009optimal}. In other words, for all measurable sets $A \in \mathcal{X}$ and $B \in \mathcal{Y}$, we have $\Gamma[A \times Y] = \mu(A)$ and $\Gamma[X \times B] = \nu(B)$. Couplings are typically used to establish some distributional properties between the two measures $\mu$ and $\nu$. For instance, they can be  employed to analyze the distance between probability measures in terms of distance metrics~\cite{Thorisson1998}, to facilitate transportation problems~\cite{villani2009optimal}, or to prove convergence properties of MCMC methods~\cite{atchade_adaptive_2005,Rosenthalminor1995}. More recently, novel couplings have been proposed to exploit variance reduction and provide unbiased estimations~\cite{jacob_unbiased_2020,heng_unbiased_2023} in the context of MCMC.

Couplings can be broadly classified into two categories: \textit{deterministic} and \textit{non-deterministic}~\cite{villani2009optimal}. In deterministic couplings, the relationship between two random variables $X$ and $Y$ defined on the two probability spaces is described by a function. Formally, this translates to the existence of a measurable function $T: X \rightarrow Y$ such that $Y = T(X)$ and $T_{\#}\mu = \nu$, where $T_{\#}$ is called the pushforward operator. The coupling is then defined by the identity map $\mathbb{I}$ on $X$ and the transport map $T$ on $Y$, i.e., $\Gamma = (\mathbb{I}, T)_{\#}\mu$. An important deterministic coupling example is the optimal transport (OT) problem which finds the optimal transport map by introducing a cost function $c(x,y)$ on $\Omega$ that can be interpreted as the work needed to transport a unit of mass from $x \in X$ to $y \in Y$. Some applications of transport maps can be found in~\cite{peherstorfer_transport-based_2019,parno_transport_2018}. Non-deterministic couplings, referred to simply as couplings, arise from invoking randomization during their construction and are described through a joint probability measure $\Gamma$ on $\Omega$. A notable example is the maximal coupling technique~\cite{Thorisson1998}, where the joint distribution $\Gamma$ is constructed such that the probability of the random variables $X$ and $Y$ being equal is maximized~\cite{Thorisson1998}.

We now extend the concept of coupling to Markov chains following the approach of~\cite{Thorisson1998,oleary_metropolis-hastings_2023}: suppose we have an ergodic MCMC kernel $p_X: X \times \mathcal{X} \rightarrow [0, 1]$ with measure $\mu$ and another ergodic MCMC kernel $p_Y: Y \times \mathcal{Y} \rightarrow [0, 1]$ with measure $\nu$. A coupling of these two kernels is a joint kernel $p: \Omega \times \mathcal{O} \rightarrow [0, 1]$  with joint measure $\Gamma$ satisfying $\int_{Y}^{}p((x,y),A \times dy) = p_X(x,A)$ and $\int_{X}^{}p((x,y),dx \times B) = p_Y(y,B)$. These equations imply that any Markov chain $\{Z^i\}_{i \in \mathbb{Z}^+} = \left(\{X^i\}_{i \in \mathbb{Z}^+}, \{Y^i\}_{i \in \mathbb{Z}^+}\right)$ constructed with the joint kernel $p$ is composed such that the sequences $\{X^i\}_{i \in \mathbb{Z}^+}$ and $\{Y^i\}_{i \in \mathbb{Z}^+}$ follow the transition dynamics of their respective kernels $p_X$ and $p_Y$. Crucially, while the marginals are fixed, the joint properties of the coupled chain can be designed to achieve specific goals, such as maximizing correlation between the two chains for variance reduction purposes.

In this work, we are dealing with MH-MCMC type methods for our inverse problem estimators, where the efficiency of the estimator depends on the quality of Markov chain coupling. As established in~\cite{oleary_metropolis-hastings_2023}, the coupling of two MH-MCMC kernels can be represented as a combination of a proposal coupling ($q_X$ and $q_Y$) followed by an acceptance coupling ($\alpha_X$ and $\alpha_Y$). Specifically, to effectively couple two Markov chains together, we first couple the proposal distributions to produce a proposed state $Z^* = \{X^*, Y^*\}$ given the current state $Z^i = \{X^i, Y^i\}$. We then perform the acceptance coupling step by marginally accepting or rejecting the proposed state on both chains using the same uniform random number in the accept-reject function (step 6 in~\cref{alg:mhmcmc}). This decomposition highlights that the key to constructing efficient Markov chain couplings lies in designing effective \textit{proposal couplings}. 

\subsection{Multilevel Markov chain Monte Carlo estimators}\label{subsec:acv-mlmc-mcmc}
Next, we integrate the MLMC estimator with MCMC methods to reduce the computational cost of solving inverse problems. The core of the integration lies in generating correlated samples from coupled Markov chains at consecutive fidelity levels. For each level $\ell=1,\ldots,L$ in the MLMC estimator (\Cref{eq:mlmc-estimate}), we generate samples for two output functionals $Q_{\ell}$ and $Q_{\ell-1}$ to estimate $\mathbb{E}\left[Q_{\ell} - Q_{\ell-1}\right]$. The samples used for $Q_{\ell}$ are drawn from a \textit{fine} chain at level $\ell$, while the samples for $Q_{\ell-1}$ are drawn from a \textit{coarse} chain at the same level. The strategy for defining the target distributions for these two chains is crucial for the efficiency of the ML-MCMC estimator. There are three potential routes: (1) First, both fine and coarse chains target the highest fidelity posterior $\pi_L$. While conceptually simple, this approach is computationally prohibitive because of the high cost of evaluating the likelihood at the finest level. (2) Second, samples from the coarse posteriors at each level approximate the fine posteriors at that level. While computationally cheaper, this approach introduces bias in the estimator. (3) Third, each chain targets its respective level-dependent posterior, i.e., the coarse chain at level $\ell$ targets $\pi_{\ell-1}$ while the fine chain targets $\pi_{\ell}$. We adopt this strategy, as it is unbiased and computationally efficient, leaving only the challenge of constructing effective couplings between the chains. 

To implement the level-dependent structure of the couplings, we introduce notation for the coupled samples. Let $\left\{\theta^i_{\ell}\right\}_{i \in \mathbb{Z}^+} \sim \pi_{\ell}$ denote the fine samples at level $\ell$, and let $\left\{\vartheta^i_{\ell-1}\right\}_{i \in \mathbb{Z}^+} \sim \pi_{\ell-1}$ denote the coarse samples at the same level. The coupled chain at MLMC level $\ell$ thus produces pairs $\left\{\left(\theta^i_{\ell}, \vartheta^i_{\ell-1}\right)\right\}_{i \in \mathbb{Z}^+}$. We emphasize that the notation $\vartheta$ distinguishes the role of the sample in the telescoping sum, not just its distribution. For example, $\theta_{0}^i$ and $\vartheta_{0}^i$ are both drawn from the same posterior $\pi_0$, but $\theta_{0}^i$ is used in the fine component of the MLMC estimator at level $0$, while $\vartheta_{0}^i$ is used in the coarse component at level $1$. Thus, the MLMC estimator in~\cref{eq:mlmc-estimate} specialized to the Monte Carlo estimate case becomes:
\begin{align}\label{eq:ml-mcmc}
    \hat{Q}^{\text{ML-MCMC}} &= \frac{1}{N_0} \sum_{i=1}^{N_0} Q_0(\theta_0^i) + \sum_{\ell=1}^{L} \frac{1}{N_{\ell}} \sum_{i=1}^{N_{\ell}} \left( Q_{\ell}(\theta_{{\ell}}^{i}) - Q_{\ell-1}(\vartheta_{{\ell-1}}^{i}) \right), \quad \theta_{\ell}^i \sim \pi_{\ell}, \vartheta_{\ell-1}^i \sim \pi_{\ell-1},
\end{align}
assuming that the Markov chains are sufficiently burned in.

The efficiency of the ML-MCMC estimator above depends on the correlation between the fine and coarse outputs at each level. This transforms the problem of variance reduction in inverse problems into one of constructing effective proposal couplings ($\Gamma_{\ell} = \{q_{\ell}, q_{\ell-1}\}$) of the level-dependent Markov chains as discussed in \cref{subsec:couplings}. Achieving efficient proposal couplings facilitates high correlations between the chains, essential for reducing the variance of the level wise (difference) estimators in \cref{eq:ml-mcmc}. Once the coupling is established, we can solve an optimization problem to find the optimal number of samples for each MLMC level depending on the covariance structure of the functional estimates~\cite{giles_multilevel_2015}.

We now provide a meta-algorithm for the ML-MCMC framework in \cref{alg:ml-mcmc}.
\begin{algorithm}[h]
  \caption{Meta algorithm for multilevel Markov chain Monte Carlo (ML-MCMC)}
  \label{alg:ml-mcmc}
    \begin{algorithmic}[1]
      \STATE{\textbf{Input}: List of target distributions, joint proposal distributions, number of samples and initial guesses for all MLMC levels $\left\{ \pi_{\ell}, \Gamma_{\ell}, N_{\ell}, \theta^0_{\ell}\right\}_{\ell=0}^{L}$}
      \STATE{\textbf{Output}: Coupled chains $\left\{\theta_0^i\right\}_{i=1}^{N_0}, \left\{\left(\theta_{\ell}^i, \vartheta_{\ell-1}^i\right)\right\}_{i=1}^{N_{\ell}}$ for $\ell = 1,2,\ldots L$}
      \IF{$\ell = 0$}
        \STATE{$\{\theta_0^i\}_{i=1}^{N_0} =$ MH-MCMC$\left(\pi_0, \Gamma_0, N_0, \theta_0^0 \right)$}\
      \ENDIF
      \FOR{$\ell = 1,2,\ldots L$}
        \FOR{$i = 0,1,\ldots N_{\ell}-1$}
          \STATE Sample $\theta_{\ell}^{*}, \vartheta_{\ell-1}^{*} \sim \Gamma_{\ell}\left((\cdot, \cdot) \mid \left(\theta_{\ell}^i, \vartheta_{\ell-1}^i\right)\right)$~\label{line: proposal-coupling-mlmcmc}
          \STATE{Sample $u \sim \mathcal{U}(0, 1)$}
          \STATE Accept or reject $\theta_{\ell}^{i+1} = \theta_{\ell}^*$ if $u < \alpha_{\ell}$, where $\alpha_{\ell}$ is given by~\cref{eq:accept-reject} for $\pi_{\ell}$ and $q_{\ell}$
          \STATE Accept or reject $\vartheta_{\ell-1}^{i+1} = \vartheta_{\ell-1}^*$ if $u < \alpha_{\ell-1}$, where $\alpha_{\ell-1}$ is given by~\cref{eq:accept-reject} for $\pi_{\ell-1}$ and $q_{\ell-1}$
        \ENDFOR
        \STATE{Save the coupled chains $\left\{ \theta_{\ell}^i, \vartheta_{\ell-1}^i \right\}_{i=1}^{N_{\ell}}$}
      \ENDFOR
      \RETURN $\left\{\theta_0^i\right\}_{i=1}^{N_0}, \left\{\left(\theta_{\ell}^i, \vartheta_{\ell-1}^i\right)\right\}_{i=1}^{N_{\ell}}$ for $\ell = 1,2,\ldots L$
    \end{algorithmic}
\end{algorithm}
The first step in the base algorithm is to sample from the lowest fidelity posterior distribution $\pi_0$ using any ergodic MCMC algorithm and a proposal distribution $\Gamma_0 = q_0$, for example, \cref{alg:mhmcmc}. Then, for the successive levels, based on the coupling method for the joint distribution, we construct a coupled Markov chain $\left\{\left(\theta^i_{\ell}, \vartheta^i_{\ell-1}\right)\right\}_{i \in \mathbb{Z}^{+}}$ for which the marginal chains at two successive levels are highly correlated and sampled from their respective target distributions. This is done by proposing a joint state from the coupled Markov kernel followed by an accept-reject step based on the same uniform random number $u$ as discussed in \cref{subsec:couplings}. The coupled chains along with the coarsest level chain are saved and returned. The samples from all the chains are then used to compute the functional quantities used in the estimator in~\cref{eq:ml-mcmc}.

\subsection{Existing coupling methods for ML-MCMC}\label{subsec:existing-methodologies}
We now review some existing non-deterministic coupling methodologies from the literature, each representing a specific choice for the proposal coupling $\Gamma_{\ell}$ in \cref{alg:ml-mcmc}

\subsubsection{Delayed acceptance coupling~\cite{christen_markov_2005}}\label{subsubsec:delayed-acceptance-coupling}
A powerful strategy to construct the coupled proposal distribution $\Gamma_{\ell}$ is inspired by the delayed acceptance method~\cite{christen_markov_2005}. Unlike standard MH couplings that separates the proposal and acceptance steps, the DA method integrates a two-stage acceptance procedure into the proposal generation itself. The key insight is that a proposal is first evaluated against a computationally cheaper coarse model before being evaluated against the more expensive fine model. This method requires less computational effort compared to standard MCMC because the fine level forward model is called only if the coarse level proposal is accepted, leading to the name delayed acceptance. This idea motivates using the \textit{coarse level posterior distribution $\pi_{\ell-1}$ as a proposal for the fine level posterior distribution $\pi_{\ell}$}. The general DA coupling strategy, which replaces Line~\ref{line: proposal-coupling-mlmcmc} in \cref{alg:ml-mcmc}, is outlined below.

\begin{algorithm}
  \caption{General delayed acceptance coupling~\cite{dodwell2015hierarchical,lykkegaard_multilevel_2023} (Line~\ref{line: proposal-coupling-mlmcmc} in \cref{alg:ml-mcmc})} 
  \label{alg:delayed-acceptance-coupling}
    \begin{algorithmic}[1]
      \STATE{Generate a coarse proposal $\vartheta_{\ell-1}^{*}$ via a method-specific procedure.}~\label{line: da-proposal-generation}
      \STATE{Set the fine proposal $\theta_{\ell}^{*} \leftarrow \vartheta_{\ell-1}^{*}$}
      \RETURN $(\theta_{\ell}^{*}, \vartheta_{\ell-1}^{*})$
    \end{algorithmic}
\end{algorithm}

The key design choice is how to generate the coarse proposal $\vartheta_{\ell-1}^{*}$ in step~\ref{line: da-proposal-generation} of \cref{alg:delayed-acceptance-coupling}. We review two implementations.

\paragraph{(1) Hierarchical ML-MCMC with subsampling~\cite{dodwell2015hierarchical}}
In this method, the coarse proposal $\vartheta_{\ell-1}^{*}$ is generated by approximating an independent draw from the coarse posterior $\pi_{\ell-1}$. This approximation is achieved through a recursive subsampling procedure. At $\ell=0$, a standard long MCMC chain is run and then subsampled to produce approximately independent samples from $\pi_0$. These samples are then used as proposals for the next level $\ell=1$. The proposals are then accepted or rejected based on the delayed acceptance framework~\cite{christen_markov_2005}. These samples are then subsampled again to produce approximately independent samples from $\pi_1$ which are then used as proposals for level $\ell=2$. This process is repeated until the finest level $L$ is reached. This recursive independent subsampling procedure is shown in~\cref{alg:recursive-independence-sampling}.
\begin{algorithm}
  \caption{Recursive independence sampling~\cite{dodwell2015hierarchical} (Line~\ref{line: da-proposal-generation} in \cref{alg:delayed-acceptance-coupling})}
  \label{alg:recursive-independence-sampling}
    \begin{algorithmic}[1]
      \STATE{\textbf{If} $\ell = 0$}: Run a long MCMC chain targeting $\pi_0$ and subsample it to produce a set of nearly independent samples.
      \STATE{\textbf{For each level $k=1,2,\ldots,\ell-1$} (recursive step):}
      \STATE{\quad Use nearly independent samples from level $k-1$ as proposals in a DA MCMC chain targeting $\pi_k$~\cite{christen_markov_2005}.}
      \STATE{\quad Subsample the resulting chain to produce nearly independent samples from $\pi_k$.}
      \STATE{Set $\vartheta_{\ell-1}^{*}$ as the final subsampled state from $\pi_{\ell-1}$.}
      \RETURN $\vartheta_{\ell-1}^{*}$
    \end{algorithmic}
\end{algorithm}
A major limitation of this recursive method is the factorial growth in computational cost with the number of levels. To produce one independent sample $\vartheta_{\ell-1}^i$ on level $\ell-1$, we need to compute $T_k:=\prod_{k^{\prime}=k}^{\ell-1}t_{k^{\prime}}$ samples on each of the levels $k = 0,1,\ldots, \ell-1$, where $t_{k}$ is the subsampling rate for level $k$. Furthermore, because the method relies on finite subsampling to approximate independent samples from the coarse posterior distributions, it introduces bias into the estimation procedure, and the coupled chains cannot be guaranteed to be ergodic~\cite{dodwell2015hierarchical,madrigal-cianci_analysis_2023}.

\paragraph{(2) Multilevel delayed acceptance (MLDA) sampler~\cite{lykkegaard_multilevel_2023}}
The MLDA sampler resolves the ergodic limitations of the hierarchial sampling approach by constructing proposals via recursive finite-length MCMC subchains. Specifically, a proposal candidate $\vartheta^*_{\ell-1}$ at level $\ell$ is generated by running a short MCMC subchain targeting $\pi_{\ell-1}$, which itself may be generated by running a short MCMC subchain targeting $\pi_{\ell-2}$, and so on, until the coarsest level $\ell=0$ is reached, where a standard MCMC chain is run. This procedure is summarized in~\cref{alg:recursive-mlda}.
\begin{algorithm}
  \caption{Recursive subchain using MLDA~\cite{lykkegaard_multilevel_2023} (Line~\ref{line: da-proposal-generation} in \cref{alg:delayed-acceptance-coupling})}~\label{alg:recursive-mlda}
    \begin{algorithmic}[1]
      \STATE{\textbf{If} $\ell = 1$:} Generate subchain targeting $\pi_0$ using any ergodic MCMC algorithm and return proposal $\vartheta_{0}^*$ for level 1.
      \STATE{\textbf{If $\ell > 1$:}} Generate subchain by recursively calling MLDA targeting $\pi_{\ell-1}$.
      \STATE{Set the final state $\vartheta_{\ell-1}^{*}$ from the subchain targeting $\pi_{\ell-1}$ as the coarse proposal.}
      \RETURN $\vartheta_{\ell-1}^{*}$
    \end{algorithmic}
\end{algorithm}
Crucially, the Metropolis-Hastings acceptance probability is correctly formulated to account for the exact transition kernel of the generating sub-chain, rather than assuming an independent proposal. This preserves the detailed balance condition and ensures that the resulting coupled ML-MCMC sampler is provably unbiased and ergodic.

Both the hierarchical ML-MCMC with subsampling and the MLDA sampler effectively utilize the delayed acceptance coupling strategy~\cite{christen_markov_2005} to generate proposals for the fine level chain from the coarse level posterior distribution. In both algorithms, we screen the fine level proposals $\theta_{\ell}^* \leftarrow \vartheta_{\ell-1}^*$ through a sequence of cheaper coarse level acceptances, which can significantly reduce the computational cost when the coarse model is a good approximation of the fine model. However, both methods suffer from a factorial growth in computational cost with the number of levels. Every proposal $\theta_{\ell}^*$ at level $\ell$ requires running subchains at all coarser levels, leading to an exponential increase in the total number of model evaluations as the number of levels increases. Furthermore, both methods rely on the assumption that the coarse and fine level posterior distributions are sufficiently similar to achieve high acceptance rates and effective correlation between the chains. When the posteriors are dissimilar, the acceptance rates drop significantly, leading to poor mixing and reduced variance reduction benefits.

To mitigate this issue of dissimilar posteriors, the authors in~\cite{lykkegaard_multilevel_2023} propose an adaptive error model (AEM). The AEM approximates the discrepancy between the coarse and fine level likelihoods using a Gaussian error model. The mean and covariance of this error model are learned adaptively using Haario's adaptive MCMC algorithm~\cite{haario_adaptive_2001} during the sampling process. The learned error model is then used to correct the coarse level likelihood, effectively aligning it more closely with the fine level likelihood. This correction improves the similarity between the coarse and fine posteriors, leading to higher acceptance rates and better sampling efficiency. However, the AEM works only if all the levels of the hierarchy use the same data. If the data or likelihood function changes between levels, the discrepancy between model outputs is no longer a function of the model error alone, but also of the data inconsistency. In such cases, we cannot learn or apply a valid error model, limiting the applicability of this approach. We demonstrate these challenges in~\Cref{subsec:prey-predator}, where we compare AEM performance against our method on a prey-predator model. In summary, while both delayed acceptance coupling methods provide effective strategies for coupling ML-MCMC chains, they face challenges related to computational cost growth and sensitivity to posterior similarity.

\subsubsection{Independent proposal coupling~\cite{madrigal-cianci_analysis_2023}}\label{subsubsec:independent-proposal}
The next method uses an independent proposal as the joint proposal distribution in~\cref{alg:ml-mcmc}. The authors in~\cite{madrigal-cianci_analysis_2023} \textit{couple the two chains with a proposal that is independent of the current state of either chain followed by an accept-reject step based on the same uniform random number}. This approach ensures the generation of marginally true chains that are correlated, assuming the chains exhibit a good overall acceptance rate and demonstrate effective mixing. They also show that under some conditions on the proposal and posterior densities, there exists a unique invariant probability measure for the coupled chain. However, achieving ergodicity requires a strong theoretical condition: the proposal distribution $q_{\ell}^*$ must have tails that decay more slowly than those of both posterior distributions $\pi_{\ell}$ and $\pi_{\ell-1}$ (Assumption A.1. in~\cite{madrigal-cianci_analysis_2023}). If the proposal decays faster than the posteriors, the coupled chain may fail to explore the tails, breaking ergodicity. 

\begin{algorithm}
  \caption{Independent proposal coupling~\cite{madrigal-cianci_analysis_2023} (Line~\ref{line: proposal-coupling-mlmcmc} in \cref{alg:ml-mcmc})}~\label{alg:independent-proposal-coupling}
    \begin{algorithmic}[1]
    \STATE{Construct independent proposal $q_{\ell}^*$}
    \STATE{Sample $\theta^{*} \sim q_{\ell}^*$ from the independent proposal}
    \STATE{Set $\theta_{\ell}^{*}, \vartheta^*_{\ell-1} = \theta^*$}
    \RETURN $(\theta_{\ell}^{*}, \vartheta_{\ell-1}^{*})$
    \end{algorithmic}
\end{algorithm}

The delayed acceptance coupling method in \cref{subsubsec:delayed-acceptance-coupling} is a special case of the independent proposal coupling method where the proposal distribution is chosen to be the coarse level posterior distribution ($q_{\ell}^* = \pi_{\ell-1}$). Therefore, the restrictive assumption on the tail decay of the proposal distribution in the independent proposal coupling method also applies to the delayed acceptance coupling method. The authors in~\cite{madrigal-cianci_analysis_2023} clearly show that the coupled chain obtained using the framework in~\cite{dodwell2015hierarchical} will not be ergodic unless the proposal distribution satisfies the condition in A.1., which is usually the case in practice. However, the independent proposal coupling method has its own challenges, particularly in selecting an appropriate proposal distribution that satisfies the tail decay assumption while maintaining sufficient correlation between the chains. Given that the posterior distributions of the coupled chains are unknown, determining a suitable proposal distribution for the tail decay assumption becomes a non-trivial task. The authors in~\cite{madrigal-cianci_analysis_2023} recommend utilizing density approximation techniques such as Kernel Density Estimation (KDE), flow-based generative models, or Laplace Approximations of the posterior distributions as viable proposal candidates. Nevertheless, these methods introduce additional computational complexity to the problem.

\subsubsection{Maximal coupling~\cite{cianci_thesis}}\label{subsubsec:maximal-coupling}
Maximal coupling is a classic coupling technique that aims to maximize the probability of two coupled random variables being equal. This concept has been primarily used to prove convergence properties of MCMC chains~\cite{Thorisson1998}. The authors in~\cite{cianci_thesis} adapt this idea to the ML-MCMC framework by \textit{maximally coupling chains at successive levels in the ML-MCMC framework to sample $\{\theta^{*}_{\ell},\vartheta^{*}_{\ell-1}\}$ from a maximal coupling of $\left\{q_{\ell}, q_{\ell-1}\right\}$ such that $\theta^{*}_{\ell} \sim q_{\ell}(\cdot \mid \theta^{i}_{\ell})$ and $\vartheta^{*}_{\ell-1} \sim q_{\ell-1}(\cdot \mid \vartheta^{i}_{\ell-1})$, with $\mathbb{P}(\theta^{*}_{\ell} \neq \vartheta^{*}_{\ell-1}) = \left\| q_{\ell}(\cdot \mid \theta^{i}_{\ell}) - q_{\ell-1}(\cdot \mid \vartheta^{i}_{\ell-1}) \right\|_{TV}$, where $\left \Vert . \right\Vert_{TV}$ is the total variation distance}. 

Note that even though we term this method as a maximal coupling of the two chains, the proposal distributions are maximally coupled and a subsequent accept-reject step is performed on the proposals $\{\theta^{*}_{\ell},\vartheta^{*}_{\ell-1}\}$ with a common uniform random number. A procedure to sample from such a maximal coupling $\Gamma^*$ is taken from~\cite{Thorisson1998} called $\gamma$-coupling and is described in~\cref{alg:maximal-coupling}. The readers are referred to~\cite{jacob_unbiased_2020,oleary_metropolis-hastings_2023} for more such sampling techniques to sample from maximally coupled proposal distributions.

\begin{algorithm}
  \caption{Maximal Coupling~\cite{cianci_thesis,jacob_unbiased_2020,Thorisson1998} (Line~\ref{line: proposal-coupling-mlmcmc} in \cref{alg:ml-mcmc})}~\label{alg:maximal-coupling}
    \begin{algorithmic}[1]
    \STATE{Sample $\theta_{\ell}^{*} \sim q_{\ell}(\cdot \mid \theta_{\ell}^{i})$ and $\psi \mid \theta_{\ell}^{*} \sim \mathcal{U}\left(0, q_{\ell}(\theta_{\ell}^{*} \mid \theta_{\ell}^{i})\right)$}
    \STATE{\textbf{If} $\psi \leq q_{\ell-1}(\theta_{\ell}^{*} \mid \vartheta_{\ell-1}^{i})$, set $\vartheta_{\ell-1}^{*} \leftarrow \theta_{\ell}^{*}$}
    \STATE{\textbf{Else}, sample $\vartheta_{\ell-1}^{*} \sim q_{\ell-1}(\cdot \mid \vartheta_{\ell-1}^{i})$ and $\phi \mid \vartheta_{\ell-1}^{*} \sim \mathcal{U}\left(0, q_{\ell-1}(\vartheta_{\ell-1}^{*} \mid \vartheta_{\ell-1}^{i})\right)$ until $\phi > q_{\ell}(\vartheta_{\ell-1}^{*} \mid \theta_{\ell}^{i})$}
    \RETURN $(\theta_{\ell}^{*}, \vartheta_{\ell-1}^{*})$
    \end{algorithmic}
\end{algorithm}

Although this method allows for state-dependent proposals in contrast to the independent proposal coupling method, and is relatively inexpensive to implement, the correlation between the chains depends on the total variation distance between the two proposal distributions, with smaller distances encouraging tighter coupling. Because the proposals are state-dependent, we want samples from each chain to be close to each other. However, achieving this proximity is infeasible in high dimensions or when the posterior distributions are dissimilar, which frequently occurs in practical applications. In those settings, the resulting weak coupling leads to poor correlation between the chains, limiting the variance reduction benefits of the ML-MCMC estimator.

\section{Proposed coupling methodology}\label{sec:proposed-coupling}
In this section, we introduce our SYNchronized step Correlation Enhancement (SYNCE) coupling method. We demonstrate how our approach overcomes the posterior-overlap limitations of the existing coupling methods, and provide a detailed framework for its implementation within the ML-MCMC estimator.

\subsection{SYNCE coupling}\label{subsubsec:synce-coupling}
Our proposed method is inspired from Pinto and Neal~\cite{pinto2001improving}, where they couple Markov chains using common random number couplings. They couple two chains --- the chain of an unknown target distribution, and a Gaussian approximation of the unknown target distribution to reduce the variance of their estimator. They state that \textit{two chains are coupled when their transitions are determined by the same random numbers}. We extend this idea to our framework by using common random numbers for the two chains that sample from the $\ell$'th and $(\ell-1)$'th level posterior distributions. Specifically, we use the same Gaussian random draw $\eta \sim \mathcal{N}(0, C_{\ell})$ for proposal generation and the same uniform random variable $u \sim \mathcal{U}(0, 1)$ for the accept-reject step at both levels. Using the same random numbers $\eta$ and $u$ for both the chains ensures that the two chains are coupled and that the samples from the two chains are highly correlated. The simplest form of our proposed approach is given in \cref{alg:synce-coupling}.

\begin{algorithm}[H]
  \caption{Baseline SYNCE coupling (Line~\ref{line: proposal-coupling-mlmcmc} in \cref{alg:ml-mcmc})}~\label{alg:synce-coupling}
    \begin{algorithmic}[1]
    \STATE{Sample $\eta \sim \mathcal{N}(0, C_{\ell})$}
    \STATE{Set $\theta^*_{\ell} = \theta_{\ell}^{i} + \eta$ and $\vartheta^*_{\ell-1} = \vartheta_{\ell-1}^{i} + \eta$}
    \RETURN $(\theta_{\ell}^{*}, \vartheta_{\ell-1}^{*})$
    \end{algorithmic}
\end{algorithm}

\begin{figure}[ht]
  \centering
  \begin{subfigure}[b]{0.245\textwidth}
      \centering
      \def\svgwidth{\textwidth}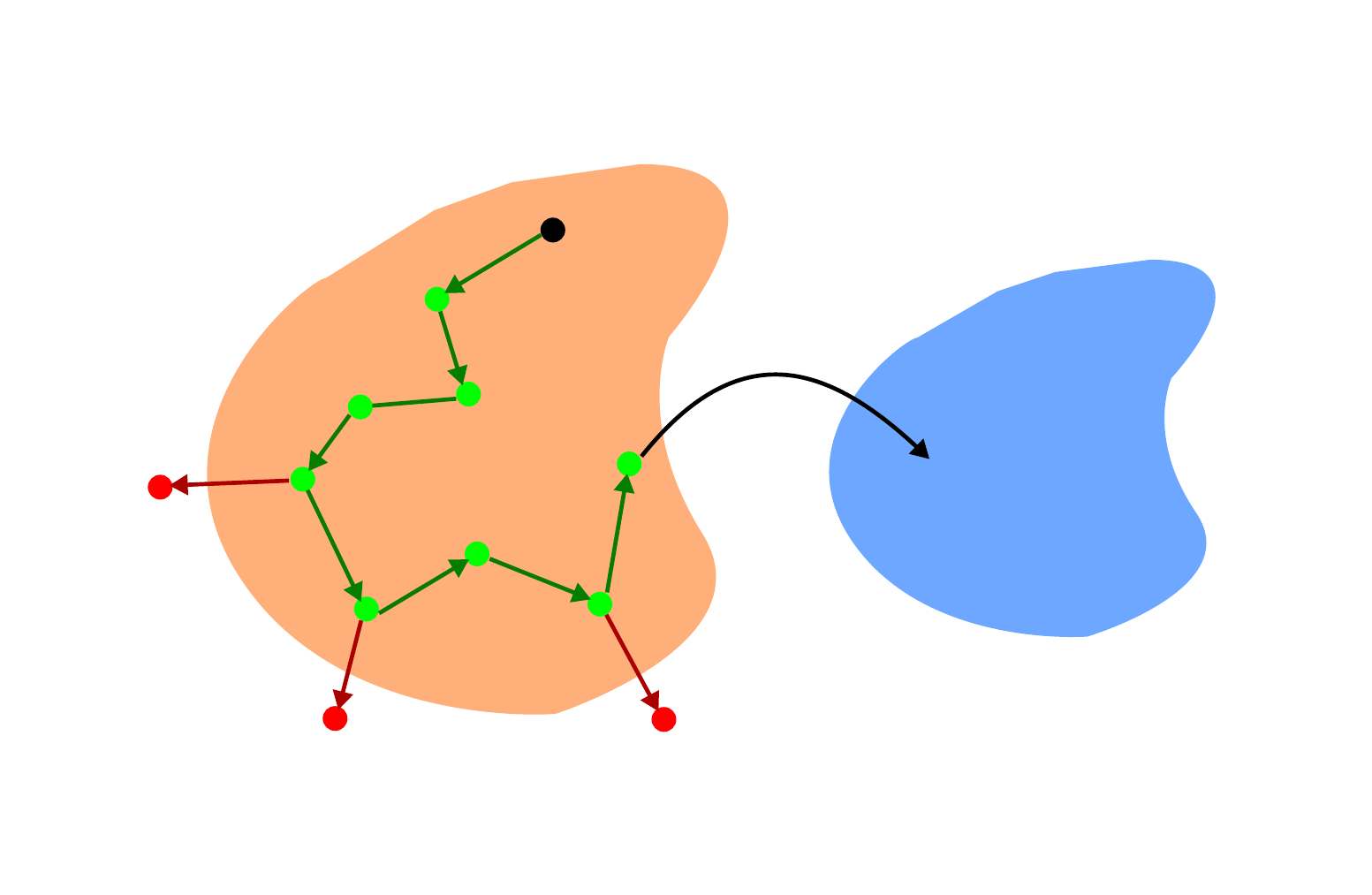
      \caption{DA coupling~\cite{dodwell2015hierarchical,lykkegaard_multilevel_2023}}
      \label{fig:delayed-acceptance-coupling}
  \end{subfigure}
  \begin{subfigure}[b]{0.245\textwidth}
      \centering
      \def\svgwidth{\textwidth}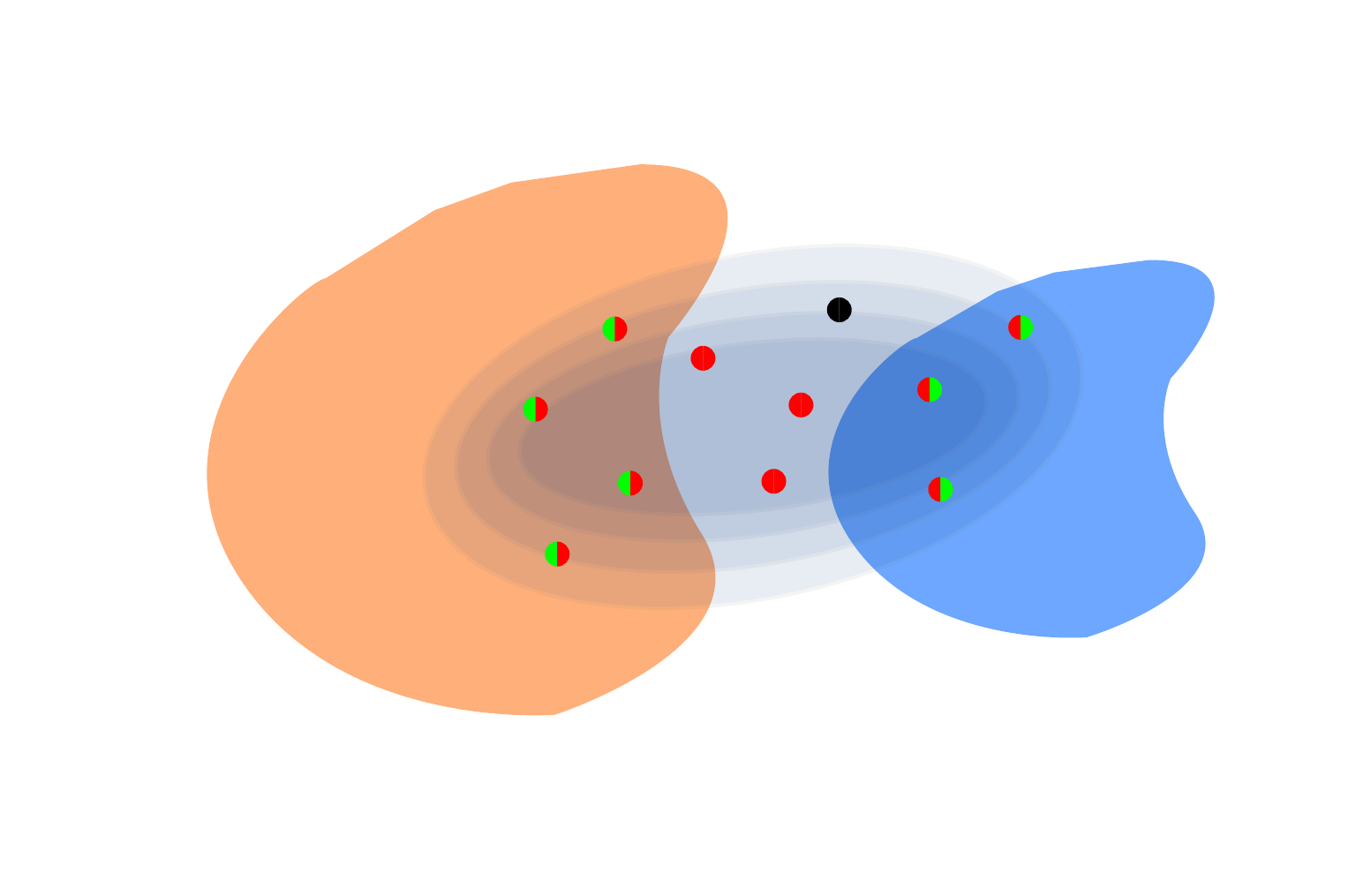
      \caption{Independent proposal coupling~\cite{madrigal-cianci_analysis_2023}}
      \label{fig:independent-proposal-coupling}
  \end{subfigure}
  \begin{subfigure}[b]{0.245\textwidth}
      \centering
      \def\svgwidth{\textwidth}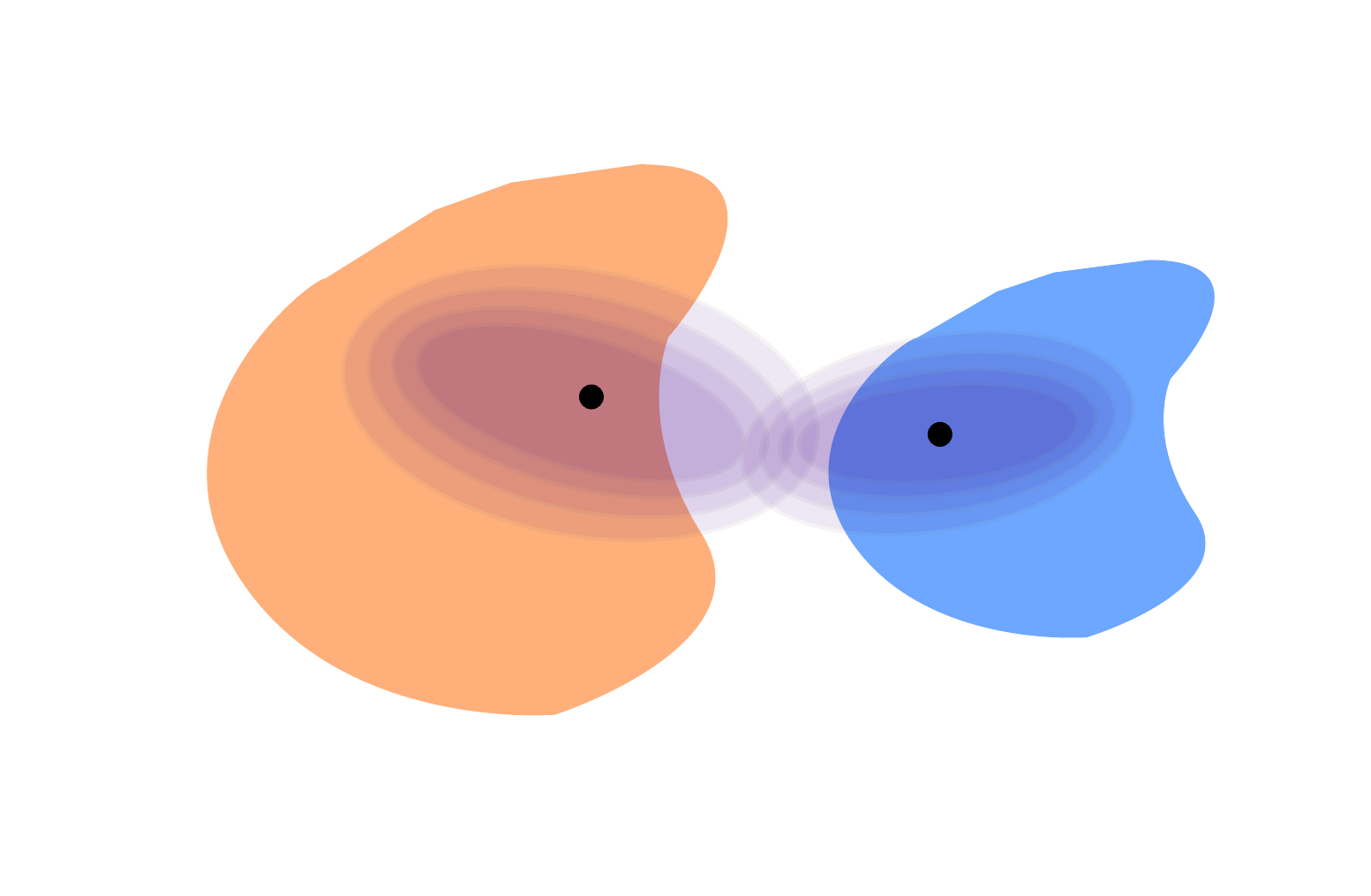
      \caption{Maximal coupling~\cite{cianci_thesis}}
      \label{fig:maximal-proposal-coupling}
  \end{subfigure}
  \begin{subfigure}[b]{0.245\textwidth}
      \centering
      \def\svgwidth{\textwidth}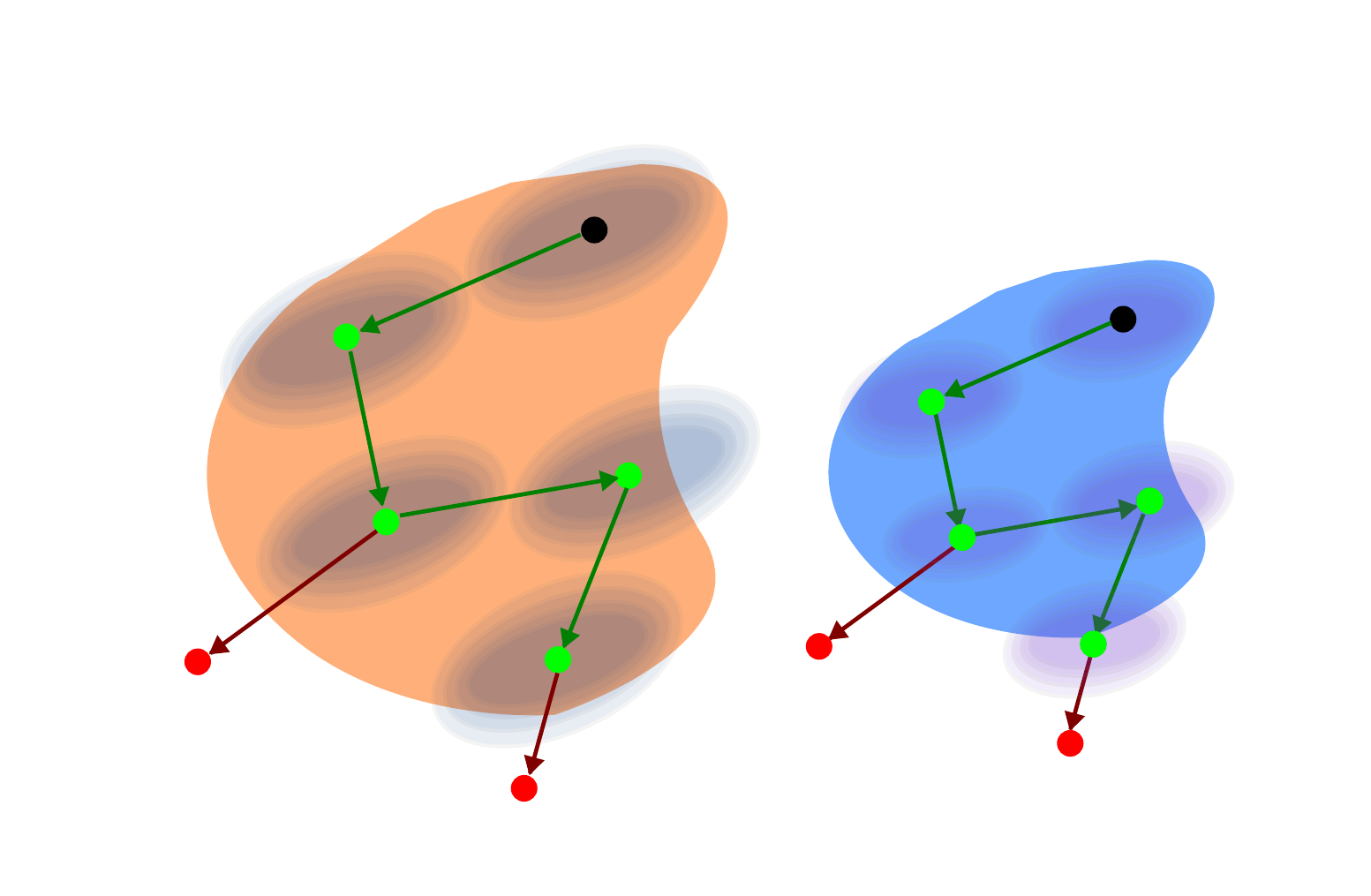
      \caption{SYNCE coupling}
      \label{fig:synce-proposal-coupling}
  \end{subfigure}
  \caption{Comparison of the four coupling methods for two posteriors at levels $\ell-1$ and $\ell$ that are different and do not overlap.}
  \label{fig:algorithm-figures}
\end{figure}

This same-step strategy is fundamentally different from the same-sample strategy employed by the existing coupling methods discussed in \cref{subsec:existing-methodologies}. Our approach ensures that the samples from both chains remain highly correlated, even when the posterior distributions are substantially different. In~\cref{fig:algorithm-figures}, we provide a visual comparison of the four coupling methods when the two posterior distributions $\pi_{\ell}$ and $\pi_{\ell-1}$ are substantially different.

In~\cref{fig:delayed-acceptance-coupling}, the fine chain uses the last accepted point from the coarse chain as its proposal. Since the posteriors are different, most of the proposals will be rejected leading to poor mixing and low correlation. In~\cref{fig:independent-proposal-coupling}, a common sample is proposed independently for both the chains. Proposals are either accepted by both chains (green dots), rejected by both chains (red dots) or accepted by one chain and rejected by the other (green/red dots). When the posteriors have no overlapping region, most of the proposals will be rejected by both chains leading to poor correlation and mixing.~\Cref{fig:maximal-proposal-coupling} depicts maximal coupling, where the probability of proposing the same sample via maximal coupling depends directly on the overlap between the two proposal distributions, which is measured by their total variation distance. When the posteriors are far apart, this overlap shrinks, reducing the probability that the chains propose identical points. Consequently, maximal coupling becomes ineffective for non-overlapping posterior distributions, reducing the coupling to independent product of the marginals. Finally,~\cref{fig:synce-proposal-coupling} illustrates the SYNCE coupling method, where the proposals are forced to have the same magnitude and direction of change. This approach produces high correlation by design and ensures good mixing even when the posteriors are different.

\subsection{Theoretical analysis}\label{subsec: Theory}
In this section, we provide a theoretical analysis of the proposed SYNCE coupling method. We first define the coupled Markov kernel $p_{\ell}^{\text{SYNCE}}$ induced by the SYNCE coupling method in~\cref{alg:synce-coupling}. We then prove that the coupled kernel admits an invariant probability measure and establish geometric ergodicity to this invariant measure. Specifically, we obtain quantitative bounds on the rate of convergence of the coupled kernel to its invariant measure.

Let $p_{\ell}^{\text{SYNCE}}: X^2 \times \mathcal{X}^2 \rightarrow [0, 1]$ denote the Markov transition kernel induced by SYNCE coupling in~\cref{alg:synce-coupling}. The kernel uses the coupled proposal to propose new states $(\theta^*_{\ell},\vartheta^*_{\ell-1}) \sim \Gamma_{\ell}\left((\cdot, \cdot) \mid (\theta^i_{\ell},\vartheta^i_{\ell-1})\right)$ for the two chains at level $\ell$ and $\ell-1$. Following the notation $\alpha^i_{\ell} := \alpha_{\ell}(\theta^i_{\ell}, \theta^*_{\ell})$, $\alpha^i_{\ell-1} := \alpha_{\ell-1}(\vartheta^i_{\ell-1}, \vartheta^*_{\ell-1})$, $\Gamma^i_{\ell} := \Gamma_{\ell}(\bm{\theta}_{\ell}^i \mid d\bm{\theta}_{\ell}^{*})$ and for any $A \in \mathcal{X}^2$,  we define the joint kernel as
\begin{align}
p_{\ell}^{\text{SYNCE}}\left(\bm{\theta}^i_{\ell}, A\right) 
&=\int_{X^2} \min(\alpha^i_{\ell}, \alpha^i_{\ell-1}) \Gamma^i_{\ell}\delta_{\theta^*_{\ell},\vartheta^*_{\ell-1}}(A) 
+ \int_{X^2}(\alpha^i_{\ell}-\alpha^i_{\ell-1})^{+} \Gamma^i_{\ell}\delta_{\theta^*_{\ell},\vartheta^i_{\ell-1}}(A) \nonumber \\
& + \int_{X^2}(\alpha^i_{\ell-1}-\alpha^i_{\ell})^{+} \Gamma^i_{\ell}\delta_{\theta^i_{\ell},\vartheta^*_{\ell-1}}(A) 
+ \left(1-\int_{X^2} \max(\alpha^i_{\ell}, \alpha^i_{\ell-1}) \Gamma^i_{\ell}\right)\delta_{\theta^*_{\ell},\vartheta^*_{\ell-1}}(A).
\label{eq:synce-kernel}
\end{align}
where $x^+ := (x + \mid x\mid)/2$, $\forall x\in\mathbb{R}$. Each of the four terms above correspond to the four possible transitions in the coupled chain. The first term corresponds to the case where both the chains accept their respectively proposed point, the second and third terms correspond to the cases where only one of the chains accepts their proposed points and the last term corresponds to the case when both the chains reject their proposed points.

Establishing ergodicity of the coupled kernel in~\cref{eq:synce-kernel} is not trivial because the joint invariant measure $\mu_{\ell,\ell-1}$ is not explicitly defined a-priori. Consequently, its existence and invariance must be derived rather than assumed. Our proof strategy builds upon the classical Meyn-Tweedie framework~\cite{meyn2012markov}: we demonstrate that the coupled kernel satisfies $\psi
$-irreducibility, aperiodicity, minorization, and a Foster-Lyapunov drift condition. These properties collectively ensure the existence of a unique invariant probability measure and guarantee geometric convergence to this measure. To facilitate our analysis, we introduce the following assumptions regarding the target densities and proposal distributions used in the marginal MH-MCMC kernels at levels $\ell$ and $\ell-1$. 

\begin{assumption}\label{assump:rw-ergodicity}
  For each marginal level $j \in \left\{\ell, \ell-1\right\}$ targeting measure $\mu_j$ with density $\pi_j$ (i.e., either $\pi_{\ell}$ or $\pi_{\ell-1}$) with a random walk proposal $q(\cdot)$,
  \begin{enumerate}[label=(A.\arabic*), ref=(A.\arabic*)]
      \item~\label{assump:bounded} [Cf. Theorem 2.1 in~\cite{jarnerGeometricErgodicityMetropolis2000}] The target density $\pi_j$ is continuous and bounded away from $0$ and $\infty$ on compact sets $S$, i.e,
      \begin{align*}
          0 < \pi_{\min} \leq \pi_j(x) \leq \pi_{\max} < \infty, \quad \forall x \in S,
      \end{align*}
      where $\pi_{\min} = \min_{x \in S}\pi_j(x)$ and $\pi_{\max} = \max_{x \in S}\pi_j(x)$. 
      \item~\label{assump:super-exp} [Cf. Eq. (25) in~\cite{jarnerGeometricErgodicityMetropolis2000}] The target density $\pi_j$ is super-exponentially light, i.e.,
      \begin{align*}
          \limsup_{|x| \rightarrow \infty} \left\langle\frac{x}{|x|}, \nabla \log \pi_j(x)\right\rangle = -\infty.
      \end{align*}
      \item~\label{assump:proposal-density} [Cf. Eq. (8) in~\cite{jarnerGeometricErgodicityMetropolis2000}] The proposal density $q(x)$ of the random walk increment is bounded away from zero in some neighborhood of the origin, i.e., there exists $\delta_q > 0$ and $\epsilon_q > 0$ such that 
      \begin{align*}
        q(x) \geq \epsilon_q, \quad \forall |x| \leq \delta_q.
      \end{align*}
  \end{enumerate}
\end{assumption}

A standard result in random walk Metropolis theory states that if the proposal is locally positive (assumption~\ref{assump:proposal-density}) and the target is continuous and bounded (assumption~\ref{assump:bounded}), then \textit{every} compact set is a small set~\cite{robertsGeneralStateSpace2004,jarnerGeometricErgodicityMetropolis2000}. A small set is a region where the chain can forget its past and regenerate with some positive probability. When the chain strays away from these compact small sets, the super-exponential tail condition (assumption~\ref{assump:super-exp}) ensures that there exists a drift function $V_j$ that pulls the chain back toward these small sets. The local mixing property (minorization) combined with the global drift condition guarantees geometric ergodicity of the chain. We formalize these ideas in the following lemma, which provides necessary ingredients for our subsequent analysis of the coupled kernel.
\begin{lemma}~\label{lemma:rw-ergodicity}
  Let $p_j, j\in \left\{\ell,\ell-1\right\}$ be the marginal MH-MCMC kernel (\Cref{eq:kernel-mhmcmc}) targeting $\pi_j$ with random walk proposal $q(\cdot)$ satisfying~\cref{assump:rw-ergodicity}. Then, there exists a small set $S_j \subset X$ and Lyapunov function $V_j$ such that:
  \begin{enumerate}[label=(B.\arabic*), ref=(B.\arabic*)]
    \item A one step minorization condition holds on $S_j$; that is, there exists a probability measure $\nu_j$ on $(X, \mathcal{X})$ and a (minorizing) constant $\delta_j \in (0,1)$ such that~\label{lemma:rw-minorization}
    \begin{align*}
      p_j(x, A) \geq \delta_j \nu_j(A), \quad \forall x \in S_j, A \in \mathcal{X}.
    \end{align*}
    \item The Lyapunov function $V_j: X \rightarrow [1, \infty)$ satisfies the drift condition:~\label{lemma:rw-drift}
    \begin{align*}
        (p_jV_j)(x) \leq \Lambda_j V_j(x) + b_j \mathbbm{1}_{x\in S_j}, \quad \forall x \in X.
    \end{align*}
    Here, $(p_jV_j)(x) := \mathbb{E}_{p_j(x, \cdot)}\left[V_j\left(X^{n+1}\right)\mid X^n=x\right]$ denotes the expected value of $V_j$ after one step of the chain starting from $x$, $\Lambda_j \in (0, 1)$, $b_j \in \mathbb{R}_{+}$ are constants, and $S_j := \{x \in X: V_j(x) \leq \Delta_j\}$ for some $\Delta_j > 0$.
  \end{enumerate}
\end{lemma}
\begin{pf}[\cref{lemma:rw-ergodicity}]
  The proof is provided in~\cite{jarnerGeometricErgodicityMetropolis2000} for~\cref{lemma:rw-minorization} and~\cite{robertsGeneralStateSpace2004} for~\cref{lemma:rw-drift}.
\end{pf} \qed

With the marginal properties now established, we now present the main theorem of this section that characterizes the ergodicity of the SYNCE coupled kernel.
\begin{theorem}[Geometric ergodicity of the SYNCE coupled kernel]\label{thm:convergence-synce}
  Suppose that~\cref{assump:rw-ergodicity} holds for the two target densities $\pi_{\ell}, \pi_{\ell-1}$ and the Gaussian random walk proposal $q = \mathcal{N}(0, C_{\ell})$. Then, for any ML-MCMC level $\ell=1,\ldots,L$, 
  \begin{enumerate}
    \item The joint kernel $p_{\ell}^{\text{SYNCE}}$ in~\cref{eq:synce-kernel} admits a unique invariant probability measure $\mu_{\ell,\ell-1}$ on $(X^2, \mathcal{X}^2)$ with marginals $\pi_{\ell}$ and $\pi_{\ell-1}$.~\label{thm:convergence-synce-invariance}
    \item The joint chain induced by the SYNCE coupled kernel is geometrically ergodic; that is, there exist a joint Lyapunov function $\hat{V}_{\ell}: X^2 \rightarrow [1, \infty)$, a constant $R_{\ell} < \infty$, and a convergence rate $\rho \in (0,1)$ such that~\label{thm:convergence-synce-geometric}
    \begin{align*}
      \left\|\left(p^{\text{SYNCE}}_{\ell}\right)^n(x, \cdot) - \mu_{\ell,\ell-1}(\cdot)\right\|_{TV} \leq R_{\ell} \hat{V}_{\ell}(x) \rho^n, \quad \forall n\geq 0, x\in X^2.
    \end{align*}
    \item The convergence rate $\rho$ satisfies the explicit bound:
    \begin{align}
      \rho = \inf_{r \in (0,1)} \max \left\{ (1 - \delta)^r (2\hat{b} + 1)^{1-r}, \left( \hat{\Lambda}_{\text{SYNCE}} + \frac{2\hat{b} + 1 - \hat{\Lambda}_{\text{SYNCE}}}{\Delta + 1} \right)^{1-r} \right\},~\label{eq:explicit-rho}
      \end{align}
      where the constants depend on the joint small set $S = S_{\ell} \times S_{\ell-1}$ constructed in~\Cref{lemma:synce-drift} as follows:
      \begin{itemize}
        \item $\delta \propto \epsilon_q\frac{\pi_{\min}}{\pi_{\max}}$ is the minorization constant from Lemma~\ref{lemma:synce-minorization}. Recall that $\epsilon_q$ is the lower bound on the proposal density from assumption~\ref{assump:proposal-density}, and $\pi_{\min}, \pi_{\max}$ are the posterior bounds from assumption~\ref{assump:bounded} within the joint small set $S$.
        \item $\hat{\Lambda}_{\text{SYNCE}} = \max\left(\Lambda_{\ell}, \Lambda_{\ell-1}\right) + \max\left(\frac{b_{\ell}}{1 + \Delta_{\ell-1}}, \frac{b_{\ell-1}}{1 + \Delta_{\ell}}\right)$ is the joint drift rate from~\Cref{lemma:synce-drift}.
        \item $\hat{b} = \frac{1}{2}(b_\ell + b_{\ell-1})$ is the drift constant on the small set $S$ from~\Cref{lemma:synce-drift}.
        \item $\Delta$ is the threshold defining the joint small set $S := {\bm{\theta} \in X^2 : \hat{V}_{\ell}(\bm{\theta}) \leq \Delta}$ from~\Cref{lemma:synce-drift}.
    \end{itemize}
  \end{enumerate}
\end{theorem}
\begin{pf}[\cref{thm:convergence-synce}]
  The proof follows by verifying the conditions of the Meyn-Tweedie framework (see Theorem 15.0.1 in~\cite{meyn2012markov}). Since $p_{\ell}^{\text{SYNCE}}$ is a coupling of standard MH-MCMC kernels, its marginal invariance is guaranteed by construction. Geometric ergodicity then follows from establishing $\psi$-irreducibility and aperiodicity (Lemma~\ref{lemma:synce-irreducibility}), a minorization condition (Lemma~\ref{lemma:synce-minorization}), and a joint drift condition (Lemma~\ref{lemma:synce-drift}). An explicit analytical bound for the convergence rate $\rho$ is provided in~\cref{eq:explicit-rho} based on Theorem 1.3.5 in~\cite{qin2024convergenceboundsmontecarlo}.
\end{pf} \qed

\begin{rmk}
  The geometric convergence rate $\rho$ for the SYNCE coupled kernel established in~\cref{thm:convergence-synce} depends exclusively on the marginal posterior bounds, tail properties, and the lower bound on the proposal. Consequently, the rate remains unaffected by the degree of overlap between the two posterior distributions $\pi_{\ell}$ and $\pi_{\ell-1}$.
\end{rmk}

\subsection{Convergence analysis and comparison}~\label{subsec:convergence-analysis-comparison}

~\cref{thm:convergence-synce} establishes that the SYNCE algorithm generates samples that converge asymptotically to the unique joint invariant measure $\mu_{\ell,\ell-1}$, preserving the correct marginals $\pi_{\ell}$ and $\pi_{\ell-1}$. The property of geometric ergodicity guarantees that convergence to this invariant measure occurs at an exponential rate. As evident from~\cref{eq:explicit-rho}, this rate is determined by two key components: the minorization constant $\delta \propto \epsilon_q \pi_{\min} / \pi_{\max}$ and the drift constant $\hat{\Lambda}_{\text{SYNCE}}$. These constants encapsulate the local mixing behavior within small sets and the global drift behavior outside these sets, respectively. This implies that the rate depends exclusively on the marginal posterior bounds, tail behavior, and the proposal bound, independent of the overlap between the posterior or proposal distributions. Even in scenarios where the two posterior distributions $\pi_{\ell}$ and $\pi_{\ell-1}$ are dissimilar or disjoint, the SYNCE coupling method maintains a robust convergence rate determined solely by the efficiency of the marginal random walk exploration. This robustness stems from the methodology of synchronizing the steps of both chains instead of attempting to force them to accept the same sample as seen in~\cref{fig:algorithm-figures}.

The theoretical analysis also provides insights towards improving its performance further through the lens of adaptation and resynchronization. To achieve fast convergence rates, we need to minimize the constant $\rho$. This requires us to maximize the proposal bound $\epsilon_q$ and minimize the drift constant $\hat{\Lambda}_{\text{SYNCE}}$. The constant $\epsilon_q$ is a property of the proposal distribution $q$, representing the minimum probability density for making small moves. A larger $\epsilon_q$ can be achieved by choosing a proposal with an appropriate step size (covariance)—not too large, not too small -- that encourages efficient local exploration. The drift constant $\hat{\Lambda}_{\text{SYNCE}}$ reflects how strongly the chain is pulled back from the tails of the distribution. The pull is strongest when the proposal distribution $q$ is well-matched to the geometry of the target posterior. A proposal that suggests moves aligned with the target's shape and scale will lead to higher acceptance rates and a stronger, faster drift back to the high-probability region, resulting in a smaller $\hat{\Lambda}_{\text{SYNCE}}$. This motivates the need for adaptation and resynchronization strategies, which we discuss in~\cref{subsec:covariance-adaptation,subsec:resynchronization}.

We compare the convergence rates of the SYNCE kernel with the three existing coupling methods. While explicitly comparing the convergence rates is challenging due to the complexity of the constants involved, we analyze the limiting factors of the convergence rate in the worst-case scenario -- when the two posterior distributions $\pi_{\ell}$ and $\pi_{\ell-1}$ are dissimilar and have little to no overlap. This regime is particularly important because it corresponds to cases with distinct model fidelities or changing data between levels.

\paragraph{Independent proposal/Delayed acceptance coupling}
For the independent proposal coupling method in~\cref{alg:independent-proposal-coupling}, the \textit{uniform}  convergence rate bound derived in~\cite{madrigal-cianci_analysis_2023} is given by
\begin{align*}
  \left\|p^n(x, \cdot) - \mu(\cdot)\right\|_{TV} \leq (1 - \rho)^n, \quad \forall n\geq 0, \forall x \in X, \rho \in (0,1),
\end{align*}
with $\rho = c\left(1 - \left\|\pi_{\ell} - \pi_{\ell-1}\right\|_{TV}\right)$. The constant $c \in (0,1)$ is related to the proposal distribution $c \leq \text{ess} \inf_{x \in X} \left(q^*_{\ell}(x) / \pi_{j}(x)\right)$. Here, $q^*_{\ell}(x)$ is the independent proposal distribution in~\cref{alg:independent-proposal-coupling} and $j \in \{\ell, \ell-1\}$. The delayed acceptance method in~\cref{alg:delayed-acceptance-coupling} is a special case with $q^*_{\ell} = \pi_{\ell-1}$. If the two posterior distributions $\pi_{\ell}, \pi_{\ell-1}$ do not overlap, then the TV distance $\left\|\pi_{\ell} - \pi_{\ell-1}\right\|_{TV}$ will be close to $1$ and the convergence will be slow because $\rho$ will be close to $0$. This dependence on posterior overlap is a significant drawback of the independent proposal coupling method and the delayed acceptance coupling method.

\paragraph{Maximal coupling}
The maximal coupling method in~\cref{alg:maximal-coupling} is designed to maximize the probability that the two chains propose the same state, $\theta^*_{\ell} = \vartheta^*_{\ell-1}$. 
As a random walk algorithm, its convergence is governed by the same theoretical framework as the SYNCE kernel established in~\cref{thm:convergence-synce}. However, a critical difference lies in the minorization constant $\delta$. For maximal coupling, the construction of the minorizing measure $\nu$ relies on the event that the chains successfully couple (see Lemma 6.3.2 in~\cite{cianci_thesis}). Specifically, the minorization constant $\delta$ for this method is bounded by the integral of the overlap between the proposal densities, $\int \min \left( q_{\ell}(\cdot \mid \theta^i_{\ell}), q_{\ell-1}(\cdot \mid \vartheta^i_{\ell-1}) \right) dx = 1 - \left\| q_{\ell}(\cdot \mid \theta^{i}_{\ell}) - q_{\ell-1}(\cdot \mid \vartheta^{i}_{\ell-1}) \right\|_{TV}$. If the posteriors $\pi_{\ell}$ and $\pi_{\ell-1}$ are dissimilar, the state-dependent proposals $q_{\ell}$ and $q_{\ell-1}$ will also have small overlap. In this limit, the minorization constant $\delta \to 0$, causing the convergence rate $\rho$ to approach $1$ (see Theorem 1.3.5 in~\cite{qin2024convergenceboundsmontecarlo}). While this method improves upon the independent proposal coupling method by utilizing state-dependent proposals, it still suffers from slow convergence when the posteriors are dissimilar.

The distinct advantages and disadvantages of each coupling strategy are summarized in \cref{tab:methods-pro-cons}, which highlights how the SYNCE coupling method effectively addresses the limitations of existing approaches by ensuring high correlation between chains for any two posterior distributions while being computationally efficient and straightforward to implement.

\begin{table}[H]
  \centering
  \begin{tabularx}{\textwidth}{|l|X|X|}
    \hline
    \textbf{Method} & \textbf{Pros} & \textbf{Cons} \\
    \hline
    DA ML-MCMC~\cite{dodwell2015hierarchical} & Good correlation for close posteriors & Factorial cost, non-ergodic \\
    \hline
    MLDA~\cite{lykkegaard_multilevel_2023} & Good correlation and ergodic & Factorial cost, requires consistent data \\
    \hline
    Independent proposal~\cite{madrigal-cianci_analysis_2023} & Good correlation and ergodic & Requires good independent proposal \\
    \hline
    Maximal coupling~\cite{cianci_thesis} & State dependent proposals & Requires overlap between proposals \\
    \hline
    SYNCE coupling [this paper] & State dependent proposals, high correlation, cheap, easy to implement & Requires proposal covariance tuning \\ 
    \hline
  \end{tabularx}
  \caption{Advantages and disadvantages of all the methods discussed. SYNCE overcomes the limitations of existing methods by ensuring high correlation between chains for any two posterior distributions while being computationally inexpensive and easy to implement.}
  \label{tab:methods-pro-cons}
\end{table}

\begin{rmk}
  While~\cref{thm:convergence-synce} establishes geometric ergodicity of the SYNCE coupled kernel, obtaining rigorous theoretical guarantees for the overall cost and convergence rates of the ML-MCMC estimator requires verifying a set of technical assumptions~\cite{cliffe_multilevel_2011,dodwell2015hierarchical,madrigal-cianci_analysis_2023}. Providing a theoretical proof that these assumptions hold in general for the SYNCE coupling method is beyond the scope of this paper. Instead, we provide numerical evidence in~\cref{sec:experiments} that the assumptions are satisfied for a subsurface flow problem taken directly from~\cite{madrigal-cianci_analysis_2023}. For completeness, we present the full set of assumptions and the resulting $\epsilon$-cost complexity theorem in the supplementary material~\cref{supp:ml-mcmc-cost-synce}.
\end{rmk}

\subsection{Adaptation}\label{subsec:covariance-adaptation}
The performance of SYNCE is highly dependent on the choice of the covariance $C_{\ell}$ for the same-step $\eta$ in~\cref{alg:synce-coupling}. In~\cref{fig:synce-proposal-coupling}, when steps of high magnitude are proposed, the proposals are likely to be rejected, especially if the posteriors are different. This leads to low acceptance rates, suboptimal mixing and small effective sample sizes. On the other extreme, when the steps are too small, samples are accepted quite often and the correlation is quite high, but the chains do not explore the posterior space effectively. To balance exploration, mixing, and correlation, we employ adaptive MCMC techniques~\cite{haario_adaptive_2001,rosenthal_optimal_nodate,andrieu_tutorial_2008}. These works provide a framework to adapt the proposal distribution such that optimality in the sense of some target measure is reached. 

Standard adaptation algorithms update a scalar $\lambda$ that scales the empirical covariance of the samples collected so far to ensure that the acceptance rate of the chain is close to some target value $\alpha^*$. The updates are given by (Algorithm 4 in~\cite{andrieu_tutorial_2008}):
\begin{align}\label{eq:adapt-covariance-updates}
  log(\lambda^{i+1}) &= log(\lambda^i) + \gamma^{i+1}\left(\alpha(\theta^i,\theta^*)-\alpha^*\right), \nonumber\\
  \mu^{i+1} &= \mu^i + \gamma^{i+1}(\theta^{i+1}-\mu^i), \nonumber\\
  \Sigma^{i+1} &= \Sigma^i + \gamma^{i+1}\left((\theta^{i+1}-\mu^i)(\theta^{i+1}-\mu^i)^T-\Sigma^i\right),
\end{align}
where $\gamma^{i+1}$'s are chosen to satisfy the diminishing adaptation condition~\cite{andrieu_tutorial_2008}. In our coupled framework, we update the scaling factor $\lambda$, empirical mean $\mu$, and covariance $\Sigma$ individually for each chain, while maintaining coupling through a shared underlying random number. Specifically, we generate a common random number $\eta$ from the standard normal distribution and multiply it with the square root of the adapted covariance and the scaling factor for each chain separately. The coupled proposals are now given by:
\begin{align}\label{eq:adapt-coupled-covariance}
  \eta \sim \mathcal{N}(0, \mathcal{I}_{d}), \quad \theta_{\ell}^{*} = \theta_{\ell}^{i} + \lambda^i_{\ell}\sqrt{\Sigma^i_{\ell}}\eta, \quad \vartheta_{\ell-1}^{*} = \vartheta_{\ell-1}^{i} + \lambda^i_{\ell-1}\sqrt{\Sigma^i_{\ell-1}}\eta,
\end{align}
where $d$ is the dimension of the parameter space. This formulation ensures that the chains are coupled, and the samples are highly correlated while exploring the individual posterior spaces effectively. This specific form has ties to the optimal transport equation for two Gaussians~\cite{villani2009optimal}, specifically for our case, transporting the sample from the standard normal Gaussian to the level-dependent marginal.

\subsection{Resynchronization}\label{subsec:resynchronization}
Covariance adaptation enables SYNCE to improve correlation between chains targeting dissimilar posteriors. However, in high-dimensional problems, we may encounter stages where the proposed step is accepted by one chain and rejected by the other. Over many iterations, this can cause the absolute states of the chains to become decorrelated, even though the underlying proposal mechanism remains synchronized. This phenomenon exposes a fundamental trade-off: SYNCE correlates the change in state, which is robust, but it does not directly force the states themselves to be correlated, which depends on the relative acceptance behavior of the two chains. To counteract this drift, we periodically resynchronize the chains to the same state. Specifically, we construct a mixture kernel~\cite{cianci_thesis}: the SYNCE coupling kernel $p_{\ell}^{\text{SYNCE}}$ and another kernel $p^{*}_{\ell}$ that proposes the same sample for both the chains, such as the independent proposal kernel or the delayed acceptance kernel. The transition kernel $p^{\text{SYNCE-R}}_{\ell}$ is then given by:
\begin{align}\label{eq:resynchronization-kernel}
  p^{\text{SYNCE-R}}_{\ell} = \omega_{\ell}p^{*}_{\ell} + (1-\omega_{\ell})p_{\ell}^{\text{SYNCE}},
\end{align}
where $\omega_{\ell} \in (0,1)$ determines the frequency of resynchronization. Crucially, resynchronization is only beneficial when the posteriors exhibit sufficient overlap. If the distributions are disjoint, resynchronization will likely lead to both chains rejecting the proposed point, resulting in no state change and wasted computational effort. Therefore, we set $\omega_{\ell}$ to 0 for small $\ell$ values and increase it as $\ell$ increases. This ensures that when the posteriors are far off from each other, no resynchronization takes place and the chains are coupled using the SYNCE coupling kernel. When the posteriors become closer with higher levels, the chains are resynchronized using both the SYNCE coupling kernel and the resynchronization kernel. We note that the optimal choice of $\omega_{\ell}$ is non-trivial. In practice, we estimate the degree of overlap using samples from the burn-in period and tune $\omega_{\ell}$ accordingly. While an adaptive strategy for $\omega_{\ell}$ could be developed, we leave this for future work.

The final adaptive proposed coupling method is given in \cref{alg:synce-coupling-adaptive}. The algorithm improves upon \cref{alg:synce-coupling} by adapting and scaling the covariance for both the chains separately and simultaneously, and resynchronizing the chains when needed for better correlation.
\begin{algorithm}
  \caption{SYNCE with adaptation and resynchronization (SYNCE-AR) (Line~\ref{line: proposal-coupling-mlmcmc} in \cref{alg:ml-mcmc})}
  \label{alg:synce-coupling-adaptive}
    \begin{algorithmic}[1]
    \STATE{\textbf{Input}: Target densities $\pi_{\ell}, \pi_{\ell-1}$; current states $\theta_{\ell}^{i}, \vartheta_{\ell-1}^{i}$; adaptive parameters $\lambda^i_{\ell}, \Sigma^i_{\ell}, \lambda^i_{\ell-1}, \Sigma^i_{\ell-1}$; resync weight $\omega_{\ell}$}
    \STATE{\textbf{Output}: Next states $\left\{\theta_{\ell}^{*}, \vartheta_{\ell-1}^{*}\right\}$}
    \STATE{Sample $w \sim \mathcal{U}(0, 1)$} 
    \STATE{Sample $\eta \sim \mathcal{N}(0, \mathcal{I}_d)$}
    \IF{$w \leq \omega_{\ell}$} 
      \STATE{$\left(\theta_{\ell}^{*}, \vartheta_{\ell-1}^{*}\right)$ = \cref{alg:independent-proposal-coupling}/\cref{alg:delayed-acceptance-coupling}} \COMMENT{resynchronize}
    \ELSIF{$w > \omega_{\ell}$}
      \STATE{Set $\theta^*_{\ell} = \theta_{\ell}^{i} + \lambda^i_{\ell}\sqrt{\Sigma^i_{\ell}}\eta$ and $\vartheta^*_{\ell-1} = \vartheta_{\ell-1}^{i} + \lambda^i_{\ell-1}\sqrt{\Sigma^i_{\ell-1}}\eta$} \COMMENT{SYNCE-A}
    \ENDIF
    \RETURN $\left(\theta_{\ell}^{*}, \vartheta_{\ell-1}^{*}\right)$ 
    \end{algorithmic}
\end{algorithm}
Using resynchronization only for the fine levels, the adaptive parameters $\lambda, \mu$ and $\Sigma$ are used to set the mean and covariance of the independent proposal distribution in \cref{alg:independent-proposal-coupling}: $\mu^{\text{IMH}}_{\ell} = 1/2\left(\mu^i_{\ell} + \mu^i_{\ell-1}\right)$ and $\Sigma^{\text{IMH}}_{\ell} = 1/2\left(\Sigma^i_{\ell} + \Sigma^i_{\ell-1}\right)$ for $\ell = 1,2,\ldots,L$. This ensures that for close and similar shaped posteriors, we obtain highly correlated chains with good mixing properties.

\section{Experimental results}\label{sec:experiments}
In this section, we validate the performance of the SYNCE coupling method through four numerical examples. The first example is a simple Gaussian taken from~\cite{madrigal-cianci_analysis_2023} to illustrate the effectiveness and sanity of the proposed methodology. The second example is a two-dimensional extension to the first Gaussian problem that highlights the need for adaptation and resynchronization in the SYNCE coupling method. These two examples are not BIP involving a likelihood function, but are useful to illustrate the performance of the proposed coupling method. The third example is a prey-predator model governed by a system of ordinary differential equations (ODEs) taken from~\cite{lykkegaard_multilevel_2023} to compare the proposed method with the AEM enhanced MLDA coupling method described in~\cref{subsubsec:delayed-acceptance-coupling}. The final example involves uncertainty quantification for Darcy flow~\cite{madrigal-cianci_analysis_2023,dodwell2015hierarchical}. This example is a well known benchmark and will be used to show that the proposed methodology satisfies the assumptions needed for the $\epsilon$-cost complexity theorem in the supplementary material~\cref{supp:ml-mcmc-cost-synce}.

\begin{figure}
  \centering
  \includegraphics[width=0.6\textwidth]{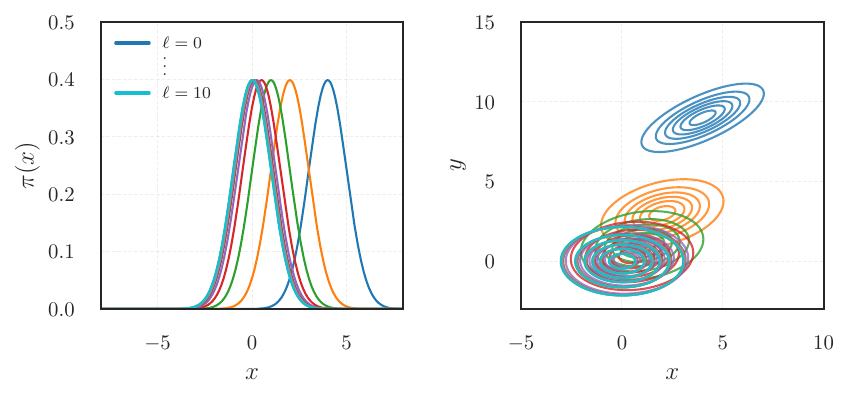}
  \caption{Posterior distributions for the Gaussian examples at different levels. The figure on the left corresponds to \cref{subsec:shifting-gaussian} and the figure on the right corresponds to \cref{subsec:rotating-shifting-gaussian}.}~\label{fig:both-gaussian-posterior}
\end{figure}

\subsection{Shifting Gaussian}\label{subsec:shifting-gaussian}
In this example, we sample from shifting Gaussians parameterized as $\pi_{\ell} = \mathcal{N}(2^{-\ell + 2}, 1)$ for $\ell = 0,1,\ldots, L$. The target posteriors are plotted on the left in \cref{fig:both-gaussian-posterior}. This setup serves as a robust test of coupling efficiency, particularly at coarser levels (e.g., $\ell=1$ and $\ell=2$) where the distributions are significantly shifted despite sharing similar shapes. We compare four methods: the delayed acceptance coupling with hierarchical subsampling (\cref{alg:recursive-independence-sampling}), the independent proposal coupling method (\cref{alg:independent-proposal-coupling}), the maximal coupling method (\cref{alg:maximal-coupling}) and the proposed SYNCE coupling method (\cref{alg:synce-coupling}). We fix $L=6$ and use the same Gaussian proposal $q_{\ell} = \mathcal{N}(2, 3)$ across all levels for the relevant algorithms. The proposal distribution for the coarsest $\ell=0$ posterior is set to $q_0 = \mathcal{N}(\cdot, 1)$ and 50,000 samples are used for all levels according to~\cite{madrigal-cianci_analysis_2023}. We use the same proposal covariance for SYNCE with $C_{\ell} = 3$ to provide a fair comparison. 

\begin{figure}[h!]
  \centering
  \begin{subfigure}[b]{0.275\textwidth}
      \centering
      \includegraphics[width=\textwidth]{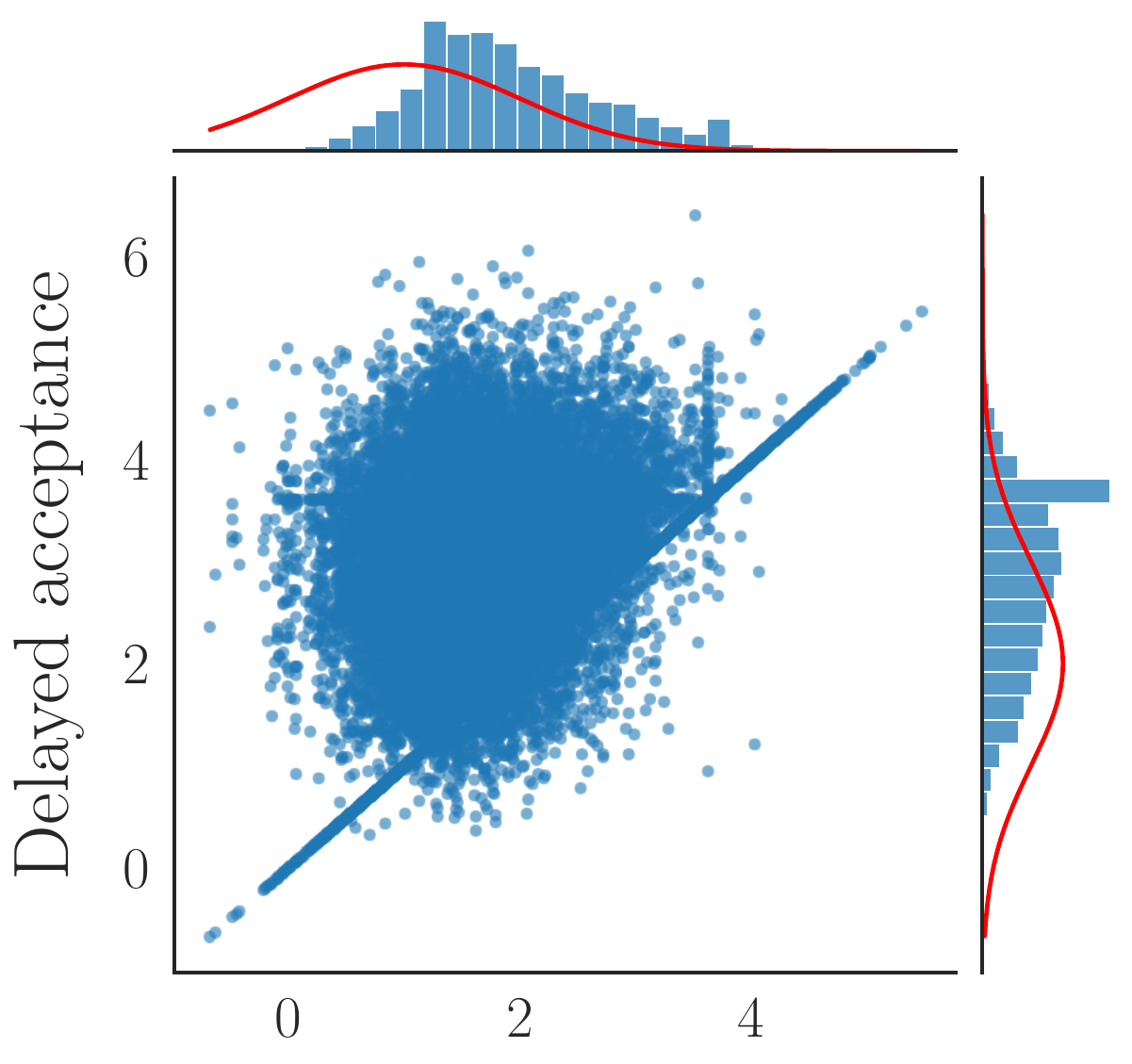}~\label{fig:shifting-gaussian-dodwell-l2}
  \end{subfigure}
  \begin{subfigure}[b]{0.25\textwidth}
      \centering
      \includegraphics[width=\textwidth]{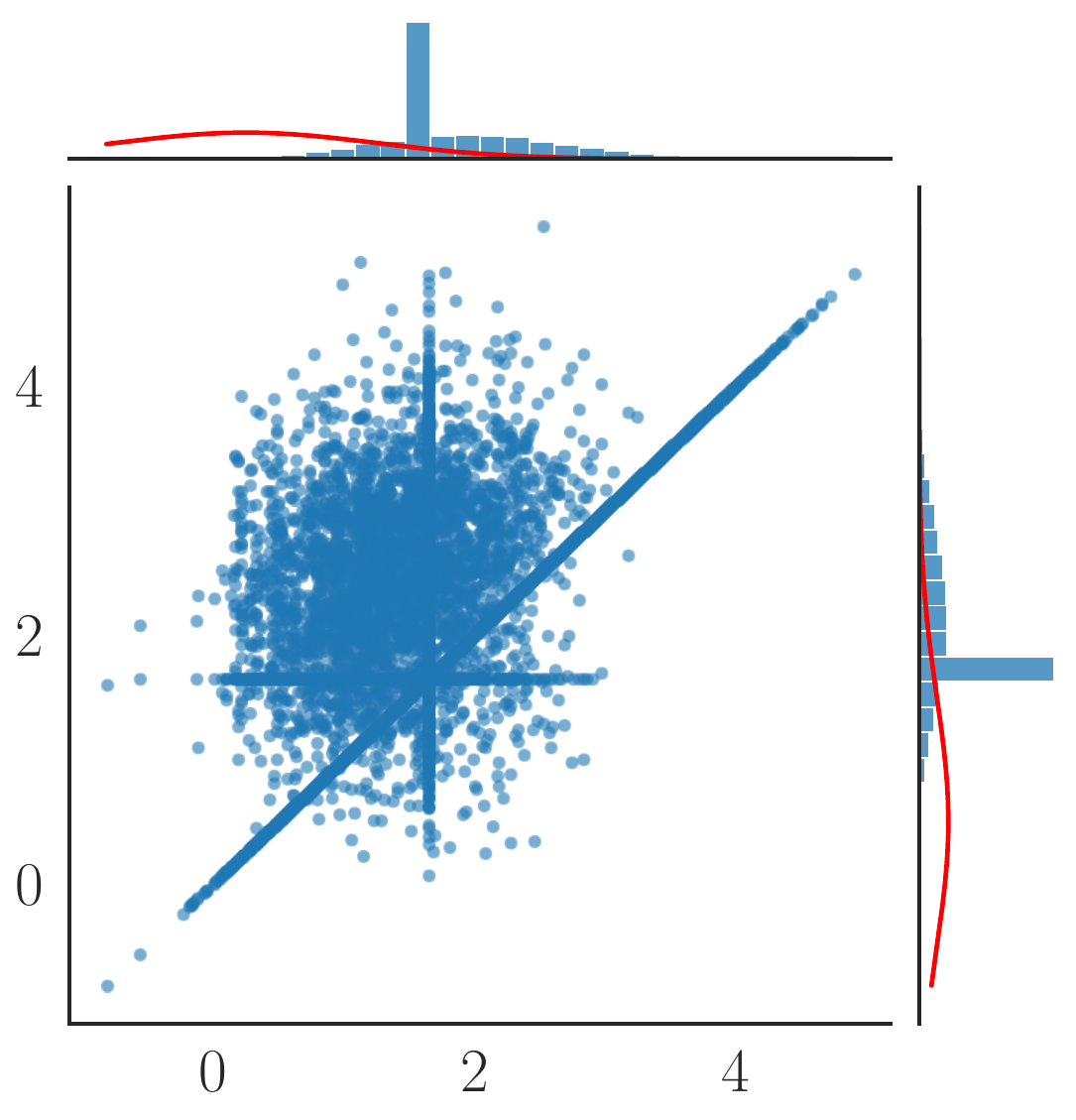}~\label{fig:shifting-gaussian-dodwell-l4}
  \end{subfigure}
  \begin{subfigure}[b]{0.25\textwidth}
      \centering
      \includegraphics[width=\textwidth]{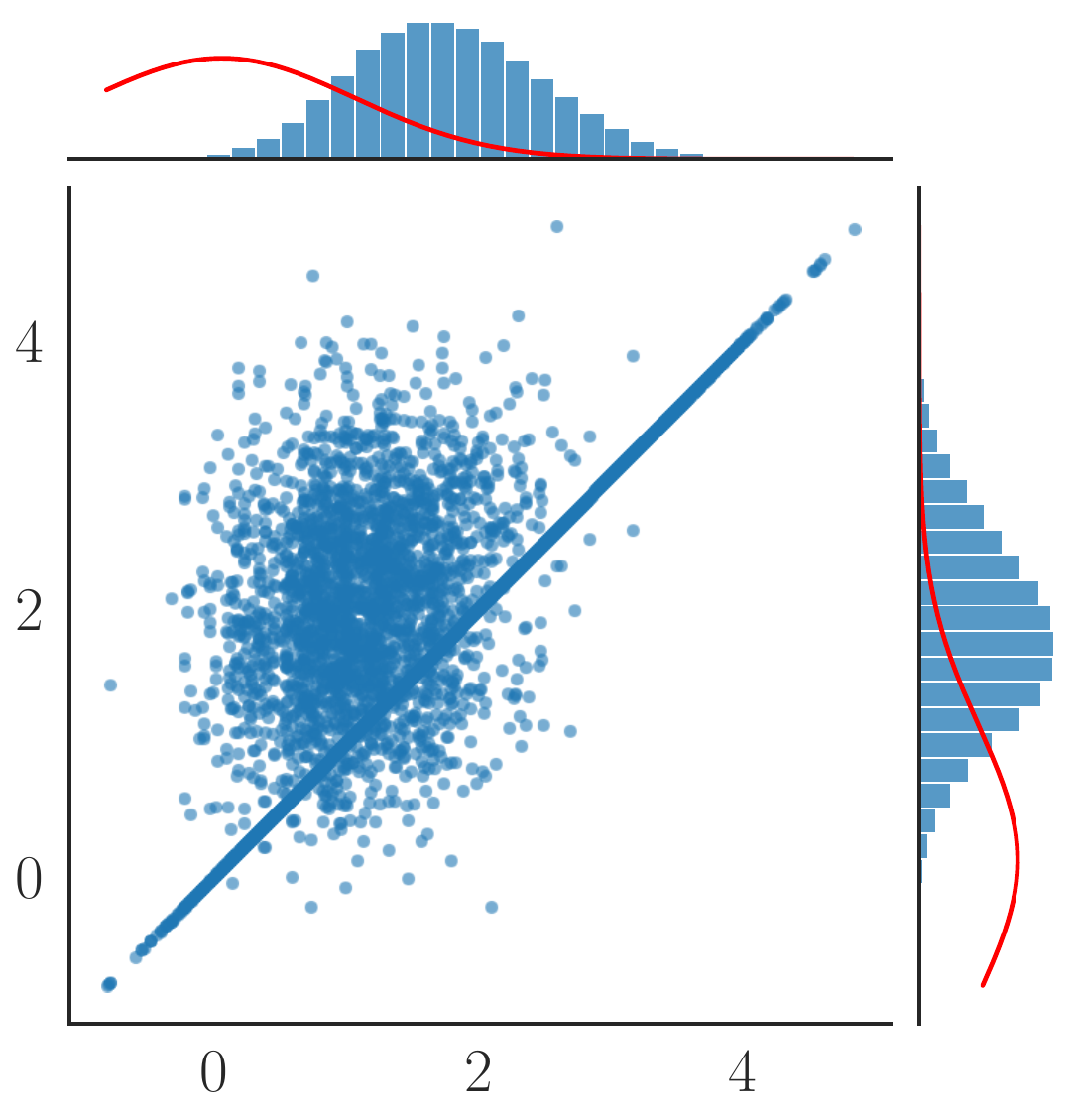}~\label{fig:shifting-gaussian-dodwell-l6}
  \end{subfigure}
  \centering
  \begin{subfigure}[b]{0.27\textwidth}
      \centering
      \includegraphics[width=\textwidth]{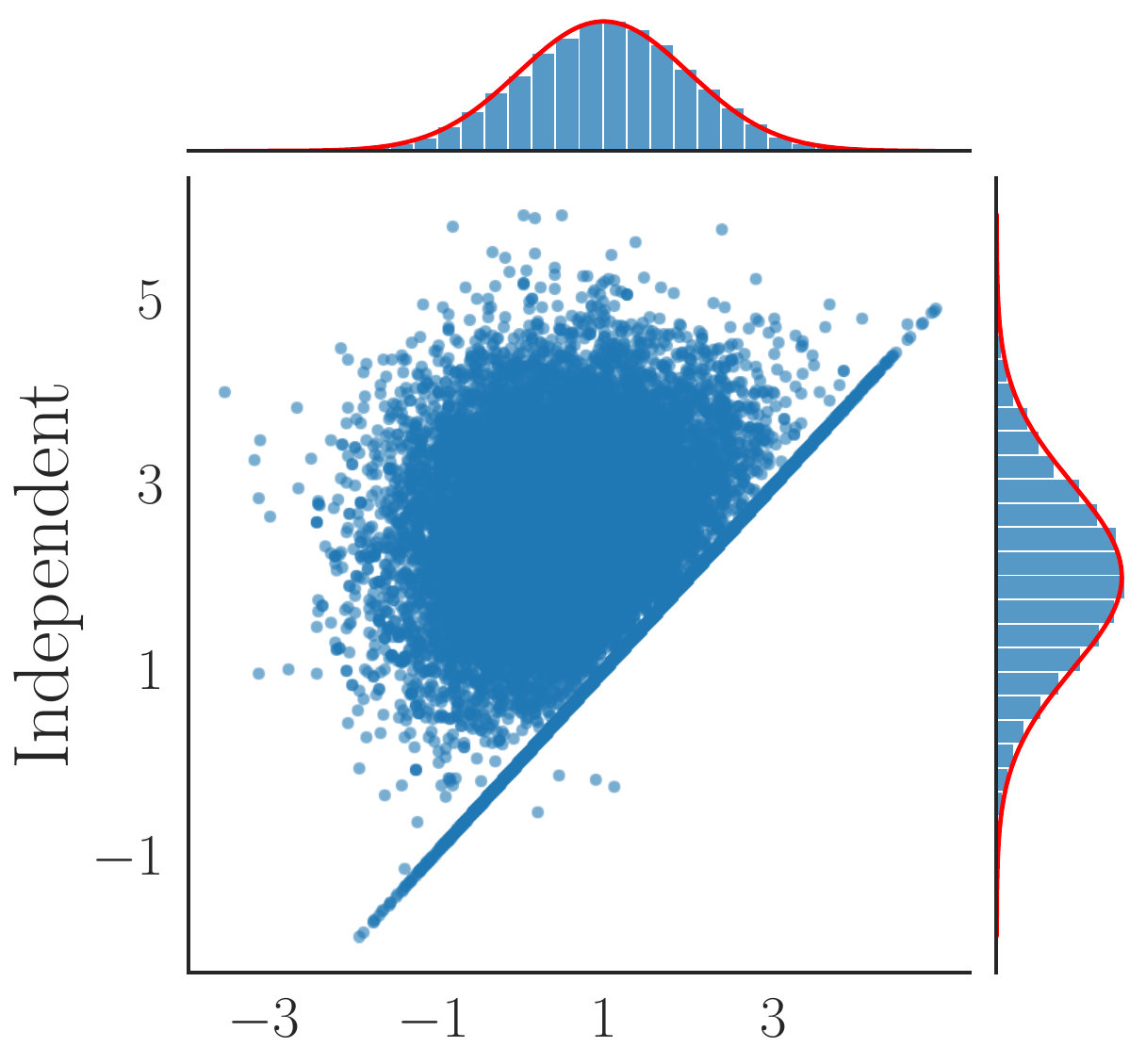}~\label{fig:shifting-gaussian-imh-l2}
  \end{subfigure}
  \begin{subfigure}[b]{0.25\textwidth}
      \centering
      \includegraphics[width=\textwidth]{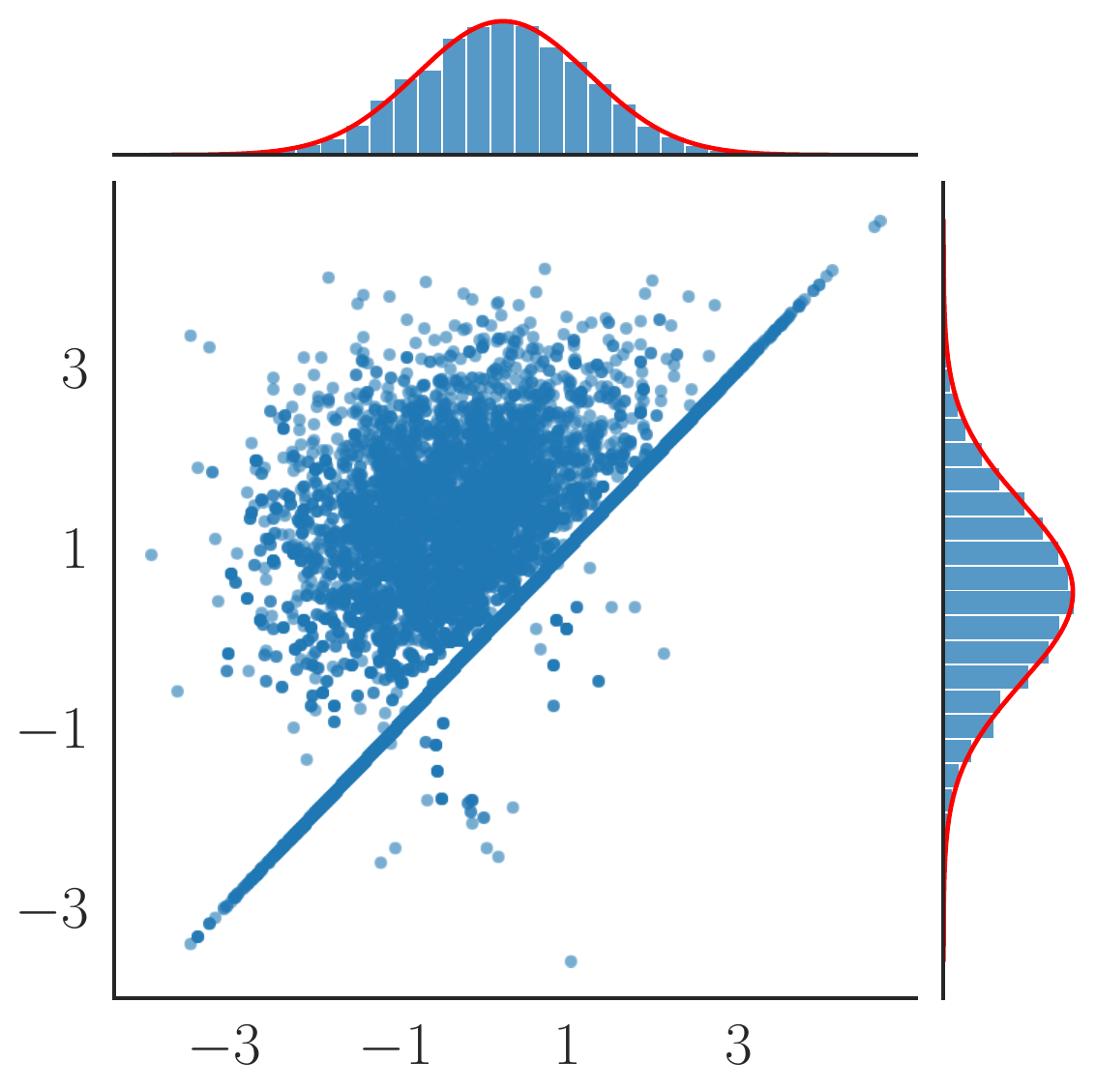}~\label{fig:shifting-gaussian-imh-l4}
  \end{subfigure}
  \begin{subfigure}[b]{0.25\textwidth}
      \centering
      \includegraphics[width=\textwidth]{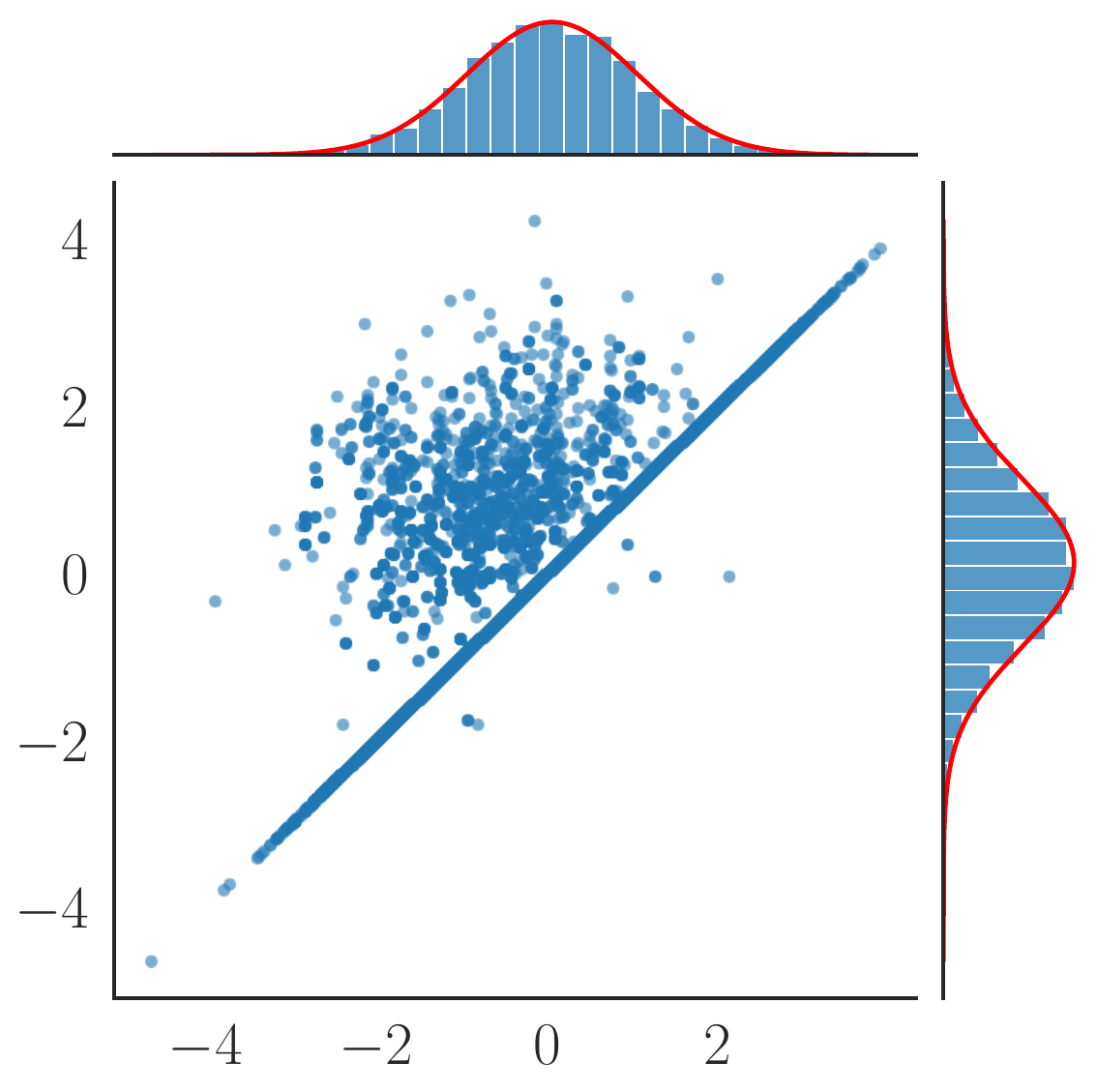}~\label{fig:shifting-gaussian-imh-l6}
  \end{subfigure}
  \centering
  \begin{subfigure}[b]{0.27\textwidth}
      \centering
      \includegraphics[width=\textwidth]{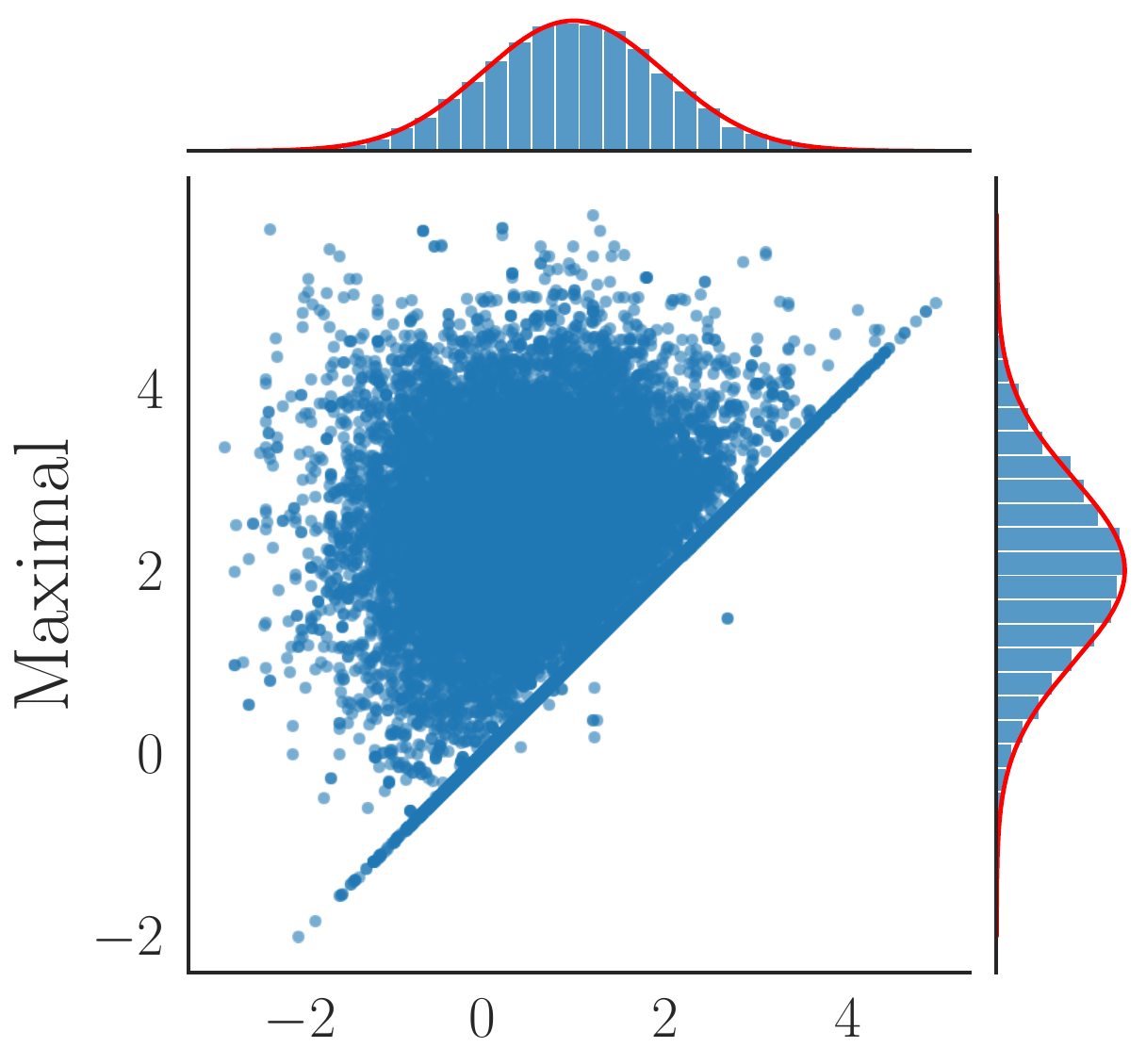}~\label{fig:shifting-gaussian-maximal-l2}
  \end{subfigure}
  \begin{subfigure}[b]{0.25\textwidth}
      \centering
      \includegraphics[width=\textwidth]{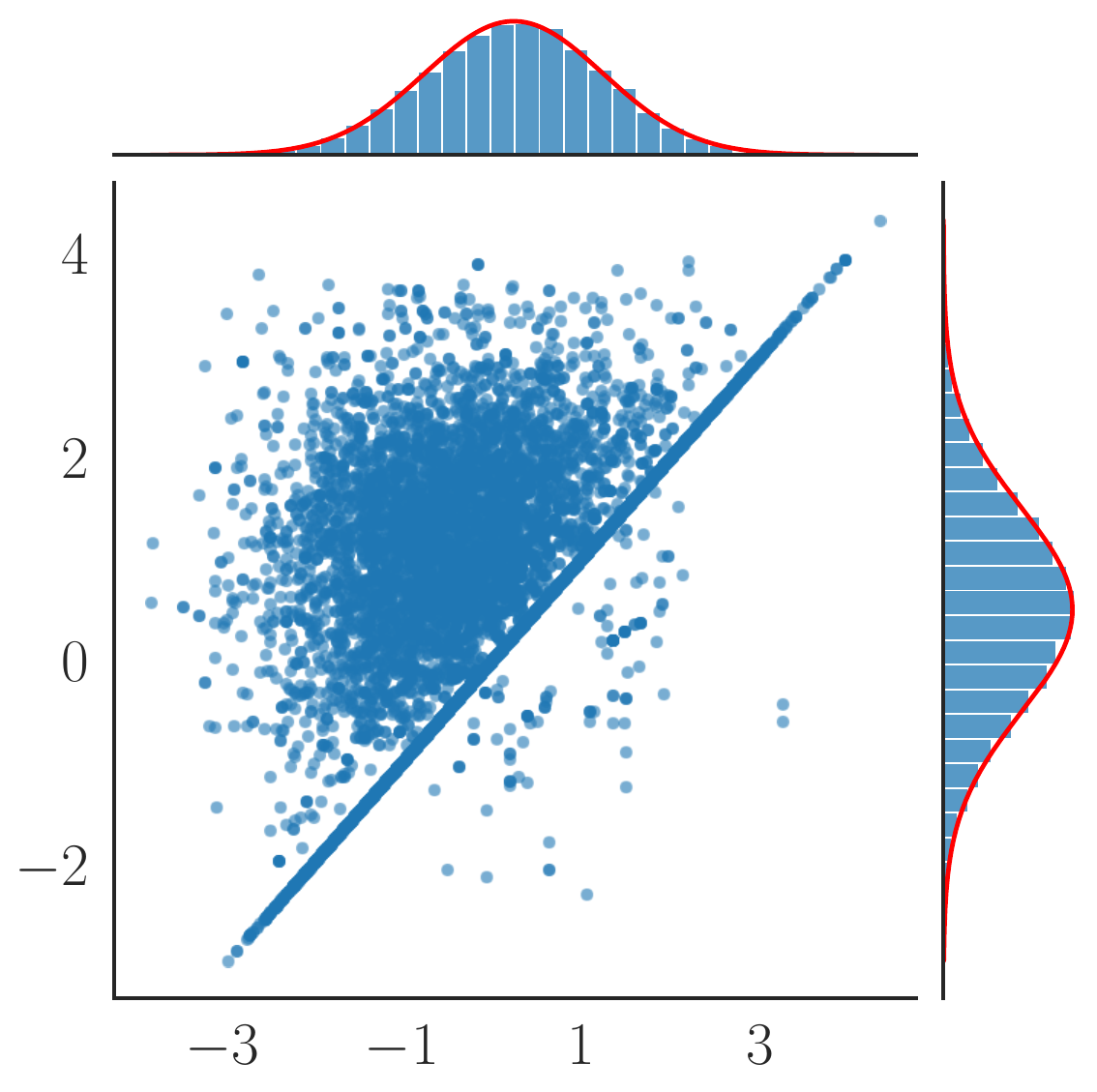}~\label{fig:shifting-gaussian-maximal-l4}
  \end{subfigure}
  \begin{subfigure}[b]{0.25\textwidth}
      \centering
      \includegraphics[width=\textwidth]{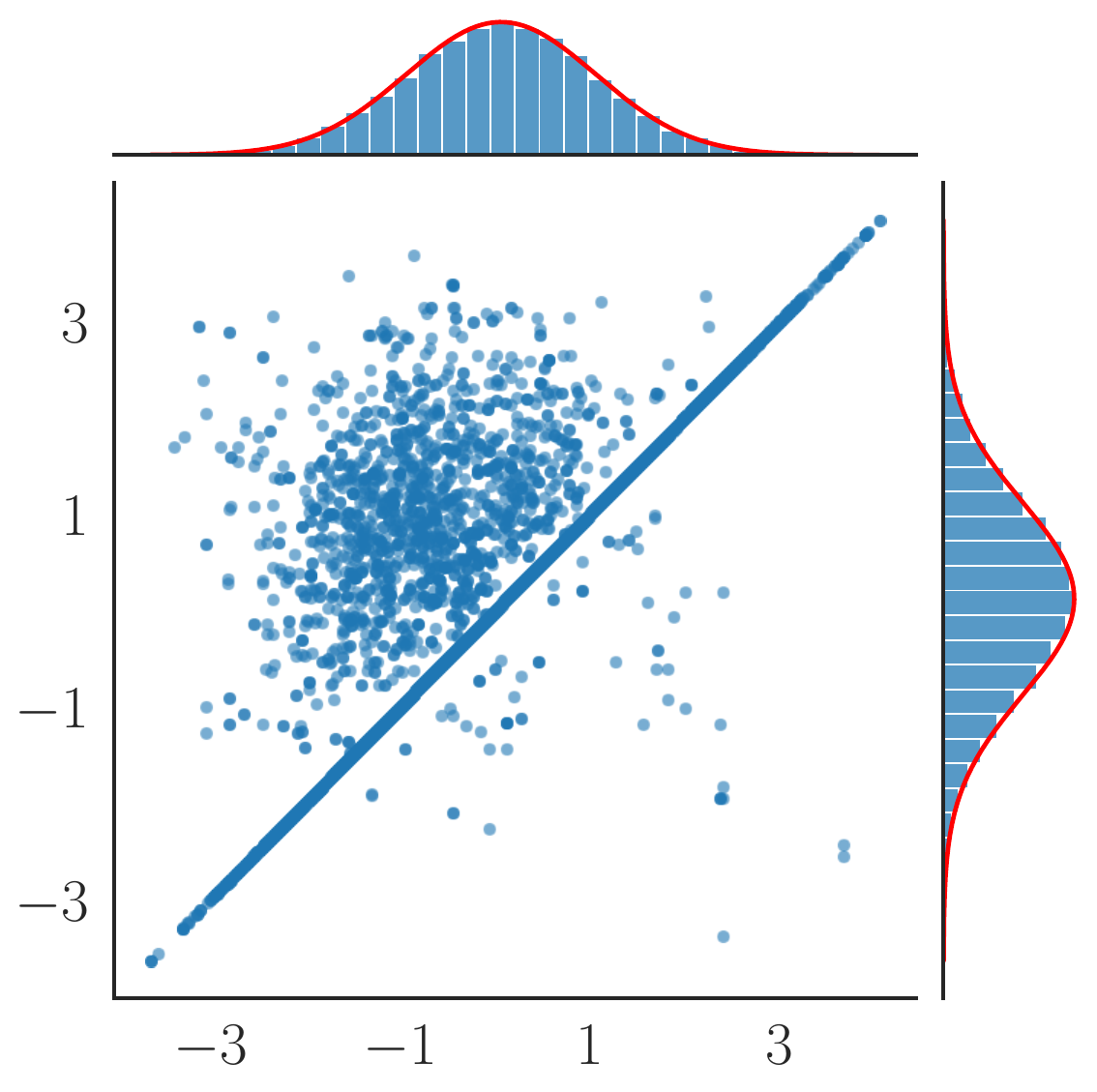}~\label{fig:shifting-gaussian-maximal-l6}
  \end{subfigure}
  \centering
  \begin{subfigure}[b]{0.27\textwidth}
      \centering
      \includegraphics[width=\textwidth]{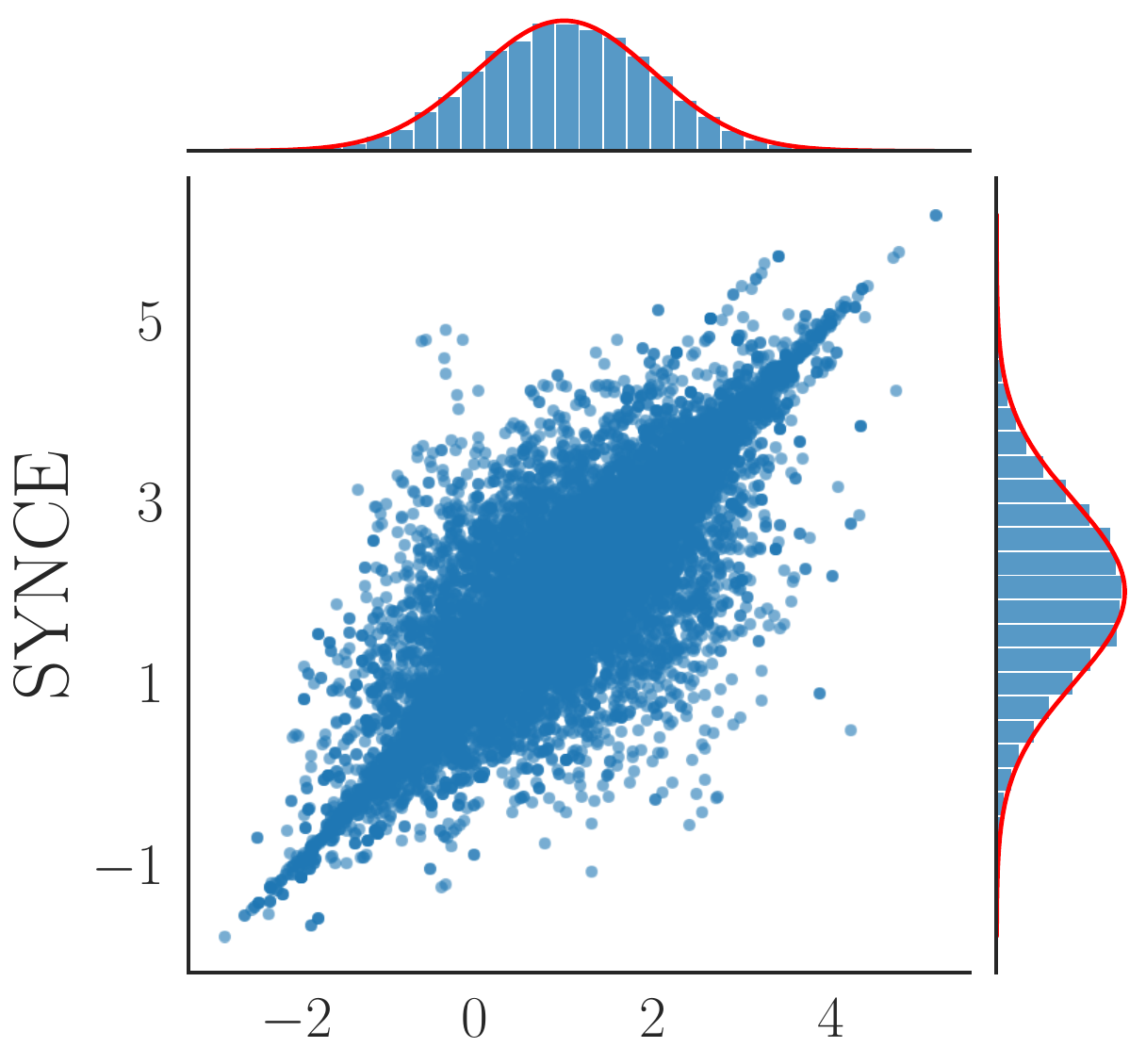}
      \caption{$\ell = 2$}~\label{fig:shifting-gaussian-pintoneal-l2}
  \end{subfigure}
  \begin{subfigure}[b]{0.25\textwidth}
      \centering
      \includegraphics[width=\textwidth]{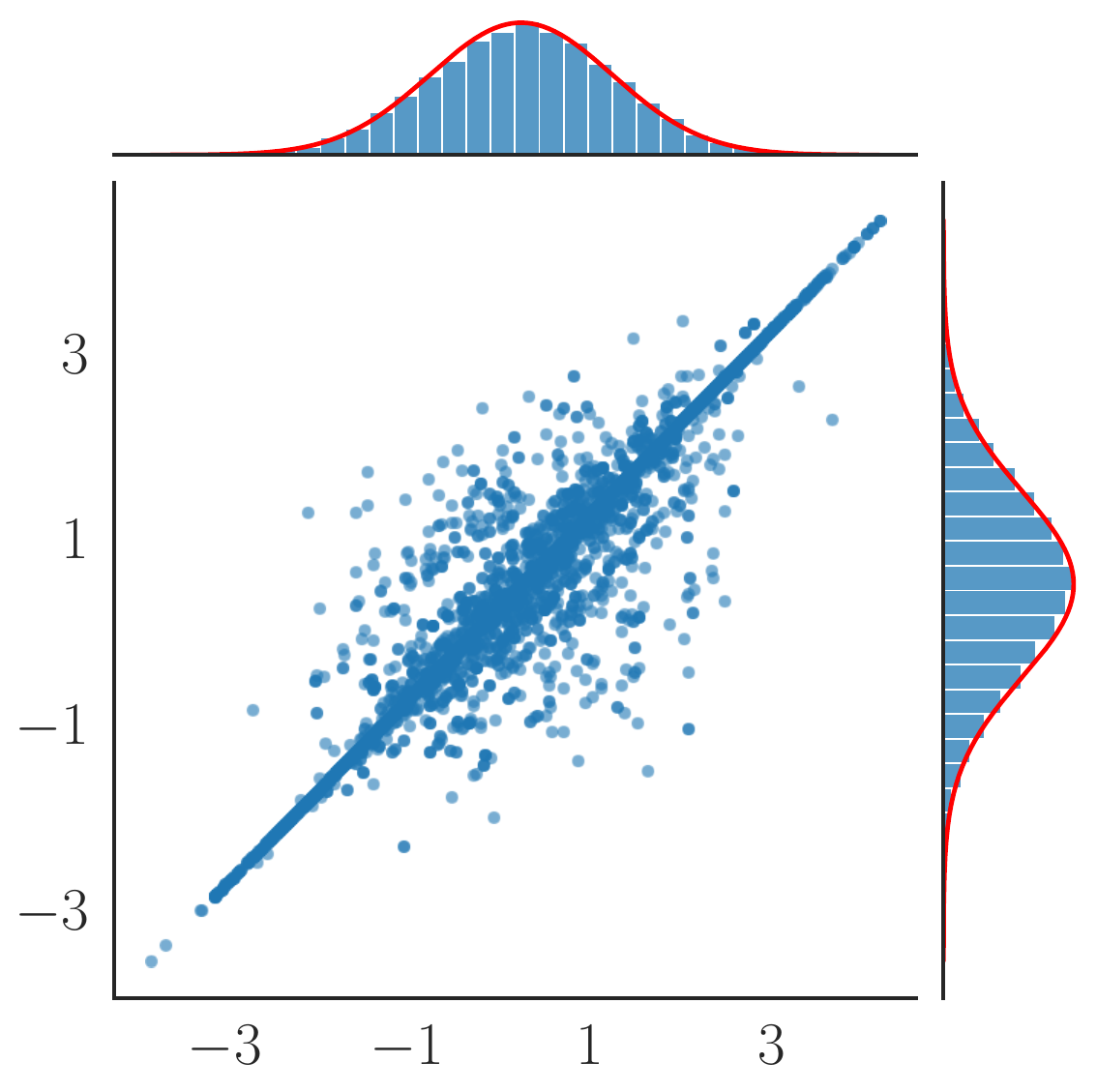}
      \caption{$\ell = 4$}~\label{fig:shifting-gaussian-pintoneal-l4}
  \end{subfigure}
  \begin{subfigure}[b]{0.25\textwidth}
      \centering
      \includegraphics[width=\textwidth]{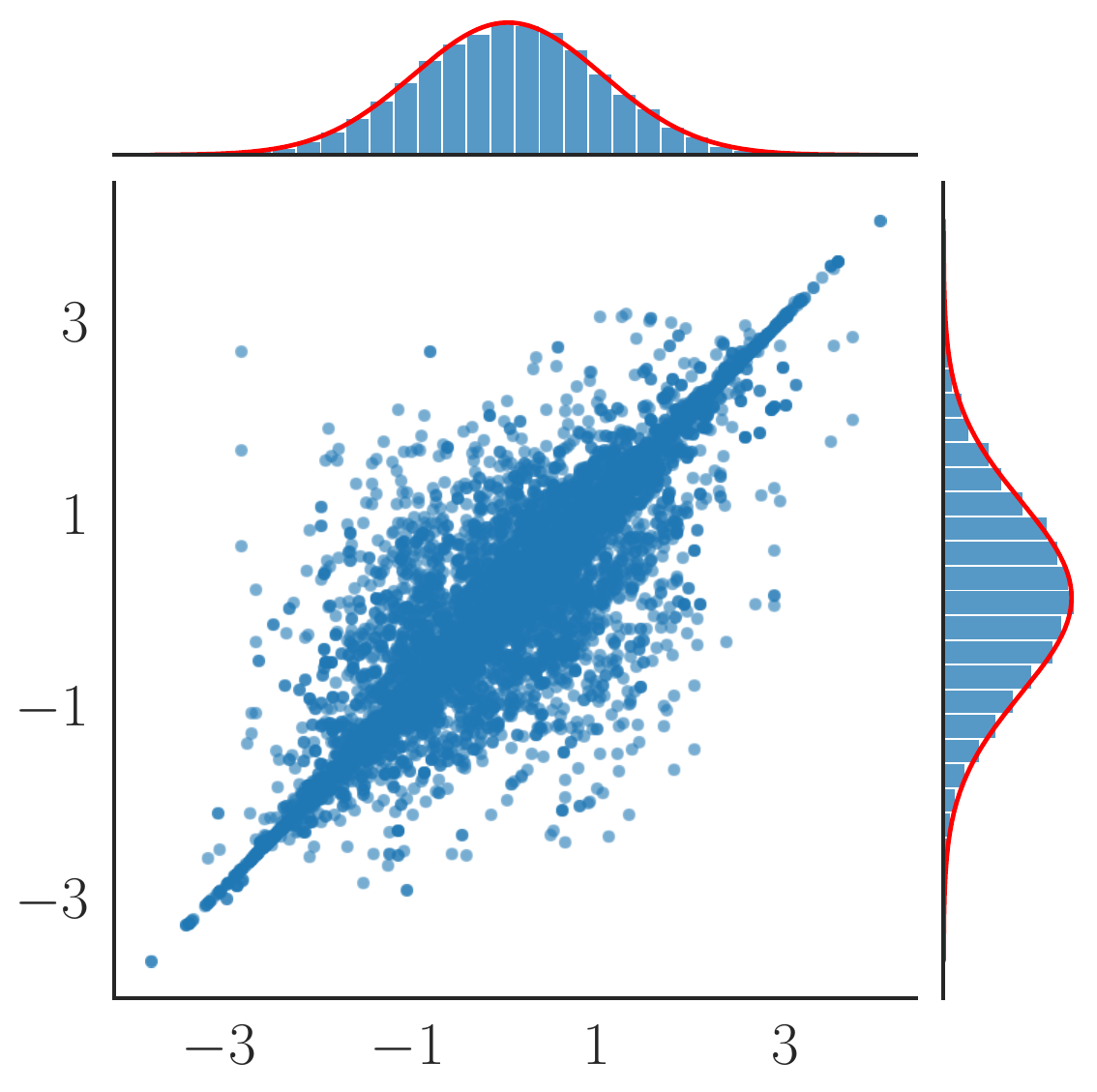}
      \caption{$\ell = 6$}~\label{fig:shifting-gaussian-pintoneal-l6}
  \end{subfigure}
     \caption{Scatter plots and histograms of samples from the different coupling algorithms for the shifting Gaussian example. For each level, the X-axis represents the samples at the $\ell$'th level and the Y-axis represents the samples at the $(\ell-1)$'th level. The red line on the histogram plots represents the true posterior distribution at the respective levels. Existing methods either fail to correctly sample from the marginals (row 1), produce diffused scatter plots with weak correlation despite correct marginals (row 2), or exhibit degrading diagonal concentration as posteriors diverge (row 3). SYNCE (row 4) consistently produces tightly clustered points along a dominant diagonal across all levels -- indicating robust correlation. The adaptive diagonal slope is not constrained to one, reflecting the relative shift between the posteriors while maintaining right marginals.}~\label{fig:shifting-gaussian-histograms}
\end{figure}

~\Cref{fig:shifting-gaussian-histograms} provides validation if the coupling algorithms are able to sample from the correct marginal distributions and demonstrate the joint distribution of levels $\ell = \left\{2,4,6\right\}$ for the four methods. In the first row of~\cref{fig:shifting-gaussian-histograms}, the DA with hierarchical subsampling method fails to correctly sample from the marginal distributions. The DA method fails because the hierarchical subsampling scheme violates the tail decay assumption required for ergodicity, as discussed in~\cref{subsubsec:independent-proposal}. The other three methods all correctly recover the true marginals, indicating that they are ergodic. However, the joint scatter plots differ significantly between the proposed SYNCE coupling method and the existing methods. The samples from the three existing methods are either perfectly coupled (lying on the line with slope one) or uncorrelated (scattered points). This behavior confirms that these methods only induce correlation when the samples fall into the shared overlapping region of the posteriors. Outside the shared region, the samples become uncorrelated, leading to scattered joint distributions above the diagonal (posteriors shifting from right to left). This scattering effect is more pronounced at coarser levels where the posteriors are more dissimilar. In contrast, the SYNCE coupling method produces a tightly clustered diagonal line with a slope that is not necessarily equal to one. This behavior indicates that SYNCE does not care about the overlap between the posteriors, effectively capturing the shift between them without any prior knowledge about their distributions.

\begin{figure}[h]
  \centering
  \includegraphics[width=0.3\textwidth]{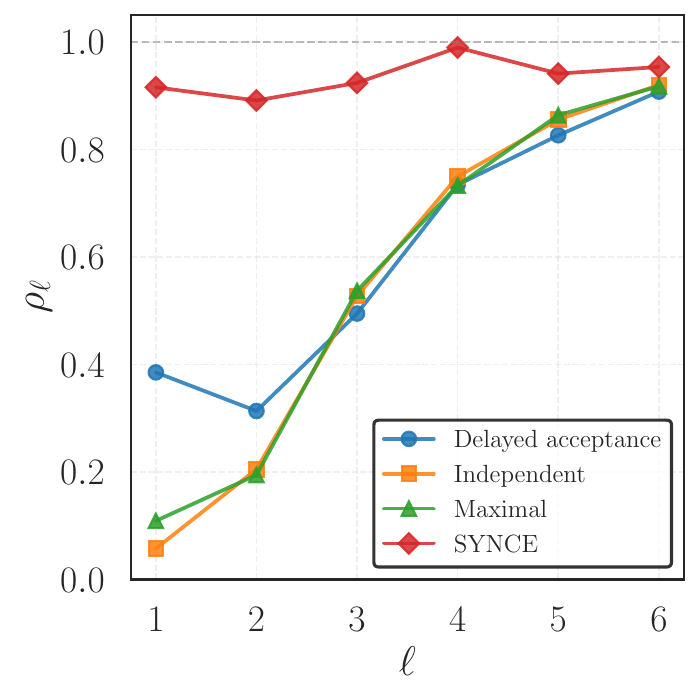}
  \caption{Pearson correlation coefficient between samples of all levels for the shifting Gaussian example. The coefficient is a linear measure of how well samples are correlated. Higher correlations yield greater variance reduction and we see that the SYNCE coupling algorithm achieves the greatest correlation, particularly for the coarser levels of the hierarchy.}~\label{fig:shifting-gaussian-correlation}
\end{figure}

In \cref{fig:shifting-gaussian-correlation}, we show how well the samples are correlated between two successive MLMC levels. We plot the Pearson correlation coefficient in~\cref{eq:pearson-correlation} between the samples at the $\ell$'th level and the $(\ell-1)$'th level for the different coupling algorithms. We see that the delayed acceptance, independent proposal and the maximal coupling methods produce samples that are not well correlated at the coarser levels. This weak correlation is a direct consequence of the fact that the posteriors at these coarse levels have minimal overlapping posterior regions. However, our method does not depend on this overlap and rather tries to couple (different) samples by proposing the same noise in the Gaussian random walk. This is why the SYNCE coupling method achieves the largest correlation, particularly for coarser levels.

\subsection{Rotating-shifting Gaussian}\label{subsec:rotating-shifting-gaussian}
Next, we consider a 2D family of Gaussians to demonstrate advantages of the SYNCE coupling method under varying covariance structures. The posterior densities are given by: 
\begin{align}\label{eq:rotating-shifting-gaussian}
  \pi_{\ell} = \mathcal{N}\left(\mu_{\ell}^{\prime} = \begin{bmatrix}
    2^{-\ell + 2} \\ 3^{-\ell + 2}
  \end{bmatrix}, \Sigma_{\ell}^{\prime} = \begin{bmatrix}
    2 & 2^{-\ell} \\ 2^{-\ell} & 1
  \end{bmatrix}\right),
\end{align}
for $\ell=0,1,\dots,L$. The posterior distributions are visualized in~\cref{fig:both-gaussian-posterior}. This example extends the first problem by introducing rotation in addition to the shift. This added complexity allows us to isolate and highlight the importance of adaptation and resynchronization discussed in~\cref{subsec:covariance-adaptation,subsec:resynchronization}. We run two different experiments with the number of levels set to $L=6$ and the number of samples in each level set to 50,000. We discard the first 20,000 samples as burn in and apply any adaptation in this period to get a good estimate of the adaptation parameters. In the first experiment, we compare the baseline SYNCE coupling method without adaptation and resynchronization (\cref{alg:synce-coupling}) with the maximal coupling method (\cref{alg:maximal-coupling}) for Gaussian random walk proposals. In the second experiment, we compare three algorithms: SYNCE-A, SYNCE-AR (\cref{alg:synce-coupling-adaptive}) and the independent proposal (\cref{alg:independent-proposal-coupling}) coupling method. SYNCE-A refers to the proposed SYNCE coupling method with only adaptation i.e, setting $\omega_{\ell}=0$ for $\ell=1,2,\ldots,L$ in \cref{alg:synce-coupling-adaptive}. 

\subsubsection{Experiment 1: Baseline SYNCE vs maximal}
In this experiment, we set a fixed proposal distribution $q_{\ell} = \mathcal{N}\left(\cdot,[[3, 0],[0, 3]]\right)$ for all the levels $\ell = 0,1,\ldots,L$. This proposal is chosen to ensure a 44\% acceptance rate for the chains at all levels, the optimal value for Gaussian random walk proposals according to~\cite{gelman_adaptively_2007}. The consistent choice of proposal distribution also ensures a fair comparison between the two algorithms.

\begin{figure}[h]
  \centering
  \begin{subfigure}[b]{0.3\textwidth}
      \centering
      \includegraphics[width=\textwidth]{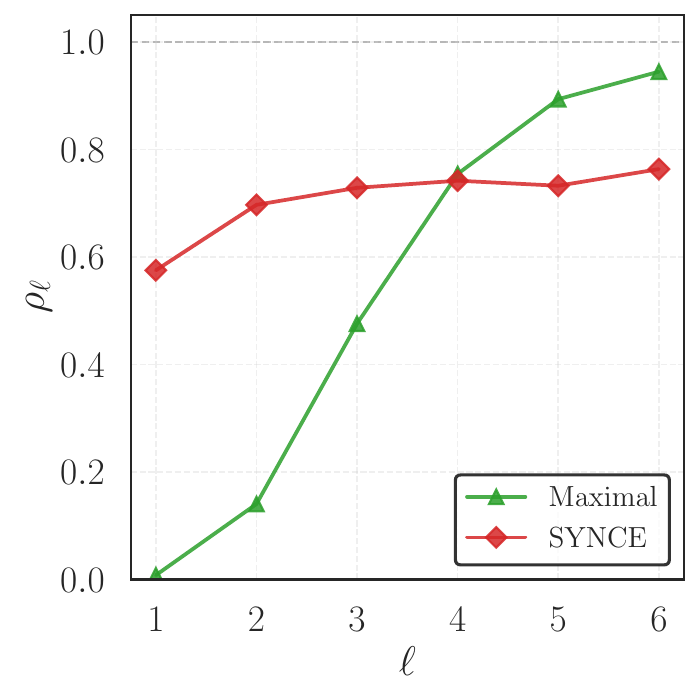}
      \caption{Correlation for each level.}~\label{fig:rot-shi-gaussian-maximal-synce-correlation}
  \end{subfigure}
  \begin{subfigure}[b]{0.3\textwidth}
      \centering
      \includegraphics[width=\textwidth]{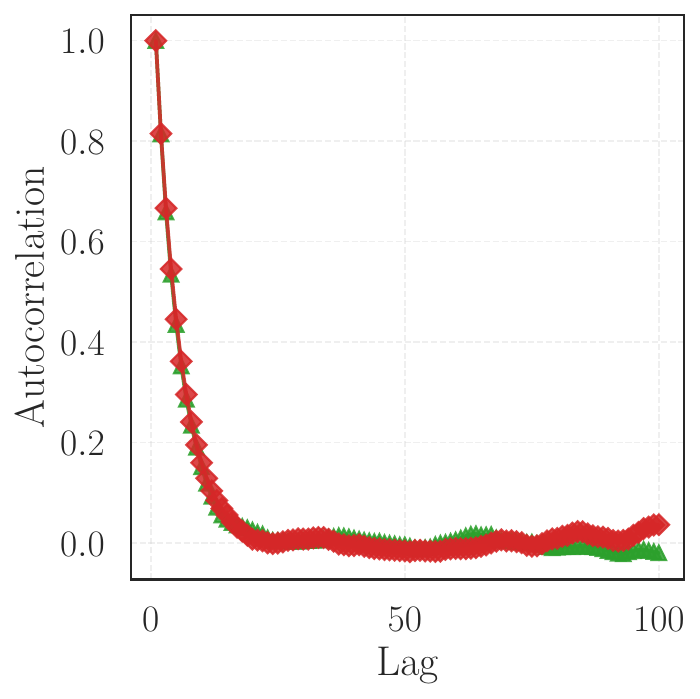}
      \caption{Autocorrelation versus lag.}~\label{fig:rot-shi-gaussian-maximal-synce-lag}
  \end{subfigure}
  \caption{Comparison of the maximal coupling method and baseline SYNCE on the rotating-shifting Gaussian example (first dimension). (a) Level-wise correlation showing SYNCE's advantage at lower levels but disadvantage at finer levels. (b) Autocorrelation at the finest level, confirming comparable sampling efficiency because of similar proposal distributions.}~\label{fig:rot-shi-gaussian-maximal-synce-correlation-lag}
\end{figure}

In \cref{fig:rot-shi-gaussian-maximal-synce-correlation}, we plot the correlation between the samples at all levels for both the methods for the first dimension. The maximal coupling method produces samples that are highly correlated at finer levels and very poorly correlated at the coarser levels. The dissimilarity between the posteriors at the coarse levels results in a high total variation distance for the coupled kernel, leading to a low probability of a common sample being proposed for the two chains. Low probability of proposing the same sample results in poor correlation between the samples for the coarse levels as seen in the figure. SYNCE however, does produce samples that are well correlated at all levels, and, there are two important observations to be made. First, there is a noticeable increasing trend in the correlation, which appears to saturate to some fixed value as we move from coarse to fine levels. At the coarsest level $\ell=1$, the correlation is significantly smaller compared to the other levels. This low correlation is attributed to the distinct shape (covariance structure) between the two posteriors $\pi_0$ and $\pi_1$ as seen in~\cref{fig:both-gaussian-posterior}. Second, the correlation at the finest level is not as high as the maximal coupling method. Even though the posteriors are quite close in shape and distance, the two chains are not exactly in sync as our method does not propose the same candidate sample for the two chains. While taking the same step is beneficial at the coarser levels, it presents drawbacks at the finer levels. In~\cref{fig:rot-shi-gaussian-maximal-synce-lag}, we plot the autocorrelation of the samples obtained from the two methods for the finest level. As expected, both the methods show similar decays in the autocorrelation with the same covariance of the proposal distribution. The moderate correlation at the finer levels suggest that we need modifications in our SYNCE algorithm. In the next experiment, we show how adaptation and resynchronization serve this purpose. 

\subsubsection{Experiment 2: SYNCE-A vs SYNCE-AR vs independent proposal coupling}\label{subsec:rot-shi-gaussian-adaptive-resync}
In this experiment, we compare the SYNCE coupling method with only adaptation (SYNCE-A), the SYNCE coupling method with adaptation and resynchronization (SYNCE-AR), and the independent proposal coupling method. For SYNCE-A, we set the resynchronization weight $\omega_{\ell}=0$ for all levels $\ell=1,2,\dots,L$ in \cref{alg:synce-coupling-adaptive}. For SYNCE-AR, we set $\omega = [0, 0, 0, 0.2, 0.3, 0.5]$ for their respective levels and choose the level wise independent proposal distribution $q_{\ell} = \mathcal{N}\left(\mu^{\textrm{IMH}}_{\ell}, [[3,0],[0,3]] \right)$ with $\mu^{\textrm{IMH}}_{\ell} = \left(\mu^{\prime}_{\ell} + \mu^{\prime}_{\ell-1}\right)/2$ as the resynchronization proposal. This proposal is chosen to satisfy the tail decay assumption and to achieve optimal mixing properties for both the coupled chains. Finally, we set the target acceptance rate to 44\% for all levels.

\begin{figure}[h]
  \centering
  \begin{subfigure}[b]{0.3\textwidth}
      \centering
      \includegraphics[width=\textwidth]{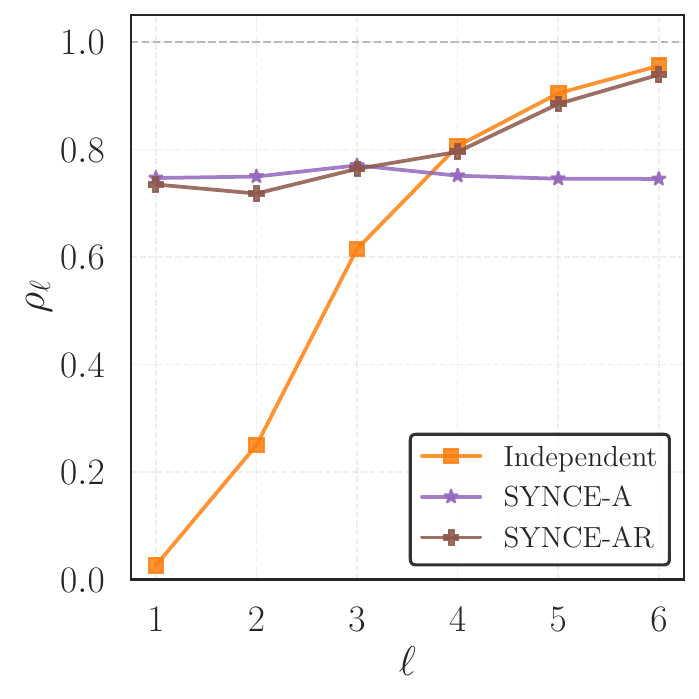}
      \caption{Correlation for each level.}~\label{fig:rot-shi-gaussian-syncear-correlation}
  \end{subfigure}
  \begin{subfigure}[b]{0.3\textwidth}
      \centering
      \includegraphics[width=\textwidth]{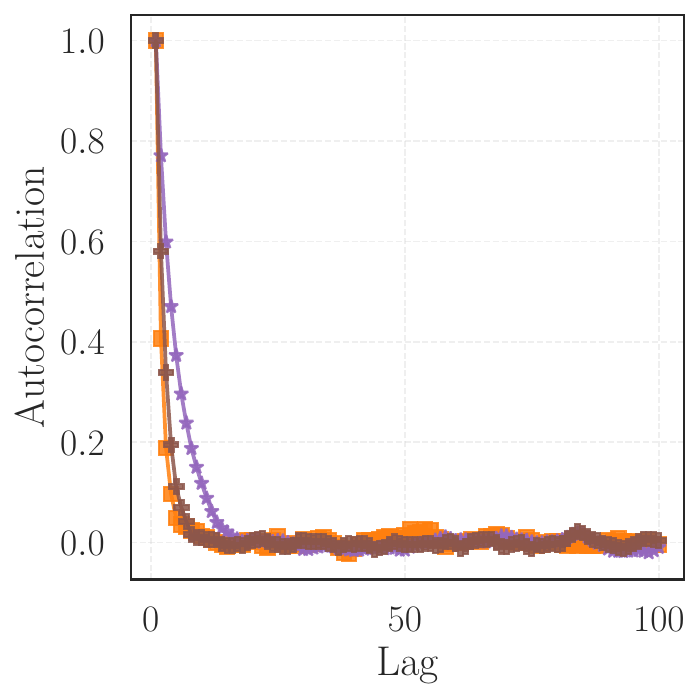}
      \caption{Autocorrelation versus lag.}~\label{fig:rot-shi-gaussian-syncear-lag}
  \end{subfigure}
  \caption{Comparison of independent, SYNCE-A, and SYNCE-AR on the rotating-shifting Gaussian example (first dimension). (a) Level-wise correlation highlighting improved correlation compared to~\Cref{fig:rot-shi-gaussian-maximal-synce-correlation} because of adaptation in the lower levels and resynchronization in the higher levels. (b) Autocorrelation at the finest level, confirming improved mixing with resynchronization.}~\label{fig:rot-shi-gaussian-syncear-correlation-lag}
\end{figure}

In~\cref{fig:rot-shi-gaussian-syncear-correlation}, we provide correlation plots as before. The effect of adaptation at the coarser levels is clear when comparing this figure with~\cref{fig:rot-shi-gaussian-maximal-synce-correlation}. Rescaling each individual chain's proposal distribution to account for the respective posterior's covariance structure ensures that the chains stay in sync as much as possible while exploring the posterior space. However, at the finer levels, we are still limited by the fact that the same sample is not proposed for the two chains. Introducing resynchronization solves this limitation, promoting better syncing of the chains, especially at the finest levels where we expect the posteriors to be quite close to each other. In~\cref{fig:rot-shi-gaussian-syncear-lag}, we plot the autocorrelation of the samples for the finest level for the first dimension. Note the improvement in the autocorrelation decay when we introduce resynchronization to the SYNCE coupling method because it promotes better mixing in fine levels. Overall, adaptation is crucial for improving correlation at coarser levels, while resynchronization enhances correlation and mixing at finer levels.

\subsection{Prey-predator problem}\label{subsec:prey-predator}
In this BIP, we adapt the Lotka-Volterra prey-predator model problem from~\cite{lykkegaard_multilevel_2023} to compare the proposed SYNCE coupling method with the AEM enhanced MLDA method described in \cref{subsubsec:delayed-acceptance-coupling}. The original problem in~\cite{lykkegaard_multilevel_2023} is not directly suitable for this comparison for two main reasons. First, the fidelity levels in the original setup are obtained by varying the time domain over which the model is solved. This choice of fidelity creates inconsistent data across levels, making it unsuitable for the AEM. Second, the resulting posterior distributions in the original problem across levels are quite similar to each other, which does not pose a challenge to multi-fidelity sampling.

To address these issues, we propose a modified version of the prey-predator problem. In this version, we maintain the same time domain of $\mathcal{T} = [0, 12]$ across all fidelity levels but introduce model discrepancy by adding some additional terms to the governing equations at coarser levels. This approach creates dissimilar posterior distributions across levels, providing a more rigorous test case (see supplementary material~\ref{supp:posterior-comparison}). The modification allows for comparison between the SYNCE method and the AEM enhanced MLDA method, implemented in the \href{https://github.com/mikkelbue/tinyDA}{\texttt{tinyDA}} library.

\begin{figure}
  \centering
  \includegraphics[width=0.75\textwidth]{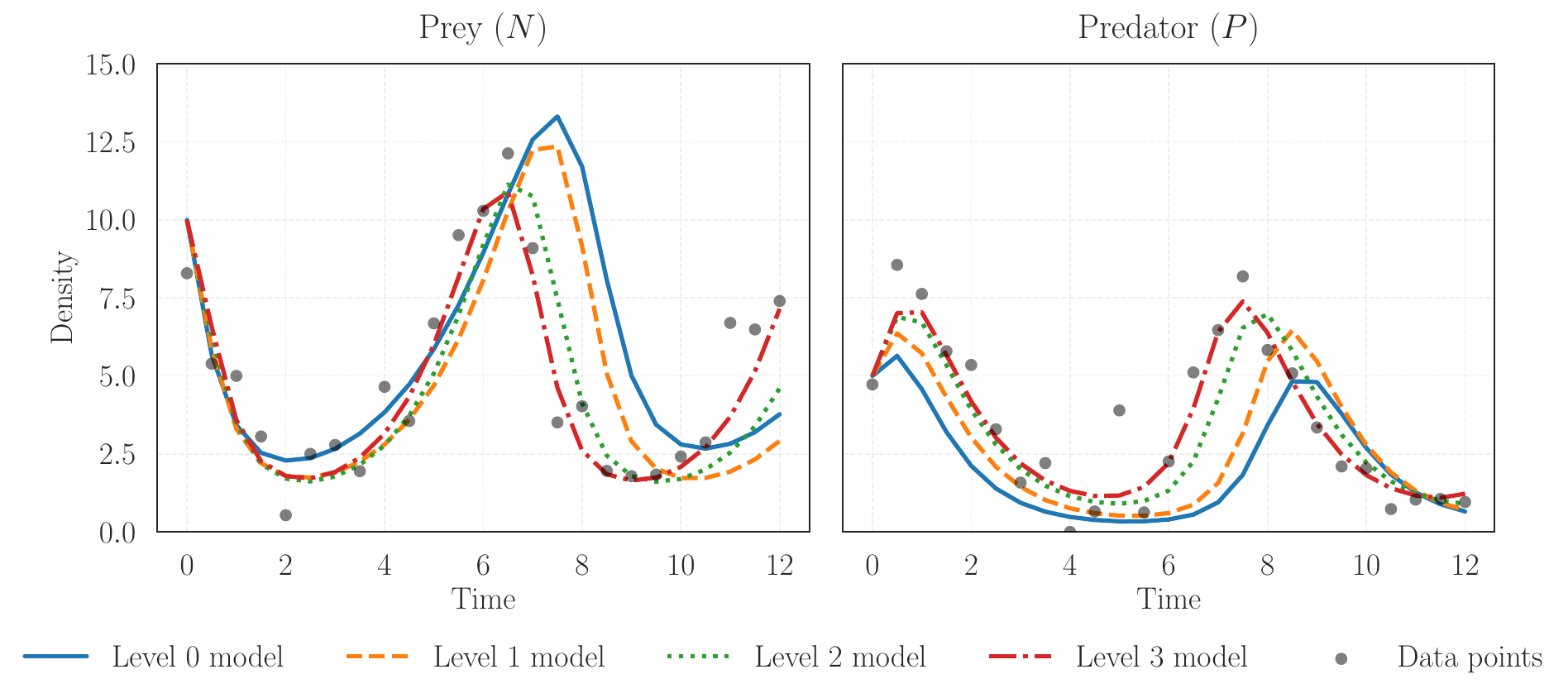}
  \caption{Model evaluations and data for the four fidelity levels of the prey-predator problem for the true parameters. Introducing model discrepancy terms creates differences between the model predictions at different fidelity levels.}~\label{fig:prey-predator-models-data}
\end{figure}

The Lotka-Volterra model describes the interaction between populations of prey ($N$) and predator ($P$) species over time~\cite{lykkegaard_multilevel_2023}. The governing equations of the \textit{general} model are given by a system of ODEs as:
\begin{align}~\label{eq:lotka-volterra-model}
  \frac{dN}{dt} &= a N - b NP - \epsilon_N N + \delta_N, \\ \nonumber
  \frac{dP}{dt} &= c NP - d P - \epsilon_P P,
\end{align}
where $a, b, c, d$ are the model parameters and $\epsilon_N, \epsilon_P, \delta_N$ represent model discrepancy terms. The initial populations of prey and predator species, denoted by $N_0$ and $P_0$ respectively, are also considered as uncertain parameters. Thus, the complete set of uncertain parameters is given by $\theta = [N_0, P_0, a, b, c, d] \in \mathbb{R}^6$. We consider $L=3$ levels and describe each level by the values of the model discrepancy terms as shown in the table below.
\begin{table}[h!]
  \centering
  \caption{Model discrepancy terms for different fidelity levels}~\label{tab:model-discrepancy}
  \begin{tabular}{c|ccc}
    \toprule
    Level ($\ell$) & $\epsilon_N$ & $\epsilon_P$ & $\delta_N$ \\
    \midrule
    3 & 0.0 & 0.0 & 0.0 \\
    2 & 0.1 & 0.0 & 0.0 \\
    1 & 0.3 & 0.1 & 0.0 \\
    0 & 0.5 & 0.2 & 0.3 \\
    \bottomrule
  \end{tabular}
\end{table}
For the BIP, we assume Gaussian prior distributions for all the parameters with means and variances as given in~\cite{lykkegaard_multilevel_2023}. The models are solved using the \texttt{SciPy} library's \texttt{solve\_ivp} function with the \texttt{RK45} method. The observed data is generated by solving the finest level model with a true set of parameters $\theta_{\text{true}} = [10, 5, 1.0, 0.3, 0.2, 1.0]$ and adding Gaussian noise with zero mean and variance $\sigma^2_{\text{noise}} = 1.0^2$ at 25 equally spaced time points in the time domain $\mathcal{T}$. The four fidelity models along with the observed data are shown in~\cref{fig:prey-predator-models-data}.

We run~\cref{alg:ml-mcmc} with both the coupling methods for 12,000 samples for the finest level and discard the first 2,000 samples as burn in. For the other three levels, we run the same 12,000 samples for SYNCE coupling method and set the subsampling rates $J_1 = J_2 = J_3 = 5$ for the AEM enhanced MLDA method~\cite{lykkegaard_multilevel_2023}. This setting results in 60,000, 300,000, and 1,500,000 samples for the coarser fidelities. The proposal distributions for both the methods are set to Gaussian random walks with covariance matrices tuned to achieve an acceptance rate of approximately 44\% for all levels. Further, we only use adaptation and no resynchronization for this problem for the SYNCE algorithm. We run 5 independent runs of the ML-MCMC algorithm for both the coupling methods and assume that the cost of evaluating each model follows the decay law: $\mathcal{C}_{\ell} = 0.001\times2^{\ell}$ with $\gamma=1$ (see supplementary material~\cref{assump:decay-errors}).
The reported minimum correlation $\rho_L$ is the median (over 5 runs) of the minimum correlation between the $L$ and $L-1$ level chains across all 6 parameter dimensions. The minimum ESS is computed similarly using the finest level samples. Total cost is $\mathcal{C}_{\text{tot}} = \sum_{\ell=0}^{L} N_{\ell} \mathcal{C}_{\ell}$. The comparisons are summarized in \cref{tab:synce-aem-comparison} for two, three and four levels.   
\begin{table}[h!]
  \centering
  \caption{Comparison of the SYNCE coupling method with the AEM enhanced MLDA method for the prey-predator problem. While MLDA-AEM achieves high correlation and ESS for few levels, its exponential cost scaling results in poor efficiency per unit cost as the number of levels increase. In contrast, SYNCE demonstrates superior scalability, achieving twice the sampling efficiency in the four-level case.}~\label{tab:synce-aem-comparison}
  \begin{tabular}{l  r r r r r}
    \toprule
    \textbf{Method} & \textbf{$\mathcal{C}_{\text{tot}}$} & \textbf{Min. $\rho_{L}$}  & \textbf{Min. ESS} & \textbf{Min. $\rho_{L}$/cost} & \textbf{Min. ESS/cost} \\
    \midrule
    \multicolumn{6}{l}{\textit{Two-level comparison}} \\
    MLDA-AEM & 84 & 0.75 & 753 & 0.0089 & 8.96 \\
    SYNCE & 36 & 0.60 & 226 & 0.0167 & 6.27 \\
    \midrule
    \multicolumn{6}{l}{\textit{Three-level comparison}} \\
    MLDA-AEM & 468 & 0.50 & 971 & 0.0010 & 2.07 \\
    SYNCE & 108 & 0.61 & 326 & 0.0056 & 3.02 \\
    \midrule
    \multicolumn{6}{l}{\textit{Four-level comparison}} \\
    MLDA-AEM & 2436 & 0.40 & 1374 & 0.0001 & 0.56 \\
    SYNCE & 252 & 0.60 & 299 & 0.0023 & 1.18 \\
    \bottomrule
  \end{tabular}
\end{table}

MLDA with AEM consistently achieves high raw ESS across all level settings, demonstrating that the AEM effectively mitigates posterior discrepancies. However, this gain in sampling efficiency comes by recursive subsampling, forcing the coarse chains to run significantly long chains to serve as good proposals for the finer levels. As shown in~\cref{tab:synce-aem-comparison}, the total computational cost for MLDA-AEM grows exponentially with the number of levels, with the four-level case being nearly an order of magnitude more expensive than SYNCE at the same level. In contrast, SYNCE relies solely on the pairwise coupling of adjacent fidelity levels, avoiding the dependence on long coarser chains. This decoupling isolates the correlation performance from the number of levels, resulting in linear cost scaling and consistent correlation across levels.

It is important to distinguish between mixing efficiency and coupling quality. The lower ESS observed in SYNCE is primarily due to the choice of proposal distributions rather than the coupling strategy itself. Since SYNCE's mixing properties depend primarily on the proposal distributions (see~\Cref{subsec:convergence-analysis-comparison}), its sampling efficiency can be improved by integrating advanced proposal mechanisms. For instance, employing transport map proposals~\cite{parno_transport_2018} could significantly enhance mixing while retaining strong correlation. We leave this exploration for future work.

Ultimately, when evaluating efficiency per unit cost, SYNCE demonstrates superior scalability compared to MLDA-AEM as the number of fidelity levels increase. In the two level case, MDLA-AEM achieves higher correlation and ESS, but at double the computational cost, rendering per unit cost metrics similar for both the methods. However, at the four-level setting, the computational overhead and posterior mismatch lead to pronounced efficiency degradation for MLDA-AEM. While MLDA-AEM excels for few closely related models, SYNCE is more suitable for problems with a large hierarchy of models and significant posterior dissimilarities. A detailed visualization of the posterior distributions for both methods is provided in~\ref{supp:posterior-comparison}.

\subsection{Groundwater problem}\label{subsec:groundwater}
To assess whether SYNCE's parameter coupling strategy achieves effective variance reduction in forward uncertainty quantification, we consider the 2D Darcy's subsurface flow equation. This is the model problem explored in great detail in~\cite{cliffe_multilevel_2011,dodwell2015hierarchical,madrigal-cianci_analysis_2023}. The exact problem is taken from~\cite{madrigal-cianci_analysis_2023} and the governing equation of the problem is given by:
\begin{align}\label{eq:darcy-pde}
  -\nabla_{x} \cdot \left(\kappa(x,\theta) \nabla_x u(x, \theta)\right) = 1, \quad x = \left(x_1, x_2\right) \in \Omega = [0, 1] \times [0, 1],
\end{align}
where $\kappa$ represents the permeability, $u$ represents the pressure head and $\theta \in \mathbb{R}^d, d=4$ represents the uncertain parameters. We close the equations by providing boundary conditions on the domain as:
\begin{align}\label{eq:darcy-bc}
  u|_{x_1 = 0} = 0, \quad u|_{x_1 = 1} = 0, \quad \partial_n u|_{x_2 = 0} = 0, \quad \partial_n u|_{x_2 = 1} = 0,
\end{align}
where $\partial_n$ is the normal derivative with $n$ pointing outwards from the domain. The first two conditions are the Dirichlet boundary conditions applied on the left and right boundary of the domain, and the last two conditions are the Neumann boundary conditions applied on the bottom and top boundary of the domain. The permeability field is given by~\cite{madrigal-cianci_analysis_2023}:
\begin{align}\label{eq:permeability}
  \kappa(x, \theta) = \exp\left(\theta_1\text{cos}(\pi x) + \frac{\theta_2}{2}\text{sin}(\pi x) + \frac{\theta_3}{3}\text{cos}(2\pi x) + \frac{\theta_4}{4}\text{sin}(2 \pi x)\right)
\end{align}

The inverse problem is to infer the uncertain parameters $\theta=[\theta_1, \theta_2, \theta_3, \theta_4]$ given some observed pressure head data. We generate this data by solving the forward problem \cref{eq:darcy-pde} on the finest grid with a random sample of $\theta_{\text{true}} \sim \mathcal{N}({0_d, \mathcal{I}_d})$ and observing the pressure head corrupted by a Gaussian noise at $4 \times 4$ equally spaced points inside the domain. Gaussian noise is added to the observed data with a zero mean and variance $\sigma^2_{\text{noise}}=0.01^2$. We set the number of levels $L = 4$ and for each level $\ell$, the solution of \cref{eq:darcy-pde} is obtained using the finite element method with a mesh size of $8 * 2^{\ell} \times 8 * 2^{\ell}$ triangular elements, using the FEniCS library~\cite{baratta_2023_10447666}.

The number of levels in the original problem in~\cite{madrigal-cianci_analysis_2023} is set to $L=3$ with the coarsest mesh at $16 \times 16$ triangular elements. We set the number of levels to $L=4$ with the coarsest mesh at $8 \times 8$ triangular elements as we can leverage the SYNCE coupling method to correlate samples even at very coarse levels. This is a direct consequence of the fact that the SYNCE coupling method does not depend on the closeness of the posteriors to correlate samples.

The quantity of interest (QoI) is the average pressure over the domain given by:
\begin{align}\label{eq:groundwater-qoi}
  Q_{\ell}(\theta_{\ell}) = \int_{x \in \Omega} u_{\ell}(x, \theta_{\ell}) dx,
\end{align}
where $\Omega$ is the domain of the problem. 

To empirically verify that SYNCE satisfies the decay rate assumptions required for ML-MCMC estimator convergence (see supplementary material~\cref{assump:decay-errors}), we run \cref{alg:ml-mcmc} with the proposed states obtained from \cref{alg:synce-coupling-adaptive}. We set the number of samples in each level to 10,000 and consider the first 4,000 samples as burn-in. We run the algorithm for 10 independent runs and plot the decay of errors. For the coarsest chain ($\ell = 0$), we set the target acceptance ratio as 44\% and run an adaptive MH-MCMC algorithm for that level. The two level wise adaptation parameters, the target acceptance ratio and the weight for resynchronization are set to 70\% for each level ($\ell > 0$) and $\omega = [0.0, 0.3, 0.5, 0.7]$. These parameters were chosen to ensure low lag and high correlation between the samples at each level. 

\begin{figure}
  \centering
  \begin{subfigure}[b]{0.4\textwidth}
      \centering
      \includegraphics[width=\textwidth]{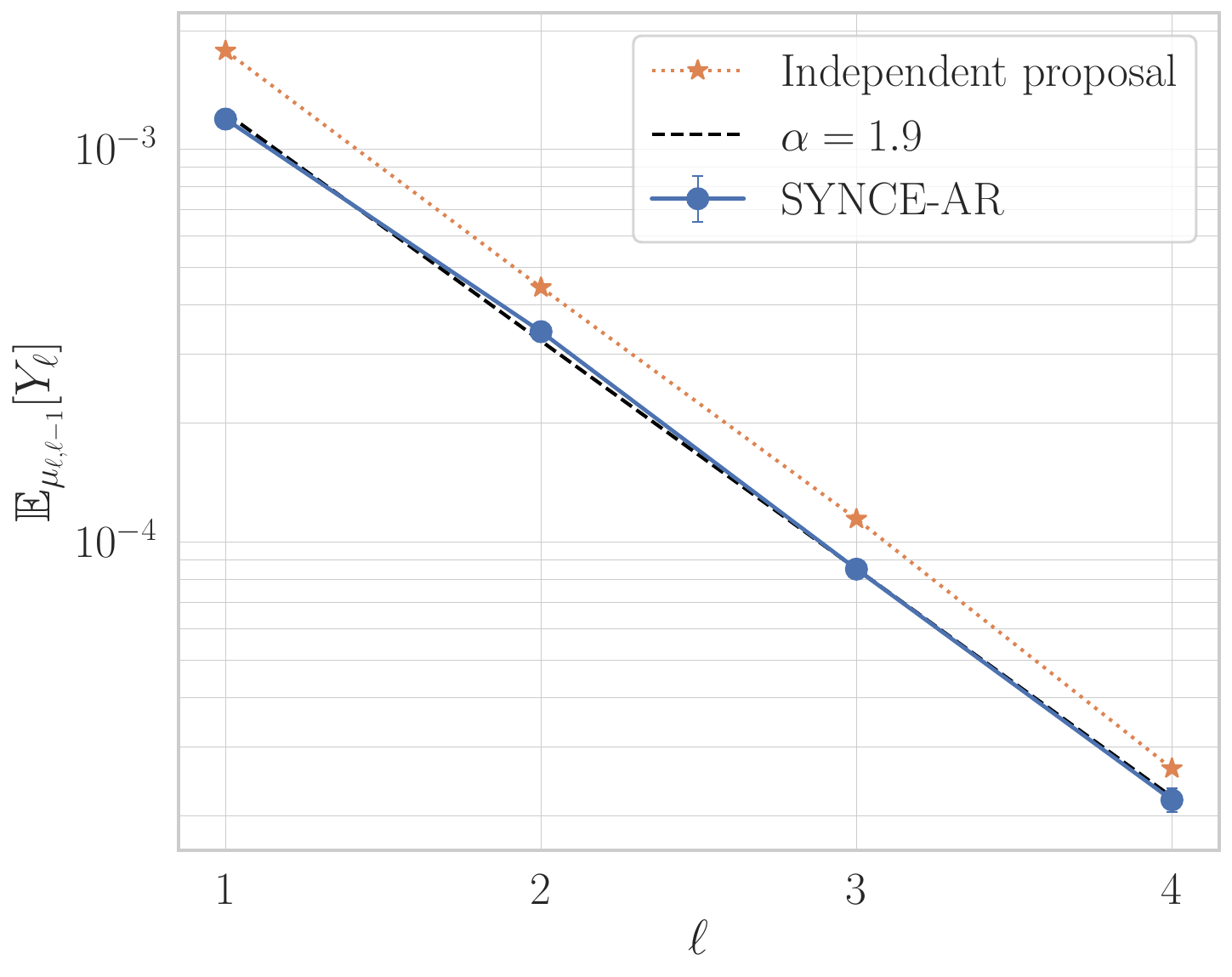}
      \caption{Mean}
      \label{fig:groundwater-mean-convergence}
  \end{subfigure}
  \begin{subfigure}[b]{0.4\textwidth}
      \centering
      \includegraphics[width=\textwidth]{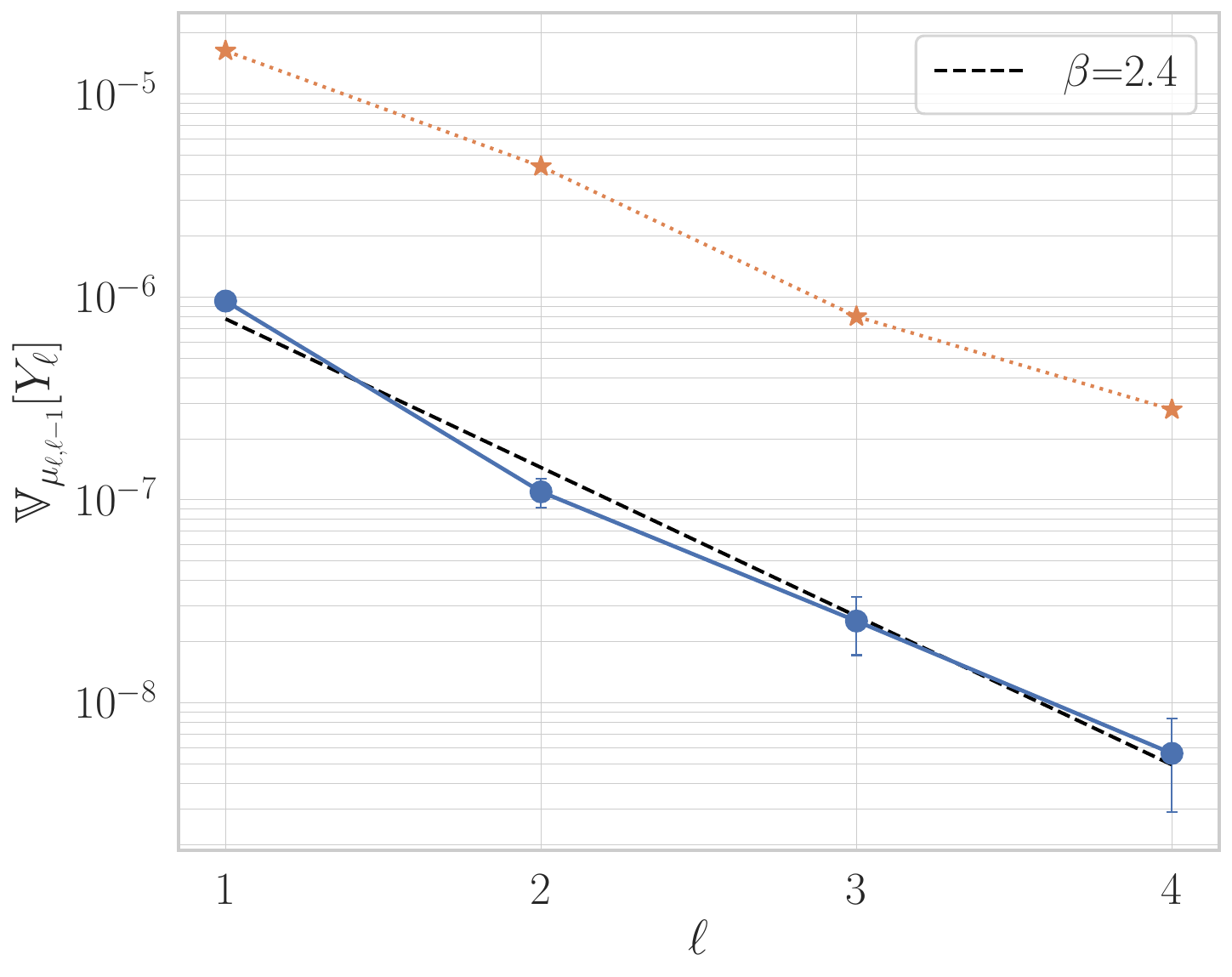}
      \caption{Variance}
      \label{fig:groundwater-variance-convergence}
  \end{subfigure}
  \caption{Decay of error plots for the groundwater problem. The independent proposal coupling and the SYNCE-AR methods are compared. Our method leads to almost an order of two gain in variance reduction compared to the result from~\cite{madrigal-cianci_analysis_2023}.}
  \label{fig:groundwater-convergence}
\end{figure}

~\Cref{fig:groundwater-convergence} compares our proposed methodology to an existing result using the independent proposal coupling method~\cite{madrigal-cianci_analysis_2023} for the same problem. The results for the independent proposal coupling method were obtained by digitizing Fig. 12 in~\cite{madrigal-cianci_analysis_2023}. We observe that the SYNCE-AR method leads to greater variance reduction for all the level wise difference estimators $\hat{Y}_{\ell}$. Even though we use a coarser level compared to the original work, we are able to achieve almost an order of two gain in variance reduction at all levels. This is a direct consequence of the proposed coupling method that is able to correlate samples even at very coarse levels. The correlation values for the output functional at the different levels are: $0.84, 0.97, 0.99, 1.0$ for $\ell=1,2,3,4$. 

\section{Conclusions}\label{sec:conclusions}

We introduced SYNCE, a new coupling methodology for multilevel Markov chain Monte Carlo estimators that achieves superior variance reduction compared to existing coupling approaches. Variance reduction in these estimators depends on the correlation between the samples at successive fidelity levels, which in turn depends on the coupling strategy used to construct the joint proposal distribution for the two chains. Existing coupling strategies depend on some form of overlap or similarity between the posteriors at different levels to correlate samples. This dependence limits the effectiveness of these coupling strategies when the posteriors are dissimilar or non-overlapping, which is often the case in practical applications. SYNCE however, does not depend on this overlap and couples chains by proposing the same noise in the Gaussian random walk proposals for both chains. We rigorously proved the existence of a joint invariant measure for the SYNCE kernel and established geometric ergodicity under standard assumptions. We also showed that the convergence rate of SYNCE is superior to existing coupling approaches when the posteriors are dissimilar or non-overlapping.

By utilizing adaptation and resynchronization, SYNCE adaptively rescales the proposal distributions and resynchronizes the chains to enhance correlation. Adaptation helps in accounting for the differences in the covariance structure of the posteriors at different levels, while resynchronization helps in promoting better mixing properties of the chains when posteriors are quite close. These enhancements enable the use of coarser model fidelities in the estimation procedure, yielding substantial computational savings. Numerical experiments on synthetic and real-world problems validate the effectiveness of the proposed methodology. 

The SYNCE coupling method is the baseline realization of the proposed coupling framework we have set up. Future work will includes exploring more general and nonlinear couplings. Transport maps~\cite{parno_transport_2018,peherstorfer_transport-based_2019} may help with mixing properties of the fine chain by learning from the coarse chain. Another possible extension to this work is to explore the use of the SYNCE coupling methodology in the context of approximate control variates. These estimators provide additional performance improvements by using a control variate weight that is estimated adaptively. Additionally, the proposed methodology can be extended in the unbiased estimation context to eliminate the burn in bias~\cite{jacob_unbiased_2020,heng_unbiased_2023}.

\section*{Author Contributions}
\textbf{Sanjan Muchandimath}: Conceptualization, Methodology, Software, Validation, Writing - original draft. 
\textbf{Alex Gorodetsky}: Conceptualization, Methodology, Writing - Review \& Editing, Supervision, Funding acquisition.

\section*{Acknowledgements}
This work was funded by the Aeronautics Research Mission Directorate at NASA through the Transformative Aeronautical Concepts Program (TACP) and the D.2 Transformational Tools and Technologies Project (TTT) under contract no.~80NSSC23M0215.


\section*{Data Availability}
Available soon \href{https://github.com/sanjan98/SYNCE-JCP-experiments#}{\texttt{here}}


\appendix

\crefalias{section}{appendix}

\renewcommand{\thedefinition}{\Alph{section}.\arabic{definition}}
\renewcommand{\thetheorem}{\Alph{section}.\arabic{theorem}}
\renewcommand{\thelemma}{\Alph{section}.\arabic{lemma}}
\renewcommand{\theassumption}{\Alph{section}.\arabic{assumption}}

\section{Proof of~\cref{thm:convergence-synce}}~\label{sec:proof-convergence-synce}
We prove~\cref{thm:convergence-synce} by establishing that the coupled kernel in~\cref{eq:synce-kernel} is $\psi$-irreducible and aperiodic, and that it satisfies minorization and drift conditions. 
\begin{lemma}[$\psi-$irreducibility]~\label{lemma:synce-irreducibility}
  Under~\cref{assump:rw-ergodicity}, the coupled kernel $p^{\text{SYNCE}}_{\ell}$ in~\cref{eq:synce-kernel} is $\psi$-irreducible.
\end{lemma}
\begin{pf}
  Let $\psi$ be the Lebesgue measure on $X^2$. There exists a rectangle $K^2 = K \times K \in A$ with positive measure for any set $A$ with $\psi(A) > 0$. Here, $K = \left\{x: |x| \leq R_K\right\}$ for some $R_K > 0$, a ball of radius $R_K$. For any starting pair $\bm{\theta}_{\ell} = \left(\theta_{\ell}, \vartheta_{\ell-1}\right) \in X^2$,
  \begin{align*}
      & p^{\text{SYNCE}}_{\ell}\left(\bm{\theta}_{\ell}, A\right) \\
      & \geq p^{\text{SYNCE}}_{\ell}\left(\bm{\theta}_{\ell}, K^2\right), \\
      & \geq \int_{X^2} \min \left(\alpha_{\ell}\left(\theta_{\ell}, \theta^*_{\ell}\right), \alpha_{\ell-1}\left(\vartheta_{\ell-1}, \vartheta^*_{\ell-1}\right)\right)  \Gamma_{\ell}\left(\bm{\theta}_{\ell} \mid d\bm{\theta}_{\ell}^{*}\right)\delta_{\boldsymbol{\theta}^*_{\ell}}(K^2), \\
      & \geq \int_{X} \min \left(\alpha_{\ell}\left(\theta_{\ell}, \theta_{\ell} + \eta\right), \alpha_{\ell-1}\left(\vartheta_{\ell-1}, \vartheta_{\ell-1} + \eta\right)\right) \mathbbm{1}_{K^2}\left(\theta_{\ell} + \eta, \vartheta_{\ell-1} + \eta\right)q(\eta)d\eta. \\
  \end{align*}
  Recall from~\cref{alg:synce-coupling} that the proposed samples are generated by $\theta^*_{\ell} = \theta_{\ell} + \eta$ and $\vartheta^*_{\ell-1} = \vartheta_{\ell-1} + \eta$ where $\eta \sim q$, $q = \mathcal{N}(0, C_{\ell})$. The third inequality arises by restricting on the event that both marginals propose into the set $K$ i.e, $\theta_{\ell} + \eta \in K$ and $\vartheta_{\ell-1} + \eta \in K$. For this condition to hold, the increment $\eta$ must satisfy $\eta \in B = \left\{x: |x| \leq R_K + \max\left\{|\theta_{\ell}|, |\vartheta_{\ell-1}|\right\}\right\}$. By assumption~\ref{assump:proposal-density}, $q(\eta)$ is bounded below on the set $B$. Also, for positive and continuous $\pi_{\ell},\pi_{\ell-1}$ as stated in assumption~\ref{assump:super-exp}, $\min \left(\alpha_{\ell}\left(\theta_{\ell}, \theta^*_{\ell}\right), \alpha_{\ell-1}\left(\vartheta_{\ell-1}, \vartheta^*_{\ell-1}\right)\right) > 0$. Therefore $p^{\text{SYNCE}}_{\ell}\left(\bm{\theta}_{\ell}, K^2\right) > 0$. Because the above holds for any set $A$ with positive Lebesgue measure and any starting point $\bm{\theta}_{\ell}$, the kernel is $\psi$-irreducible for $n=1$. Furthermore, aperiodicity follows because the kernel is $\psi$-irreducible and the chain can stay in the same state with positive probability.
\end{pf} \qed

\begin{lemma}[Minorization condition]~\label{lemma:synce-minorization}
    Under~\cref{assump:rw-ergodicity}, the coupled kernel $p^{\text{SYNCE}}_{\ell}$ in~\cref{eq:synce-kernel} satisfies a one step minorization condition on any compact set $S \in \mathcal{X}^2$.
\end{lemma}
\begin{pf}
    Let $C_{\ell}, C_{\ell-1} \in X$ be compact sets, both contained in $B(0, R)$ and define $C = C_{\ell} \times C_{\ell-1}$ as the set of starting points. Let $B$ be the set from assumption~\ref{assump:proposal-density} i.e, $B := \left\{\eta: |\eta| \leq \delta_q\right\}$. 
    Define the intersection set $D$ as the larger compact set containing all possible proposals $D := \left\{(\theta_{\ell} + \eta, \vartheta_{\ell-1} + \eta): \eta \in B\right\}$. Then, for any starting point $\bm{\theta}_{\ell} \in C$ and any set $A \in \mathcal{X}^2$, we have
    \begin{align*}
        & p^{\text{SYNCE}}_{\ell}\left(\bm{\theta}_{\ell}, A\right) \\
        & \geq \int_{X^2} \min \left(\alpha_{\ell}\left(\theta_{\ell}, \theta^*_{\ell}\right), \alpha_{\ell-1}\left(\vartheta_{\ell-1}, \vartheta^*_{\ell-1}\right)\right)  \Gamma_{\ell}\left(\bm{\theta}_{\ell} \mid  d\bm{\theta}_{\ell}^{*}\right)\delta_{\bm{\theta}^*_{\ell}}(A). \\
        & \geq \int_{B} \min \left(\alpha_{\ell}\left(\theta_{\ell}, \theta_{\ell} + \eta\right), \alpha_{\ell-1}\left(\vartheta_{\ell-1}, \vartheta_{\ell-1} + \eta\right)\right) \mathbbm{1}_{A}\left(\theta_{\ell} + \eta, \vartheta_{\ell-1} + \eta\right)q(\eta)d\eta,
    \end{align*}
    From assumption~\ref{assump:super-exp}, the posterior densities $\pi_{\ell}, \pi_{\ell-1}$ are bounded above and below away from zero on any compact set. Therefore, 
    \begin{align*}
      & \min \left(\alpha_{\ell}\left(\theta_{\ell}, \theta_{\ell} + \eta\right), \alpha_{\ell-1}\left(\vartheta_{\ell-1}, \vartheta_{\ell-1} + \eta\right)\right) \geq \frac{\pi_{\min}}{\pi_{\max}}.
    \end{align*}
    We also use assumption~\ref{assump:proposal-density} to bound $q(\eta)$ below by $\epsilon_q > 0$ on the set $B$. Therefore,
    \begin{align*}
      p^{\text{SYNCE}}_{\ell}\left(\bm{\theta}_{\ell}, A\right) \geq \epsilon_q\frac{\pi_{\min}}{\pi_{\max}}\int_{B}  \mathbbm{1}_{A}\left(\theta_{\ell} + \eta, \vartheta_{\ell-1} + \eta\right)d\eta.
    \end{align*}
    The integral term represents the volume of the set $A$ reachable by the synchronized proposal from $\bm{\theta}_{\ell}$. Because we are proposing the same increment $\eta$ to both chains, this volume is non-zero only when $A$ contains points along the line defined by $\left\{(\theta_{\ell} + \eta, \vartheta_{\ell-1} + \eta): \eta \in B\right\}$, which is not the case in general. Hence this term does not define a non-trivial probability measure. However, since the chain is $\psi$-irreducible (Lemma~\ref{lemma:synce-irreducibility}) and possesses a continuous density, we can utilize Theorem 6.2.5 from~\cite{meyn2012markov} to guarantee that any compact set $C$ is small. Intuitively, this theorem states that as long as the kernel has a continuous density (the same step move) and is $\psi$-irreducible (eventually visit any state), then it guarantees that minorization holds on any compact set. Therefore, this theorem guarantees the existence of a non-trivial measure $\nu(A)$ and a constant $\delta > 0$ for the minorization condition:
    \begin{align}~\label{eq:minorizing-synce}
      p^{\text{SYNCE}}_{\ell}\left(\bm{\theta}_{\ell}, A\right) \geq \delta \nu(A).
    \end{align}
    Furthermore, the minorization constant scales as $\delta \propto \epsilon_q\pi_{\min}/\pi_{\max}$, where the proportionality constant depends on the overlap of the proposal distribution with the compact set $C$~\cite{jarnerGeometricErgodicityMetropolis2000}. 
\end{pf} \qed

\begin{lemma}[Drift condition]~\label{lemma:synce-drift}
    Under~\cref{assump:rw-ergodicity}, the coupled kernel satisfies a drift condition with Lyapunov function $\hat{V}_{\ell} = 1/2\left(V_{\ell} + V_{\ell-1}\right)$, where $V_j$ are the marginal Lyapunov functions for $j=\left\{\ell,\ell-1\right\}$.
\end{lemma}
\begin{pf}
    The proof adapts the strategy from~\cite{cianci_thesis}. Under~\cref{assump:rw-ergodicity}, the marginal random walk chains with kernels $p_{\ell}, p_{\ell-1}$ satisfy drift conditions (\cref{lemma:rw-ergodicity}~\cref{lemma:rw-drift}). Define the joint Lyapunov function as
    \begin{align}~\label{eq:coupled-drift}
      \hat{V}_{\ell}\left(\bm{\theta}_{\ell}\right) = \frac{1}{2}\left(V_{\ell}\left(\theta_{\ell}\right) + V_{\ell-1}\left(\vartheta_{\ell-1}\right)\right),
    \end{align}
    and $S = S_{\ell} \times S_{\ell-1}$ where $S_{\ell}$ and $S_{\ell-1}$ are the marginal small sets. Precisely, $S_{\ell} := \{\theta_{\ell} \in X: V_{\ell}(\theta_{\ell}) \leq \Delta_{\ell}\}$ and $S_{\ell-1} := \{\vartheta_{\ell-1} \in X: V_{\ell-1}(\vartheta_{\ell-1}) \leq \Delta_{\ell-1}\}$. The goal is to show that $p_{\ell}^{\text{SYNCE}}$ satisfies the drift condition with joint Lyapunov function $\hat{V}_{\ell}$ and joint small set $S$. 
    
    The state $\bm{\theta}_{\ell}$ can either lie in $S$ or outside $S$. If $\bm{\theta}_{\ell} \notin S$, then there are three possibilities --- $\theta_{\ell} \in S_{\ell}, \vartheta_{\ell-1} \notin S_{\ell-1}$; $\theta_{\ell} \notin S_{\ell}, \vartheta_{\ell-1} \in S_{\ell-1}$ and $\theta_{\ell} \notin S_{\ell}, \vartheta_{\ell-1} \notin S_{\ell-1}$
    In the first case, we have
    \begin{align*}
      p_{\ell}^{\text{SYNCE}}\hat{V}_{\ell}\left(\bm{\theta}_{\ell}\right) \leq \frac{1}{2}\left(\Lambda_{\ell}V_{\ell}\left(\theta_{\ell}\right) + \Lambda_{\ell-1}V_{\ell-1}\left(\vartheta_{\ell-1}\right)\right) + \frac{b_{\ell}}{2},
    \end{align*}
    Since $\vartheta_{\ell-1} \notin S_{\ell-1}$, we have that $V_{\ell-1}\left(\vartheta_{\ell-1}\right) \geq \Delta_{\ell-1}$. Therefore, 
    \begin{align*}
      \hat{V}_{\ell}\left(\bm{\theta}_{\ell}\right) &= \frac{1}{2}\left(V_{\ell}\left(\theta_{\ell}\right) + V_{\ell-1}\left(\vartheta_{\ell-1}\right)\right) \geq \frac{1}{2}\left(1 + \Delta_{\ell-1}\right),
    \end{align*}
    which implies that $1/2 \leq \hat{V}_{\ell}\left(\bm{\theta}_{\ell}\right) / (1 + \Delta_{\ell-1})$. Using this in the previous inequality, we get
    \begin{align*}
      p_{\ell}^{\text{SYNCE}}\hat{V}_{\ell}\left(\bm{\theta}_{\ell}\right) &\leq \frac{1}{2}\left(\Lambda_{\ell}V_{\ell}\left(\theta_{\ell}\right) + \Lambda_{\ell-1}V_{\ell-1}\left(\vartheta_{\ell-1}\right)\right) + \frac{b_{\ell}}{1 + \Delta_{\ell-1}}\hat{V}_{\ell}\left(\bm{\theta}_{\ell}\right) \leq \left(\max\left(\Lambda_{\ell},\Lambda_{\ell-1}\right) + \frac{b_{\ell}}{1 + \Delta_{\ell-1}}\right)\hat{V}_{\ell}\left(\bm{\theta}_{\ell}\right).
    \end{align*}
    Note that $\max\left(\Lambda_{\ell},\Lambda_{\ell-1}\right) + b_{\ell}/ (1 + \Delta_{\ell-1}) < 1$ because $\Lambda_j < 1$ and $0 < b_j < \infty$ for $j=\left\{\ell,\ell-1\right\}$. The second case is similar and leads to a similar inequality,
    \begin{align*}
      p_{\ell}^{\text{SYNCE}}\hat{V}_{\ell}\left(\bm{\theta}_{\ell}\right) \leq \left(\max\left(\Lambda_{\ell},\Lambda_{\ell-1}\right) + \frac{b_{\ell-1}}{1 + \Delta_{\ell}}\right)\hat{V}_{\ell}\left(\bm{\theta}_{\ell}\right).
    \end{align*}
    The third case leads to
    \begin{align*}
      p_{\ell}^{\text{SYNCE}}\hat{V}_{\ell}\left(\bm{\theta}_{\ell}\right) \leq \left(\max \left(\Lambda_{\ell}, \Lambda_{\ell-1}\right)\right)\hat{V}_{\ell}\left(\bm{\theta}_{\ell}\right).
    \end{align*}
    Combining all three cases, we have that for $\bm{\theta}_{\ell} \notin S$,
    \begin{align}~\label{eq:synce-drift-1}
      & p_{\ell}^{\text{SYNCE}}\hat{V}_{\ell}\left(\bm{\theta}_{\ell}\right) \leq \hat{\Lambda} \hat{V}_{\ell}\left(\bm{\theta}_{\ell}\right), \\
      & \hat{\Lambda}_{\text{SYNCE}} = \max\left(\Lambda_{\ell}, \Lambda_{\ell-1}\right) + \max\left(\frac{b_{\ell}}{1 + \Delta_{\ell-1}}, \frac{b_{\ell-1}}{1 + \Delta_{\ell}}\right) < 1. \nonumber
    \end{align}
    Finally, if $\bm{\theta}_{\ell} \in S$, then
    \begin{align}~\label{eq:synce-drift-2}
      p_{\ell}^{\text{SYNCE}}\hat{V}_{\ell}\left(\bm{\theta}_{\ell}\right) &\leq \frac{1}{2}\left(\Lambda_{\ell}V_{\ell}\left(\theta_{\ell}\right) + \Lambda_{\ell-1}V_{\ell-1}\left(\vartheta_{\ell-1}\right)\right) + \frac{1}{2}\left(b_{\ell} + b_{\ell-1}\right) \leq \hat{\Lambda}_{\text{SYNCE}} \hat{V}_{\ell}\left(\bm{\theta}_{\ell}\right) + \hat{b}, \nonumber
    \end{align}
    where $\hat{b} = 1/2\left(b_{\ell} + b_{\ell-1}\right)$. Combining~\cref{eq:synce-drift-1,eq:synce-drift-2}, we have that the coupled kernel $p_{\ell}^{\text{SYNCE}}$ satisfies a drift condition with Lyapunov function $\hat{V}_{\ell}$ given by
    \begin{align}
      p_{\ell}^{\text{SYNCE}}\hat{V}_{\ell}\left(\bm{\theta}_{\ell}\right) \leq \hat{\Lambda}_{\text{SYNCE}} \hat{V}_{\ell}\left(\bm{\theta}_{\ell}\right) + \hat{b} \mathbbm{1}_{S}\left(\bm{\theta}_{\ell}\right).
    \end{align}
\end{pf} \qed

\section*{Supplementary Material}
\setcounter{subsection}{0} 
\renewcommand{\thesubsection}{S.\arabic{subsection}}
\subsection{Convergence and cost analysis of ML-MCMC with SYNCE coupling}\label{supp:ml-mcmc-cost-synce}
For completeness, this section presents the standard ML-MCMC complexity theorem from~\cite{dodwell2015hierarchical}, which establishes a bound on the $\epsilon$-cost of the ML-MCMC estimator under certain assumptions. We briefly summarize some notations and definitions before stating the assumptions and the theorem. 
\begin{itemize}
  \item $\mathbb{E}_{\mu_{\ell}}$ and $\mathbb{V}\text{ar}_{\mu_{\ell}}$ denote expectation and variance with respect to the invariant measure $\mu_{\ell}$ at level $\ell$.
  \item $\mathbb{E}_{\Theta_{\ell}}$ and $\mathbb{V}\text{ar}_{\Theta_{\ell}}$ denote expectation and variance with respect to the distribution of the random variables $\theta_{\ell}^i \in \Theta_{\ell}$ with density $\pi_{\ell}$. 
  \item $\mathbb{E}_{\text{ML-MCMC}}$ and $\mathbb{V}\mathrm{ar}_{\text{ML-MCMC}}$ denote expectation and variance with respect to the whole ML-MCMC estimator defined by \cref{eq:ml-mcmc}, which uses coupled samples from the SYNCE method in~\cref{alg:synce-coupling} across all levels.
  \item $e_{\ell,\ell-1} = \mathbb{V}\mathrm{ar}_{\Theta_{\ell,\ell-1}}\left[\hat{Q}_{\ell} - \hat{Q}_{\ell-1}\right] + \left(\mathbb{E}_{\Theta_{\ell,\ell-1}}\left[\hat{Q}_{\ell} - \hat{Q}_{\ell-1}\right] - \mathbb{E}_{\mu_{\ell,\ell-1}}[Q_{\ell}-Q_{\ell-1}]\right)^2$ denotes the mean square error (MSE) of the estimator for the difference in QOIs at levels $\ell$ and $\ell-1$ using $N_{\ell}$ samples from the coupled distribution $\pi_{\ell,\ell-1}$.
  \item $e_{\text{ML-MCMC}} = \mathbb{E}_{\text{ML-MCMC}}\left[\left(\hat{Q}_{\text{ML-MCMC}} - \mathbb{E}_{\mu}[Q]\right)^2\right]$ denotes the mean square error (MSE) of the ML-MCMC estimator.
  \item $\mathcal{C}_{\ell}$ denotes the computational cost of generating a single sample of the QOI $Q_{\ell}$ at level $\ell$.
  \item $\mathcal{C}_{\epsilon}\left(\hat{Q}^{\text{ML-MCMC}}\right)$ denotes the total computational cost of the ML-MCMC estimator to achieve an MSE of less than $\epsilon^2$.
\end{itemize}
The assumptions are given in \cref{assump:decay-errors} and the complexity theorem is stated next.
\begingroup 
\renewcommand{\theassumption}{B.\arabic{assumption}}
\begin{assumption}[Decay of errors]\label{assump:decay-errors}
  The following assumptions hold for some positive constants $\alpha, \beta, \gamma$ and for all $\ell = 0,1,\ldots,L$:
  \begin{enumerate}[label=(B.\arabic*), ref=(B.\arabic*)]
      \item~\label{assump:weak-convergence} $\left|\mathbb{E}_{\mu_{\ell}}\left[Q_{\ell}\right] - \mathbb{E}_{\mu}\left[Q\right]\right| \leq C_1s^{-\alpha \ell}$.
      \item~\label{assump:variance-convergence} $\mathbb{V}\text{ar}_{\mu_{\ell,\ell-1}}\left[Q_{\ell} - Q_{\ell-1}\right] \leq C_2s^{-\beta \ell}$.
      \item~\label{assump:mse-convergence} $e_{\ell,\ell-1} \leq C_3N_{\ell}^{-1}\mathbb{V}\text{ar}_{\mu_{\ell,\ell-1}}\left[Q_{\ell} - Q_{\ell-1}\right]$.
      \item~\label{assump:cost-per-sample} $\mathcal{C}_{\ell} \leq C_4s^{\gamma \ell}$.
  \end{enumerate}
  Recall that $s > 1$ is the refinement factor between two consecutive levels. Also, $C_1, C_2, C_3, C_4$ are positive constants independent of $\ell$.
\end{assumption}
\endgroup

These four assumptions are standard in the MLMC literature~\cite{giles_multilevel_2015,teckentrup2015multilevel}. Assumption~\ref{assump:weak-convergence} states that the bias in the QOI decays at a rate of $\alpha$ as we refine the discretization level. Assumption~\ref{assump:variance-convergence} is the key to ML-MCMC's efficiency, requiring the variance of the difference in QOIs between two consecutive levels to decay. A strong coupling method makes this variance decay faster, leading to lower number of samples required at finer levels. Assumption~\ref{assump:mse-convergence} states that the MSE of the estimator for the difference in QOIs at levels $\ell$ and $\ell-1$ decays inversely with the number of samples used, which is standard for Monte Carlo estimators. Finally, assumption~\ref{assump:cost-per-sample} states that the cost of generating a single sample at level $\ell$ increases at a rate of $\gamma$ as we refine the discretization level. The overall complexity is determined by the interplay between these rates $\alpha, \beta, \gamma$ as shown in the following theorem.

\begingroup
\renewcommand{\thetheorem}{B.\arabic{theorem}}
\begin{theorem}[$\epsilon$-cost of ML-MCMC with SYNCE coupling]\label{thm:ml-mcmc-cost-synce}
  Suppose that~\cref{assump:decay-errors} holds with $\alpha > 1/2\min(\beta, \gamma)$. Then, for any $\epsilon < e^{-1}$, there exists a level $L$ and a sequence of sample sizes $\left\{N_{\ell}\right\}_{\ell=0}^{L}$ such that the ML-MCMC estimator $\hat{Q}^{\text{ML-MCMC}}$ in~\cref{eq:ml-mcmc} using the SYNCE coupling method satisfies $e_{\text{ML-MCMC}} < \epsilon^2$ with a computational cost bounded by
  \begin{align*}
      \mathcal{C}_{\epsilon}\left(\hat{Q}^{\text{ML-MCMC}}\right) \leq C_5 \begin{cases}
          \epsilon^{-2}\left|\log \epsilon\right|, & \text{if } \beta > \gamma, \\
          \epsilon^{-2}\left|\log\epsilon\right|^3, & \text{if } \beta = \gamma, \\
          \epsilon^{-\left(2 + (\gamma - \beta) / \alpha\right)}\left|\log \epsilon\right|, & \text{if } \beta < \gamma.
      \end{cases}
  \end{align*}
\end{theorem}
\endgroup

\begin{pf}[Proof of~\cref{thm:ml-mcmc-cost-synce}]
  The proof follows directly from Theorem 3.4 in~\cite{dodwell2015hierarchical} provided the SYNCE coupling method satisfies~\cref{assump:decay-errors}.
\end{pf}

\subsection{Posterior distributions for the prey-predator problem}~\label{supp:posterior-comparison}
To better understand the fundamental difference between the MLDA-AEM and SYNCE coupling methods, we visualize the posterior distributions obtained from both methods for the four level prey-predator problem in~\cref{fig:prey-predator-aem-scatter,fig:prey-predator-synce-scatter}, respectively.

The MLDA-AEM method uses the adaptive error model to correct the coarse level posteriors so that they align with the fine-level posterior. This results in all four posteriors showing strong overlap, leading to good mixing and efficient sampling across levels. However, the MLDA-AEM method no longer samples from the exact posterior at each level, but rather from artificially corrected posteriors intended to mimic the fine-level posterior. In contrast, the SYNCE coupling method samples from the exact posterior at each level, reflecting the actual behavior of that model given the data. The clear visual separation between the posteriors at different levels in the SYNCE method does not deter the overall sampling and correlation generating properties of the method, as seen in~\cref{tab:synce-aem-comparison}. SYNCE leverages the correlation structure between different posteriors without requiring overlap, thus making the method more robust and scalable.

\begingroup
\renewcommand{\thefigure}{B.\arabic{figure}}
\setcounter{figure}{0}
\begin{figure}[h!]
  \centering
  \begin{subfigure}[b]{0.75\textwidth}
    \centering
    \includegraphics[width=\textwidth]{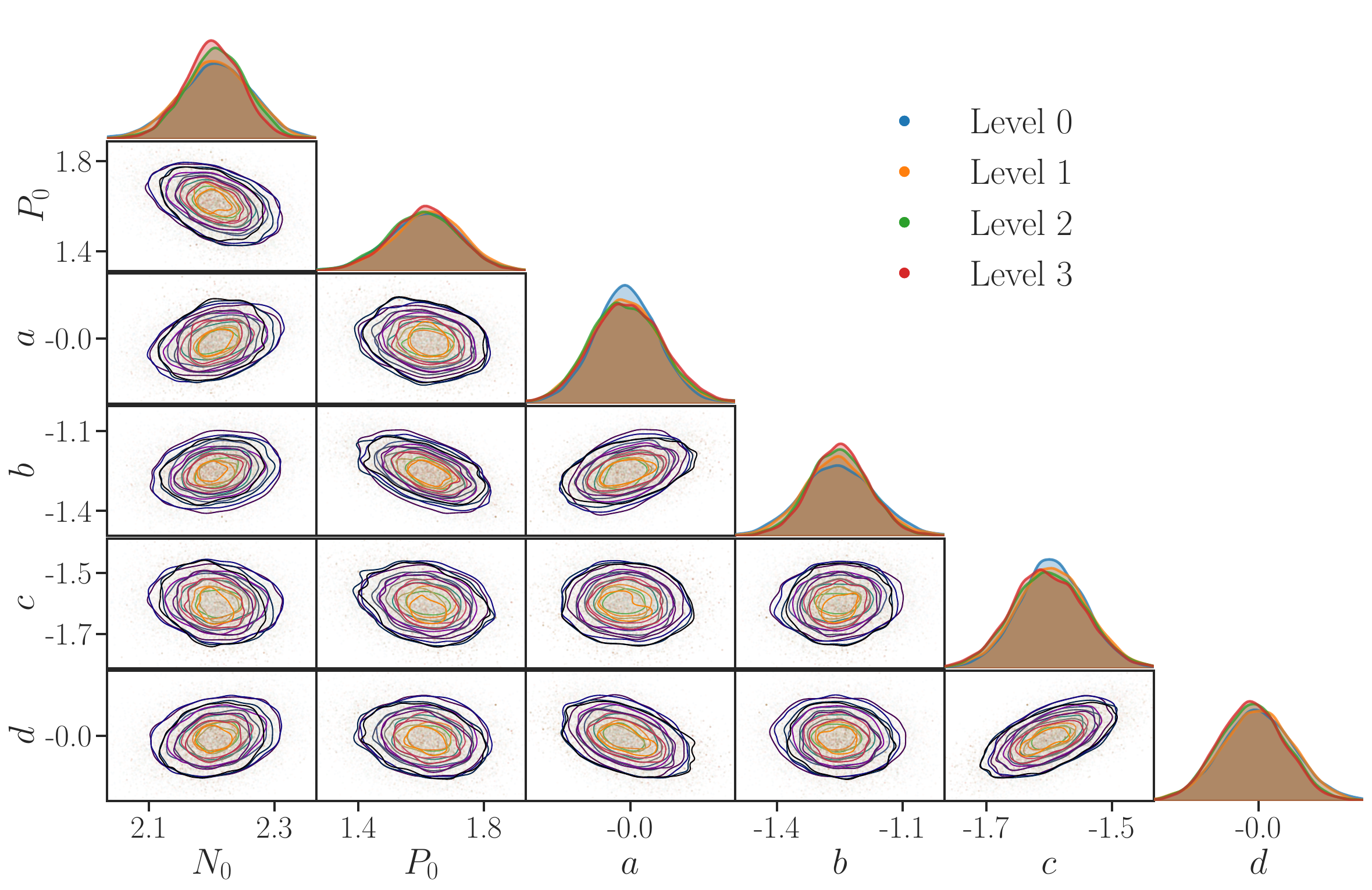}
    \caption{MLDA-AEM coupling method~\cite{lykkegaard_multilevel_2023}}
    \label{fig:prey-predator-aem-scatter}
  \end{subfigure}
  
  \vspace{0.5cm} 
  
  \begin{subfigure}[b]{0.75\textwidth}
    \centering
    \includegraphics[width=\textwidth]{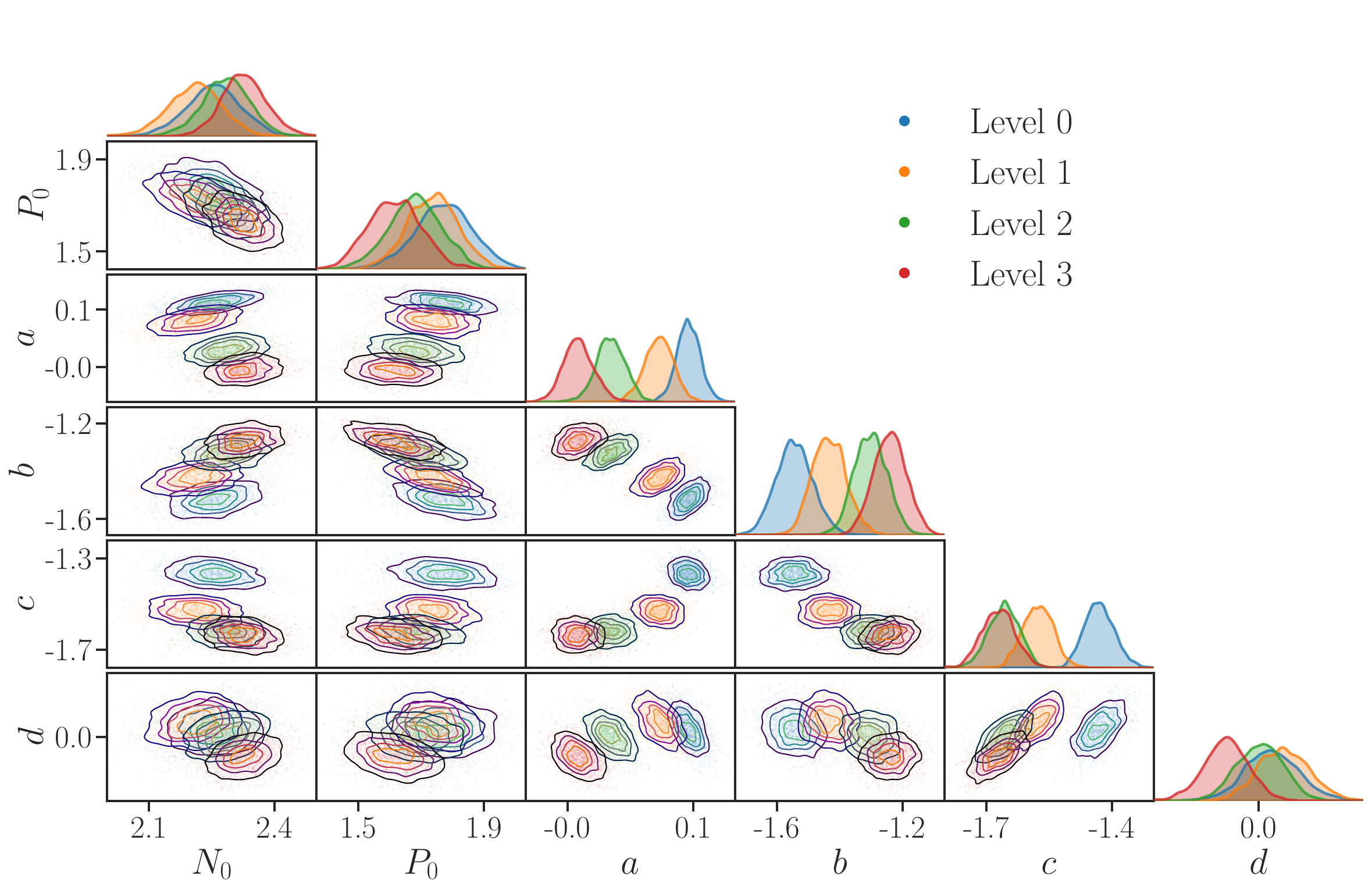}
    \caption{SYNCE coupling method}
    \label{fig:prey-predator-synce-scatter}
  \end{subfigure}
  \caption{Comparison of posterior distributions for the four-level prey-predator problem. (a) The MLDA-AEM method forces all posteriors to align with the fine-level posterior. (b) The SYNCE method preserves the natural posterior at each fidelity level.}
  \label{fig:prey-predator-comparison}
\end{figure}
\endgroup

\newpage

\bibliographystyle{elsarticle-num-names} 
\bibliography{references}

\end{document}

%% file: Images/coarse_proposal.pdf_tex
\begingroup%
  \makeatletter%
  \providecommand\color[2][]{%
    \errmessage{(Inkscape) Color is used for the text in Inkscape, but the package 'color.sty' is not loaded}%
    \renewcommand\color[2][]{}%
  }%
  \providecommand\transparent[1]{%
    \errmessage{(Inkscape) Transparency is used (non-zero) for the text in Inkscape, but the package 'transparent.sty' is not loaded}%
    \renewcommand\transparent[1]{}%
  }%
  \providecommand\rotatebox[2]{#2}%
  \newcommand*\fsize{\dimexpr\f@size pt\relax}%
  \newcommand*\lineheight[1]{\fontsize{\fsize}{#1\fsize}\selectfont}%
  \ifx\svgwidth\undefined%
    \setlength{\unitlength}{744.99998474bp}%
    \ifx\svgscale\undefined%
      \relax%
    \else%
      \setlength{\unitlength}{\unitlength * \real{\svgscale}}%
    \fi%
  \else%
    \setlength{\unitlength}{\svgwidth}%
  \fi%
  \global\let\svgwidth\undefined%
  \global\let\svgscale\undefined%
  \makeatother%
  \begin{picture}(1,0.65234901)%
    \lineheight{1}%
    \setlength\tabcolsep{0pt}%
    \put(0,0){\includegraphics[width=\unitlength,page=1]{Images/coarse_proposal.pdf}}%
    \put(0.32385739,0.02){\color[rgb]{0.50196078,0.50196078,0.50196078}\transparent{0.5}\makebox(0,0)[lt]{\lineheight{1.25}\smash{\begin{tabular}[t]{l}$\pi_{\ell-1}$\end{tabular}}}}%
    \put(0.72830177,0.02){\color[rgb]{0.50196078,0.50196078,0.50196078}\transparent{0.5}\makebox(0,0)[lt]{\lineheight{1.25}\smash{\begin{tabular}[t]{l}$\pi_{\ell}$\end{tabular}}}}%
  \end{picture}%
\endgroup%

%% file: Images/independent_proposal.pdf_tex
\begingroup%
  \makeatletter%
  \providecommand\color[2][]{%
    \errmessage{(Inkscape) Color is used for the text in Inkscape, but the package 'color.sty' is not loaded}%
    \renewcommand\color[2][]{}%
  }%
  \providecommand\transparent[1]{%
    \errmessage{(Inkscape) Transparency is used (non-zero) for the text in Inkscape, but the package 'transparent.sty' is not loaded}%
    \renewcommand\transparent[1]{}%
  }%
  \providecommand\rotatebox[2]{#2}%
  \newcommand*\fsize{\dimexpr\f@size pt\relax}%
  \newcommand*\lineheight[1]{\fontsize{\fsize}{#1\fsize}\selectfont}%
  \ifx\svgwidth\undefined%
    \setlength{\unitlength}{744.99998474bp}%
    \ifx\svgscale\undefined%
      \relax%
    \else%
      \setlength{\unitlength}{\unitlength * \real{\svgscale}}%
    \fi%
  \else%
    \setlength{\unitlength}{\svgwidth}%
  \fi%
  \global\let\svgwidth\undefined%
  \global\let\svgscale\undefined%
  \makeatother%
  \begin{picture}(1,0.65369127)%
    \lineheight{1}%
    \setlength\tabcolsep{0pt}%
    \put(0,0){\includegraphics[width=\unitlength,page=1]{Images/independent_proposal.pdf}}%
    \put(0.32385739,0.02){\color[rgb]{0.50196078,0.50196078,0.50196078}\transparent{0.5}\makebox(0,0)[lt]{\lineheight{1.25}\smash{\begin{tabular}[t]{l}$\pi_{\ell-1}$\end{tabular}}}}%
    \put(0.72830177,0.02){\color[rgb]{0.50196078,0.50196078,0.50196078}\transparent{0.5}\makebox(0,0)[lt]{\lineheight{1.25}\smash{\begin{tabular}[t]{l}$\pi_{\ell}$\end{tabular}}}}%
  \end{picture}%
\endgroup%

%% file: Images/maximal_proposal.pdf_tex
\begingroup%
  \makeatletter%
  \providecommand\color[2][]{%
    \errmessage{(Inkscape) Color is used for the text in Inkscape, but the package 'color.sty' is not loaded}%
    \renewcommand\color[2][]{}%
  }%
  \providecommand\transparent[1]{%
    \errmessage{(Inkscape) Transparency is used (non-zero) for the text in Inkscape, but the package 'transparent.sty' is not loaded}%
    \renewcommand\transparent[1]{}%
  }%
  \providecommand\rotatebox[2]{#2}%
  \newcommand*\fsize{\dimexpr\f@size pt\relax}%
  \newcommand*\lineheight[1]{\fontsize{\fsize}{#1\fsize}\selectfont}%
  \ifx\svgwidth\undefined%
    \setlength{\unitlength}{744.99998474bp}%
    \ifx\svgscale\undefined%
      \relax%
    \else%
      \setlength{\unitlength}{\unitlength * \real{\svgscale}}%
    \fi%
  \else%
    \setlength{\unitlength}{\svgwidth}%
  \fi%
  \global\let\svgwidth\undefined%
  \global\let\svgscale\undefined%
  \makeatother%
  \begin{picture}(1,0.65369127)%
    \lineheight{1}%
    \setlength\tabcolsep{0pt}%
    \put(0,0){\includegraphics[width=\unitlength,page=1]{Images/maximal_proposal.pdf}}%
    \put(0.32385739,0.02){\color[rgb]{0.50196078,0.50196078,0.50196078}\transparent{0.5}\makebox(0,0)[lt]{\lineheight{1.25}\smash{\begin{tabular}[t]{l}$\pi_{\ell-1}$\end{tabular}}}}%
    \put(0.72830177,0.02){\color[rgb]{0.50196078,0.50196078,0.50196078}\transparent{0.5}\makebox(0,0)[lt]{\lineheight{1.25}\smash{\begin{tabular}[t]{l}$\pi_{\ell}$\end{tabular}}}}%
    \put(0,0){\includegraphics[width=\unitlength,page=2]{Images/maximal_proposal.pdf}}%
  \end{picture}%
\endgroup%

%% file: Images/synce_proposal.pdf_tex
\begingroup%
  \makeatletter%
  \providecommand\color[2][]{%
    \errmessage{(Inkscape) Color is used for the text in Inkscape, but the package 'color.sty' is not loaded}%
    \renewcommand\color[2][]{}%
  }%
  \providecommand\transparent[1]{%
    \errmessage{(Inkscape) Transparency is used (non-zero) for the text in Inkscape, but the package 'transparent.sty' is not loaded}%
    \renewcommand\transparent[1]{}%
  }%
  \providecommand\rotatebox[2]{#2}%
  \newcommand*\fsize{\dimexpr\f@size pt\relax}%
  \newcommand*\lineheight[1]{\fontsize{\fsize}{#1\fsize}\selectfont}%
  \ifx\svgwidth\undefined%
    \setlength{\unitlength}{744.99998474bp}%
    \ifx\svgscale\undefined%
      \relax%
    \else%
      \setlength{\unitlength}{\unitlength * \real{\svgscale}}%
    \fi%
  \else%
    \setlength{\unitlength}{\svgwidth}%
  \fi%
  \global\let\svgwidth\undefined%
  \global\let\svgscale\undefined%
  \makeatother%
  \begin{picture}(1,0.65369127)%
    \lineheight{1}%
    \setlength\tabcolsep{0pt}%
    \put(0,0){\includegraphics[width=\unitlength,page=1]{Images/synce_proposal.pdf}}%
    \put(0.32385739,0.02){\color[rgb]{0.50196078,0.50196078,0.50196078}\transparent{0.5}\makebox(0,0)[lt]{\lineheight{1.25}\smash{\begin{tabular}[t]{l}$\pi_{\ell-1}$\end{tabular}}}}%
    \put(0.72830177,0.02){\color[rgb]{0.50196078,0.50196078,0.50196078}\transparent{0.5}\makebox(0,0)[lt]{\lineheight{1.25}\smash{\begin{tabular}[t]{l}$\pi_{\ell}$\end{tabular}}}}%
  \end{picture}%
\endgroup%

%% file: main_article.bbl
\begin{thebibliography}{45}
\expandafter\ifx\csname natexlab\endcsname\relax\def\natexlab#1{#1}\fi
\providecommand{\url}[1]{\texttt{#1}}
\providecommand{\href}[2]{#2}
\providecommand{\path}[1]{#1}
\providecommand{\DOIprefix}{doi:}
\providecommand{\ArXivprefix}{arXiv:}
\providecommand{\URLprefix}{URL: }
\providecommand{\Pubmedprefix}{pmid:}
\providecommand{\doi}[1]{\href{http://dx.doi.org/#1}{\path{#1}}}
\providecommand{\Pubmed}[1]{\href{pmid:#1}{\path{#1}}}
\providecommand{\bibinfo}[2]{#2}
\ifx\xfnm\relax \def\xfnm[#1]{\unskip,\space#1}\fi
\bibitem[{Robert and Casella(2004)}]{robert_monte_2004}
\bibinfo{author}{C.~P. Robert}, \bibinfo{author}{G.~Casella}, \bibinfo{title}{Monte Carlo Statistical Methods}, Springer Texts in Statistics, \bibinfo{publisher}{Springer New York}, \bibinfo{year}{2004}.
\bibitem[{Hammersley and Handscomb(1961)}]{hammersley1961monte}
\bibinfo{author}{J.~M. Hammersley}, \bibinfo{author}{D.~C. Handscomb},
\newblock \bibinfo{title}{Monte carlo methods},
\newblock in: \bibinfo{booktitle}{Proceedings}, volume~\bibinfo{volume}{7}, \bibinfo{organization}{US Army Research Office.}, \bibinfo{year}{1961}, p.~\bibinfo{pages}{17}.
\bibitem[{Peherstorfer et~al.(2018)Peherstorfer, Willcox, and Gunzburger}]{peherstorfer_survey_2018}
\bibinfo{author}{B.~Peherstorfer}, \bibinfo{author}{K.~Willcox}, \bibinfo{author}{M.~Gunzburger},
\newblock \bibinfo{title}{Survey of multifidelity methods in uncertainty propagation, inference, and optimization},
\newblock \bibinfo{journal}{SIAM Review} \bibinfo{volume}{60} (\bibinfo{year}{2018}) \bibinfo{pages}{550--591}. \DOIprefix\doi{10.1137/16M1082469}.
\bibitem[{nel(1987)}]{nelson1990control}
\bibinfo{title}{On control variate estimators},
\newblock \bibinfo{journal}{Computers \& Operations Research} \bibinfo{volume}{14} (\bibinfo{year}{1987}) \bibinfo{pages}{219--225}. \DOIprefix\doi{https://doi.org/10.1016/0305-0548(87)90024-4}.
\bibitem[{Emsermann and Simon(2002)}]{emsermann2002improving}
\bibinfo{author}{M.~Emsermann}, \bibinfo{author}{B.~Simon},
\newblock \bibinfo{title}{Improving simulation efficiency with quasi control variates},
\newblock \bibinfo{journal}{Stochastic Models} \bibinfo{volume}{18} (\bibinfo{year}{2002}) \bibinfo{pages}{425--448}. \DOIprefix\doi{10.1081/STM-120014220}.
\bibitem[{Geraci et~al.(2015)Geraci, Eldred, and Iaccarino}]{geraci2015multifidelity}
\bibinfo{author}{G.~Geraci}, \bibinfo{author}{M.~Eldred}, \bibinfo{author}{G.~Iaccarino},
\newblock \bibinfo{title}{A multifidelity control variate approach for the multilevel monte carlo technique},
\newblock \bibinfo{journal}{Center for Turbulence Research Annual Research Briefs}  (\bibinfo{year}{2015}) \bibinfo{pages}{169--181}.
\bibitem[{Gorodetsky et~al.(2020)Gorodetsky, Geraci, Eldred, and Jakeman}]{gorodetsky_generalized_2020}
\bibinfo{author}{A.~A. Gorodetsky}, \bibinfo{author}{G.~Geraci}, \bibinfo{author}{M.~S. Eldred}, \bibinfo{author}{J.~D. Jakeman},
\newblock \bibinfo{title}{A generalized approximate control variate framework for multifidelity uncertainty quantification},
\newblock \bibinfo{journal}{Journal of Computational Physics} \bibinfo{volume}{408} (\bibinfo{year}{2020}) \bibinfo{pages}{109257}. \DOIprefix\doi{https://doi.org/10.1016/j.jcp.2020.109257}.
\bibitem[{Pham and Gorodetsky(2022)}]{pham2022ensemble}
\bibinfo{author}{T.~Pham}, \bibinfo{author}{A.~A. Gorodetsky},
\newblock \bibinfo{title}{Ensemble approximate control variate estimators: Applications to multifidelity importance sampling},
\newblock \bibinfo{journal}{SIAM/ASA Journal on Uncertainty Quantification} \bibinfo{volume}{10} (\bibinfo{year}{2022}) \bibinfo{pages}{1250--1292}. \DOIprefix\doi{10.1137/21M1390426}.
\bibitem[{Dixon et~al.(2024)Dixon, Warner, Bomarito, and Gorodetsky}]{thomasgroupedACV2024}
\bibinfo{author}{T.~O. Dixon}, \bibinfo{author}{J.~E. Warner}, \bibinfo{author}{G.~F. Bomarito}, \bibinfo{author}{A.~A. Gorodetsky},
\newblock \bibinfo{title}{Covariance expressions for multifidelity sampling with multioutput, multistatistic estimators: Application to approximate control variates},
\newblock \bibinfo{journal}{SIAM/ASA Journal on Uncertainty Quantification} \bibinfo{volume}{12} (\bibinfo{year}{2024}) \bibinfo{pages}{1005--1049}. \DOIprefix\doi{10.1137/23M1607994}.
\bibitem[{Heinrich(2001)}]{heinrich2001multilevel}
\bibinfo{author}{S.~Heinrich},
\newblock \bibinfo{title}{Multilevel monte carlo methods},
\newblock in: \bibinfo{editor}{S.~Margenov}, \bibinfo{editor}{J.~Wa{\'{s}}niewski}, \bibinfo{editor}{P.~Yalamov} (Eds.), \bibinfo{booktitle}{Large-Scale Scientific Computing}, \bibinfo{publisher}{Springer Berlin Heidelberg}, \bibinfo{address}{Berlin, Heidelberg}, \bibinfo{year}{2001}, pp. \bibinfo{pages}{58--67}.
\bibitem[{Giles(2015)}]{giles_multilevel_2015}
\bibinfo{author}{M.~B. Giles},
\newblock \bibinfo{title}{Multilevel monte carlo methods},
\newblock \bibinfo{journal}{Acta Numerica} \bibinfo{volume}{24} (\bibinfo{year}{2015}) \bibinfo{pages}{259–328}. \DOIprefix\doi{10.1017/S096249291500001X}.
\bibitem[{Cliffe et~al.(2011)Cliffe, Giles, Scheichl, and Teckentrup}]{cliffe_multilevel_2011}
\bibinfo{author}{K.~A. Cliffe}, \bibinfo{author}{M.~B. Giles}, \bibinfo{author}{R.~Scheichl}, \bibinfo{author}{A.~L. Teckentrup},
\newblock \bibinfo{title}{Multilevel {{Monte Carlo}} methods and applications to elliptic {{PDEs}} with random coefficients},
\newblock \bibinfo{journal}{Comput. Visual Sci.} \bibinfo{volume}{14} (\bibinfo{year}{2011}) \bibinfo{pages}{3--15}. \DOIprefix\doi{10.1007/s00791-011-0160-x}.
\bibitem[{Nobile and Tesei(2015)}]{nobile2015multi}
\bibinfo{author}{F.~Nobile}, \bibinfo{author}{F.~Tesei},
\newblock \bibinfo{title}{A multi level monte carlo method with control variate for elliptic pdes with log-normal coefficients},
\newblock \bibinfo{journal}{Stochastic Partial Differential Equations: Analysis and Computations} \bibinfo{volume}{3} (\bibinfo{year}{2015}) \bibinfo{pages}{398--444}. \DOIprefix\doi{10.1007/s40072-015-0055-9}.
\bibitem[{Schaden and Ullmann(2020)}]{schaden2020multilevel}
\bibinfo{author}{D.~Schaden}, \bibinfo{author}{E.~Ullmann},
\newblock \bibinfo{title}{On multilevel best linear unbiased estimators},
\newblock \bibinfo{journal}{SIAM/ASA Journal on Uncertainty Quantification} \bibinfo{volume}{8} (\bibinfo{year}{2020}) \bibinfo{pages}{601--635}. \DOIprefix\doi{10.1137/19M1263534}.
\bibitem[{Dodwell et~al.(2015)Dodwell, Ketelsen, Scheichl, and Teckentrup}]{dodwell2015hierarchical}
\bibinfo{author}{T.~J. Dodwell}, \bibinfo{author}{C.~Ketelsen}, \bibinfo{author}{R.~Scheichl}, \bibinfo{author}{A.~L. Teckentrup},
\newblock \bibinfo{title}{A hierarchical multilevel markov chain monte carlo algorithm with applications to uncertainty quantification in subsurface flow},
\newblock \bibinfo{journal}{SIAM/ASA Journal on Uncertainty Quantification} \bibinfo{volume}{3} (\bibinfo{year}{2015}) \bibinfo{pages}{1075--1108}. \DOIprefix\doi{10.1137/130915005}.
\bibitem[{Christen and Fox(2005)}]{christen_markov_2005}
\bibinfo{author}{J.~A. Christen}, \bibinfo{author}{C.~Fox},
\newblock \bibinfo{title}{Markov chain monte carlo using an approximation},
\newblock \bibinfo{journal}{Journal of Computational and Graphical Statistics} \bibinfo{volume}{14} (\bibinfo{year}{2005}) \bibinfo{pages}{795--810}. \DOIprefix\doi{10.1198/106186005X76983}.
\bibitem[{Lykkegaard et~al.(2023)Lykkegaard, Dodwell, Fox, Mingas, and Scheichl}]{lykkegaard_multilevel_2023}
\bibinfo{author}{M.~B. Lykkegaard}, \bibinfo{author}{T.~J. Dodwell}, \bibinfo{author}{C.~Fox}, \bibinfo{author}{G.~Mingas}, \bibinfo{author}{R.~Scheichl},
\newblock \bibinfo{title}{Multilevel delayed acceptance mcmc},
\newblock \bibinfo{journal}{SIAM/ASA Journal on Uncertainty Quantification} \bibinfo{volume}{11} (\bibinfo{year}{2023}) \bibinfo{pages}{1--30}. \DOIprefix\doi{10.1137/22M1476770}.
\bibitem[{Cui et~al.(2024)Cui, Detommaso, and Scheichl}]{DILIMultilevel2024}
\bibinfo{author}{T.~Cui}, \bibinfo{author}{G.~Detommaso}, \bibinfo{author}{R.~Scheichl},
\newblock \bibinfo{title}{Multilevel dimension-independent likelihood-informed mcmc for large-scale inverse problems},
\newblock \bibinfo{journal}{Inverse Problems} \bibinfo{volume}{40} (\bibinfo{year}{2024}) \bibinfo{pages}{035005}. \DOIprefix\doi{10.1088/1361-6420/ad1e2c}.
\bibitem[{Cui et~al.(2016)Cui, Law, and Marzouk}]{DILI2016}
\bibinfo{author}{T.~Cui}, \bibinfo{author}{K.~J. Law}, \bibinfo{author}{Y.~M. Marzouk},
\newblock \bibinfo{title}{Dimension-independent likelihood-informed mcmc},
\newblock \bibinfo{journal}{Journal of Computational Physics} \bibinfo{volume}{304} (\bibinfo{year}{2016}) \bibinfo{pages}{109--137}. \DOIprefix\doi{https://doi.org/10.1016/j.jcp.2015.10.008}.
\bibitem[{Madrigal-Cianci et~al.(2023)Madrigal-Cianci, Nobile, and Tempone}]{madrigal-cianci_analysis_2023}
\bibinfo{author}{J.~P. Madrigal-Cianci}, \bibinfo{author}{F.~Nobile}, \bibinfo{author}{R.~Tempone},
\newblock \bibinfo{title}{Analysis of a class of multilevel markov chain monte carlo algorithms based on independent metropolis–hastings},
\newblock \bibinfo{journal}{SIAM/ASA Journal on Uncertainty Quantification} \bibinfo{volume}{11} (\bibinfo{year}{2023}) \bibinfo{pages}{91--138}. \DOIprefix\doi{10.1137/21M1420927}.
\bibitem[{Madrigal~Cianci(2022)}]{cianci_thesis}
\bibinfo{author}{J.~P. Madrigal~Cianci}, \bibinfo{title}{Hierarchical Markov chain Monte Carlo methods for Bayesian inverse problems}, Ph.D. thesis, EPFL, \bibinfo{address}{Lausanne}, \bibinfo{year}{2022}. \DOIprefix\doi{10.5075/epfl-thesis-8951}.
\bibitem[{Jacob et~al.(2020)Jacob, O’Leary, and Atchadé}]{jacob_unbiased_2020}
\bibinfo{author}{P.~E. Jacob}, \bibinfo{author}{J.~O’Leary}, \bibinfo{author}{Y.~F. Atchadé},
\newblock \bibinfo{title}{Unbiased markov chain monte carlo methods with couplings},
\newblock \bibinfo{journal}{Journal of the Royal Statistical Society Series B: Statistical Methodology} \bibinfo{volume}{82} (\bibinfo{year}{2020}) \bibinfo{pages}{543--600}. \DOIprefix\doi{10.1111/rssb.12336}.
\bibitem[{Heng et~al.(2023)Heng, Jasra, Law, and Tarakanov}]{heng_unbiased_2023}
\bibinfo{author}{J.~Heng}, \bibinfo{author}{A.~Jasra}, \bibinfo{author}{K.~J.~H. Law}, \bibinfo{author}{A.~Tarakanov},
\newblock \bibinfo{title}{On unbiased estimation for discretized models},
\newblock \bibinfo{journal}{{SIAM}/{ASA} J. Uncertainty Quantification} \bibinfo{volume}{11} (\bibinfo{year}{2023}) \bibinfo{pages}{616--645}.
\bibitem[{Juntao et~al.(2025)Juntao, Adie, See, Gualandi, and Mengaldo}]{Juntao2025hybrid}
\bibinfo{author}{Y.~Juntao}, \bibinfo{author}{J.~Adie}, \bibinfo{author}{S.~See}, \bibinfo{author}{A.~Gualandi}, \bibinfo{author}{G.~Mengaldo},
\newblock \bibinfo{title}{A hybrid two-level mcmc framework to accelerate posterior mean estimation with deep learning surrogates for bayesian inverse problems},
\newblock \bibinfo{journal}{Journal of Computational Physics}  (\bibinfo{year}{2025}) \bibinfo{pages}{114502}. \DOIprefix\doi{https://doi.org/10.1016/j.jcp.2025.114502}.
\bibitem[{Geraci et~al.(2018)Geraci, Eldred, Gorodetsky, and Jakeman}]{geraci2018leveraging}
\bibinfo{author}{G.~Geraci}, \bibinfo{author}{M.~S. Eldred}, \bibinfo{author}{A.~A. Gorodetsky}, \bibinfo{author}{J.~D. Jakeman},
\newblock \bibinfo{title}{Leveraging active directions for efficient multifidelity uncertainty quantification},
\newblock in: \bibinfo{booktitle}{6th European Conference on Computational Mechanics (ECCM 6)}, \bibinfo{year}{2018}, pp. \bibinfo{pages}{2735--2746}.
\bibitem[{Lam et~al.(2020)Lam, Zahm, Marzouk, and Willcox}]{lam2020multifidelity}
\bibinfo{author}{R.~R. Lam}, \bibinfo{author}{O.~Zahm}, \bibinfo{author}{Y.~M. Marzouk}, \bibinfo{author}{K.~E. Willcox},
\newblock \bibinfo{title}{Multifidelity dimension reduction via active subspaces},
\newblock \bibinfo{journal}{SIAM Journal on Scientific Computing} \bibinfo{volume}{42} (\bibinfo{year}{2020}) \bibinfo{pages}{A929--A956}. \DOIprefix\doi{10.1137/18M1214123}.
\bibitem[{Pinto and Neal(2001)}]{pinto2001improving}
\bibinfo{author}{R.~L. Pinto}, \bibinfo{author}{R.~M. Neal}, \bibinfo{title}{Improving Markov chain Monte Carlo estimators by coupling to an approximating chain}, \bibinfo{type}{Technical Report}, Technical Report, \bibinfo{year}{2001}.
\bibitem[{Hastings(1970)}]{hastings_monte_1970}
\bibinfo{author}{W.~K. Hastings},
\newblock \bibinfo{title}{Monte carlo sampling methods using markov chains and their applications},
\newblock \bibinfo{journal}{Biometrika} \bibinfo{volume}{57} (\bibinfo{year}{1970}) \bibinfo{pages}{97--109}. \DOIprefix\doi{10.2307/2334940}.
\bibitem[{Villani et~al.(2009)}]{villani2009optimal}
\bibinfo{author}{C.~Villani}, et~al., \bibinfo{title}{Optimal transport: old and new}, volume \bibinfo{volume}{338}, \bibinfo{publisher}{Springer}, \bibinfo{year}{2009}.
\bibitem[{Thorisson(1998)}]{Thorisson1998}
\bibinfo{author}{H.~Thorisson}, \bibinfo{title}{Coupling}, \bibinfo{publisher}{Springer New York}, \bibinfo{address}{New York, NY}, \bibinfo{year}{1998}, pp. \bibinfo{pages}{319--339}.
\bibitem[{Atchadé and Rosenthal(2005)}]{atchade_adaptive_2005}
\bibinfo{author}{Y.~F. Atchadé}, \bibinfo{author}{J.~S. Rosenthal},
\newblock \bibinfo{title}{On adaptive markov chain monte carlo algorithms},
\newblock \bibinfo{journal}{Bernoulli} \bibinfo{volume}{11} (\bibinfo{year}{2005}) \bibinfo{pages}{815--828}.
\bibitem[{Rosenthal(1995)}]{Rosenthalminor1995}
\bibinfo{author}{J.~S. Rosenthal},
\newblock \bibinfo{title}{Minorization conditions and convergence rates for markov chain monte carlo},
\newblock \bibinfo{journal}{Journal of the American Statistical Association} \bibinfo{volume}{90} (\bibinfo{year}{1995}) \bibinfo{pages}{558--566}. \DOIprefix\doi{10.1080/01621459.1995.10476548}.
\bibitem[{Peherstorfer and Marzouk(2019)}]{peherstorfer_transport-based_2019}
\bibinfo{author}{B.~Peherstorfer}, \bibinfo{author}{Y.~Marzouk},
\newblock \bibinfo{title}{A transport-based multifidelity preconditioner for markov chain monte carlo},
\newblock \bibinfo{journal}{Adv Comput Math} \bibinfo{volume}{45} (\bibinfo{year}{2019}) \bibinfo{pages}{2321--2348}. \DOIprefix\doi{10.1007/s10444-019-09711-y}.
\bibitem[{Parno and Marzouk(2018)}]{parno_transport_2018}
\bibinfo{author}{M.~D. Parno}, \bibinfo{author}{Y.~M. Marzouk},
\newblock \bibinfo{title}{Transport map accelerated markov chain monte carlo},
\newblock \bibinfo{journal}{SIAM/ASA Journal on Uncertainty Quantification} \bibinfo{volume}{6} (\bibinfo{year}{2018}) \bibinfo{pages}{645--682}. \DOIprefix\doi{10.1137/17M1134640}.
\bibitem[{O’Leary and Wang(2024)}]{oleary_metropolis-hastings_2023}
\bibinfo{author}{J.~O’Leary}, \bibinfo{author}{G.~Wang},
\newblock \bibinfo{title}{{Metropolis–Hastings transition kernel couplings}},
\newblock \bibinfo{journal}{Annales de l'Institut Henri Poincaré, Probabilités et Statistiques} \bibinfo{volume}{60} (\bibinfo{year}{2024}) \bibinfo{pages}{1101 -- 1124}. \DOIprefix\doi{10.1214/22-AIHP1360}.
\bibitem[{Haario et~al.(2001)Haario, Saksman, and Tamminen}]{haario_adaptive_2001}
\bibinfo{author}{H.~Haario}, \bibinfo{author}{E.~Saksman}, \bibinfo{author}{J.~Tamminen},
\newblock \bibinfo{title}{{An adaptive Metropolis algorithm}},
\newblock \bibinfo{journal}{Bernoulli} \bibinfo{volume}{7} (\bibinfo{year}{2001}) \bibinfo{pages}{223 -- 242}.
\bibitem[{Meyn and Tweedie(2012)}]{meyn2012markov}
\bibinfo{author}{S.~P. Meyn}, \bibinfo{author}{R.~L. Tweedie}, \bibinfo{title}{Markov chains and stochastic stability}, \bibinfo{publisher}{Springer Science \& Business Media}, \bibinfo{year}{2012}. \DOIprefix\doi{10.1017/CBO9780511626630}.
\bibitem[{Jarner and Hansen(2000)}]{jarnerGeometricErgodicityMetropolis2000}
\bibinfo{author}{S.~F. Jarner}, \bibinfo{author}{E.~Hansen},
\newblock \bibinfo{title}{Geometric ergodicity of {{Metropolis}} algorithms},
\newblock \bibinfo{journal}{Stochastic Processes and their Applications} \bibinfo{volume}{85} (\bibinfo{year}{2000}) \bibinfo{pages}{341--361}. \DOIprefix\doi{10.1016/S0304-4149(99)00082-4}.
\bibitem[{Roberts and Rosenthal(2004)}]{robertsGeneralStateSpace2004}
\bibinfo{author}{G.~O. Roberts}, \bibinfo{author}{J.~S. Rosenthal},
\newblock \bibinfo{title}{General state space {{Markov}} chains and {{MCMC}} algorithms},
\newblock \bibinfo{journal}{Probab. Surveys} \bibinfo{volume}{1} (\bibinfo{year}{2004}). \DOIprefix\doi{10.1214/154957804100000024}. \href{http://arxiv.org/abs/math/0404033}{{\tt arXiv:math/0404033}}.
\bibitem[{Qin(2024)}]{qin2024convergenceboundsmontecarlo}
\bibinfo{author}{Q.~Qin}, \bibinfo{title}{Convergence bounds for monte carlo markov chains}, \bibinfo{year}{2024}. \href{http://arxiv.org/abs/2409.14656}{{\tt arXiv:2409.14656}}.
\bibitem[{Rosenthal et~al.(2011)}]{rosenthal_optimal_nodate}
\bibinfo{author}{J.~S. Rosenthal}, et~al.,
\newblock \bibinfo{title}{Optimal proposal distributions and adaptive mcmc},
\newblock \bibinfo{journal}{Handbook of Markov Chain Monte Carlo} \bibinfo{volume}{4} (\bibinfo{year}{2011}).
\bibitem[{Andrieu and Thoms(2008)}]{andrieu_tutorial_2008}
\bibinfo{author}{C.~Andrieu}, \bibinfo{author}{J.~Thoms},
\newblock \bibinfo{title}{A tutorial on adaptive {{MCMC}}},
\newblock \bibinfo{journal}{Stat Comput} \bibinfo{volume}{18} (\bibinfo{year}{2008}) \bibinfo{pages}{343--373}. \DOIprefix\doi{10.1007/s11222-008-9110-y}.
\bibitem[{Pasarica and Gelman(2010)}]{gelman_adaptively_2007}
\bibinfo{author}{C.~Pasarica}, \bibinfo{author}{A.~Gelman},
\newblock \bibinfo{title}{Adaptively scaling the metropolis algorithm using expected squared jumped distance},
\newblock \bibinfo{journal}{Statistica Sinica} \bibinfo{volume}{20} (\bibinfo{year}{2010}) \bibinfo{pages}{343--364}.
\bibitem[{Baratta et~al.(2023)Baratta, Dean, Dokken, Habera, Hale, Richardson, Rognes, Scroggs, Sime, and Wells}]{baratta_2023_10447666}
\bibinfo{author}{I.~A. Baratta}, \bibinfo{author}{J.~P. Dean}, \bibinfo{author}{J.~S. Dokken}, \bibinfo{author}{M.~Habera}, \bibinfo{author}{J.~S. Hale}, \bibinfo{author}{C.~N. Richardson}, \bibinfo{author}{M.~E. Rognes}, \bibinfo{author}{M.~W. Scroggs}, \bibinfo{author}{N.~Sime}, \bibinfo{author}{G.~N. Wells}, \bibinfo{title}{{DOLFINx: The next generation FEniCS problem solving environment}}, \bibinfo{year}{2023}. \DOIprefix\doi{10.5281/zenodo.10447665}.
\bibitem[{Teckentrup et~al.(2015)Teckentrup, Jantsch, Webster, and Gunzburger}]{teckentrup2015multilevel}
\bibinfo{author}{A.~L. Teckentrup}, \bibinfo{author}{P.~Jantsch}, \bibinfo{author}{C.~G. Webster}, \bibinfo{author}{M.~Gunzburger},
\newblock \bibinfo{title}{A multilevel stochastic collocation method for partial differential equations with random input data},
\newblock \bibinfo{journal}{SIAM/ASA Journal on Uncertainty Quantification} \bibinfo{volume}{3} (\bibinfo{year}{2015}) \bibinfo{pages}{1046--1074}. \DOIprefix\doi{10.1137/140969002}.

\end{thebibliography}
